\documentclass[twocolumn]{aastex631}
\usepackage{}
\usepackage{amsmath}
\usepackage{threeparttable}
\usepackage{graphicx}	
\usepackage{comment}

\newcommand{\tess}{\emph{TESS}}

\newcommand\teff{$T_{\rm eff}$}
\newcommand\logg{log\,{\it g}}

\newcommand{\kms}{\,km\,s$^{-1}$} 
\newcommand{\ms}{\,m\,s$^{-1}$} 
\newcommand\target{TOI~4633}
\newcommand{\triceratops}{\texttt{TRICERATOPS}}

\newcommand{\Tzeroc}[1][days]{ $ 1864.8265_{-0.0083}^{+0.0088} $~#1 } 
\newcommand{\Pc}[1][days]{ $ 271.9445_{-0.0040}^{+0.0039} $~#1 } 
\newcommand{\esinc}[1][ ]{ $ -0.06_{-0.27}^{+0.23} $~#1 } 
\newcommand{\ecosc}[1][ ]{ $ 0.04_{-0.35}^{+0.31} $~#1 } 
\newcommand{\bc}[1][ ]{ $ 0.33_{-0.21}^{+0.24} $~#1 } 
 
\newcommand{\rrc}[1][ ]{ $ 0.02111_{-0.00069}^{+0.00076} $~#1 } 
\newcommand{\kc}[1][${\rm m\,s^{-1}}$]{ $ 4.58_{-2.29}^{+2.56} $~#1 } 
\newcommand{\mpc}[1][$M_{\oplus}$]{ $ 47.8_{-23.8}^{+27.6} $~#1 } 
\newcommand{\rpc}[1][$R_{\oplus}$]{ $ 2.42_{-0.14}^{+0.15} $~#1 } 
 
\newcommand{\ec}[1][ ]{ $ 0.117_{-0.085}^{+0.186} $~#1 } 
\newcommand{\wc}[1][deg]{ $ -21_{-108}^{+131} $~#1 } 
 
\newcommand{\ic}[1][deg]{ $ 89.888_{-0.064}^{+0.069} $~#1 } 
 
\newcommand{\ac}[1][AU]{ $ 0.847 \pm 0.061 $~#1 } 
 
\newcommand{\insolationc}[1][${\rm F_{\oplus}}$]{ $ 1.56_{-0.16}^{+0.20}$~#1} 
 
\newcommand{\denstrc}[1][${\rm g\,cm^{-3}}$]{ $ 1.35 \pm 0.21 $~#1 }

\newcommand{\ttotc}[1][hours]{ $ 11.45_{-0.28}^{+0.46} $~#1 }

\newcommand{\Tzerob}[1][days]{ $ 2796.64_{-1.19}^{+1.10} $~#1 } 
\newcommand{\Pb}[1][days]{ $ 34.15 \pm 0.15 $~#1 } 
\newcommand{\esinb}[1][ ]{ $ -0.12_{-0.22}^{+0.26} $~#1 } 
\newcommand{\ecosb}[1][ ]{ $ 0.07_{-0.24}^{+0.22} $~#1 } 
\newcommand{\kb}[1][${\rm m\,s^{-1}}$]{ $ 19.97_{-2.30}^{+2.29} $~#1 } 
\newcommand{\mpb}[1][$M_{\oplus}$]{ $ 106.8_{-12.8}^{+13.0} $~#1 } 
 
\newcommand{\eb}[1][ ]{ $ 0.096_{-0.065}^{+0.102} $~#1 } 
\newcommand{\wb}[1][deg]{ $ -43.9_{-72.8}^{+104.8} $~#1 } 
\newcommand{\qone}[1][]{ $ 0.39_{-0.23}^{+0.37} $~#1 } 
\newcommand{\qtwo}[1][]{ $ 0.34_{-0.22}^{+0.31} $~#1 }

\newcommand{\OHP}[1][${\rm km\,s^{-1}}$]{ $ -0.0007 \pm 0.0034 $~#1 } 
\newcommand{\HIRES}[1][${\rm km\,s^{-1}}$]{ $ 0.0023_{-0.0020}^{+0.0019} $~#1 } 
\newcommand{\jOHP}[1][${\rm m\,s^{-1}}$]{ $ 13.40_{-2.39}^{+3.10} $~#1 } 
\newcommand{\jHIRES}[1][${\rm m\,s^{-1}}$]{ $ 12.36_{-1.20}^{+1.36} $~#1 } 
\newcommand{\jtr}[1][]{ $ 204_{-15}^{+14} $~#1 } 

\newcommand{\rpccorr}[1][R$_{\oplus}$] {$3.2 _{ - 0.19 } ^ { + 0.20 }$~#1} 
\newcommand{\oscar}[1]{{\color{black}{{#1}}}}

\usepackage{newunicodechar}
\DeclareRobustCommand{\okina}{%
 \raisebox{\dimexpr\fontcharht\font`A-\height}{%
 \scalebox{0.8}{`}%
 }%
}

\newunicodechar{ʻ}{\okina}

\begin{document}


\title{Planet Hunters \textit{TESS} V: a planetary system around a binary star, including a mini-Neptune in the habitable zone}

\author[0000-0002-0786-7307]{Nora L. Eisner}
\altaffiliation{Flatiron Research Fellow}
\altaffiliation{Henry Norris Russel Fellow}
\affiliation{Center for Computational Astrophysics, Flatiron Institute, 162 Fifth Avenue, New York, NY 10010, USA}
\affiliation{Department of Astrophysical Sciences, Princeton University, Princeton, NJ 08544, USA}

\author[0000-0003-4976-9980]{Samuel K. Grunblatt}
\affiliation{Department of Physics and Astronomy, Johns Hopkins University, 3400 N Charles St, Baltimore, MD 21218, USA}

\author[0000-0003-0563-0493]{Oscar Barrag\'an}
\affiliation{Oxford Astrophysics, Denys Wilkinson Building, University of Oxford, OX1 3RH, Oxford, UK}

\author[0000-0003-3799-3635]{Thea H. Faridani}
\affiliation{Department of Physics and Astronomy, University of California, Los Angeles, CA 90095, USA}

\author[0000-0001-5578-359X]{Chris Lintott}
\affiliation{Oxford Astrophysics, Denys Wilkinson Building, University of Oxford, OX1 3RH, Oxford, UK}

\author[0000-0003-1453-0574]{Suzanne Aigrain}
\affiliation{Oxford Astrophysics, Denys Wilkinson Building, University of Oxford, OX1 3RH, Oxford, UK}

\author[0000-0003-0563-0493]{Cole Johnston}
\affiliation{Radboud University Nijmegen, Department of Astrophysics, IMAPP, P.O. Box 9010, 6500 GL Nijmegen, The Netherlands}

\author{Ian R. Mason}
\affiliation{Citizen Scientist, Zooniverse c/o University of Oxford, Keble Road, Oxford OX1 3RH, UK}

\author[0000-0002-3481-9052]{Keivan G. Stassun}
\affiliation{Vanderbilt University, Department of Physics \& Astronomy, 6301 Stevenson Center Ln., Nashville, TN 37235, USA}

\author[0000-0001-9907-7742]{Megan Bedell}
\affiliation{Center for Computational Astrophysics, Flatiron Institute, 162 Fifth Avenue, New York, NY 10010, USA}

\author[0000-0001-6037-2971]{Andrew W. Boyle}
\affiliation{Department of Astronomy, California Institute of Technology, 1200 E. California Boulevard, Pasadena, CA 91125, USA}

\author[0000-0002-5741-3047]{David R. Ciardi}
\affiliation{NASA Exoplanet Science Institute-Caltech/IPAC, Pasadena, CA USA 91350}

\author[0000-0002-2361-5812]{Catherine A. Clark}
\affiliation{Jet Propulsion Laboratory, California Institute of Technology, Pasadena, CA 91109 USA}
\affiliation{NASA Exoplanet Science Institute, IPAC, California Institute of Technology, Pasadena, CA 91125 USA}

\author{Guillaume Hebrard}
\affiliation{Institut d'Astrophysique de Paris, UMR 7095 CNRS, Université Pierre \& Marie Curie, 98 bis boulevard Arago, F-75014 Paris, France}
\affiliation{Observatoire de Haute-Provence, Université d'Aix-Marseille \& CNRS, F-04870 Saint Michel l'Observatoire, France}

\author[0000-0003-2866-9403]{David W. Hogg}
\affiliation{Center for Computational Astrophysics, Flatiron Institute, 162 Fifth Avenue, New York, NY 10010, USA}

\author[0000-0002-2532-2853]{Steve~B.~Howell}
\affil{NASA Ames Research Center, Moffett Field, CA 94035, USA}

\author[0000-0003-0637-5236]{Baptiste Klein}
\affiliation{Oxford Astrophysics, Denys Wilkinson Building, University of Oxford, OX1 3RH, Oxford, UK}

\author{Joe Llama}
\affiliation{Lowell Observatory, 1400 Mars Hill Road, Flagstaff, AZ 86001, USA}

\author[0000-0002-4265-047X]{Joshua N.\ Winn}
\affiliation{Department of Astrophysical Sciences, Princeton University, Princeton, NJ 08544, USA}

\author[0000-0002-3852-3590]{Lily L. Zhao}
\affiliation{Center for Computational Astrophysics, Flatiron Institute, 162 Fifth Avenue, New York, NY 10010, USA}
%
%
%
\author[0000-0001-8898-8284]{Joseph M. Akana Murphy}
\altaffiliation{NSF Graduate Research Fellow}
\affiliation{Department of Astronomy and Astrophysics, University of California, Santa Cruz, CA 95064, USA}

\author[0000-0001-7708-2364]{Corey Beard}
\altaffiliation{NASA FINESST Fellow}
\affiliation{Department of Physics \& Astronomy, The University of California, Irvine, Irvine, CA 92697, USA}

\author[0000-0002-4480-310X]{Casey L. Brinkman}
\affiliation{Institute for Astronomy, University of Hawaiʻi at M\=anoa, 2680 Woodlawn Drive, Honolulu, HI 96822, USA}

\author[0000-0003-1125-2564]{Ashley Chontos}
\altaffiliation{Henry Norris Russel Fellow}
\affiliation{Department of Astrophysical Sciences, Princeton University, Princeton, NJ 08544, USA}

\author[0000-0002-6174-4666]{Pia Cortes-Zuleta}
\affiliation{Aix Marseille Univ, CNRS, CNES, LAM, Marseille, France}

\author{Xavier Delfosse}
\affiliation{University Grenoble Alpes, CNRS, IPAG, F-38000 Grenoble, France}

\author[0000-0002-8965-3969]{Steven Giacalone}
\affiliation{Department of Astronomy, University of California Berkeley, Berkeley, CA 94720, USA}

\author[0000-0002-0388-8004]{Emily A. Gilbert}
\affiliation{Jet Propulsion Laboratory, California Institute of Technology, 4800 Oak Grove Drive, Pasadena, CA 91109, USA}

\author{Neda Heidari}
\affiliation{Aix Marseille Univ, CNRS, CNES, LAM, Marseille, France}

\author[0000-0002-5034-9476]{Rae Holcomb}
\affiliation{Department of Physics \& Astronomy, The University of California, Irvine, Irvine, CA 92697, USA}

\author{Jon M. Jenkins}
\affiliation{NASA Ames Research Center, Moffett Field, CA 94035, USA}

\author[0000-0001-9129-4929]{Flavien Kiefer}
\affiliation{LESIA, Observatoire de Paris, Université PSL, CNRS, Sorbonne Université, Université Paris Cité, 5 place Jules Janssen, 92195 Meudon, France}
\affiliation{American University of Paris, 5, boulevard de La Tour-Maubourg 75007 Paris, France}

\author[0000-0001-8342-7736]{Jack Lubin}
\affiliation{Department of Physics \& Astronomy, University of California Irvine, Irvine, CA 92697, USA}

\author[0000-0002-5084-168X]{Eder Martioli}
\affiliation{Laborat\'{o}rio Nacional de Astrof\'{i}sica, Rua Estados Unidos 154, 37504-364, Itajub\'{a} - MG, Brazil }
\affiliation{Institut d'Astrophysique de Paris, CNRS, UMR 7095, Sorbonne Universit\'{e}, 98 bis bd Arago, 75014 Paris, France}

\author[0000-0001-7047-8681]{Alex S. Polanski}
\affiliation{Department of Physics and Astronomy, University of Kansas, Lawrence, KS, USA}

\author[0000-0003-2657-3889]{Nicholas Saunders}
\altaffiliation{NSF Graduate Research Fellow}
\affiliation{Institute for Astronomy, University of Hawaiʻi at M\=anoa, 2680 Woodlawn Drive, Honolulu, HI 96822, USA}

\author[0000-0002-6892-6948]{Sara Seager}
\affiliation{Department of Physics and Kavli Institute for Astrophysics and Space Research, Massachusetts Institute of Technology, Cambridge, MA 02139, USA}
\affiliation{Department of Earth, Atmospheric and Planetary Sciences, Massachusetts Institute of Technology, Cambridge, MA 02139, USA}
\affiliation{Department of Aeronautics and Astronautics, MIT, 77 Massachusetts Avenue, Cambridge, MA 02139, USA}

\author[0000-0002-1836-3120]{Avi~Shporer} 
\affiliation{Department of Physics and Kavli Institute for Astrophysics and Space Research, Massachusetts Institute of Technology, Cambridge, MA 02139, USA}

\author[0000-0003-0298-4667]{Dakotah Tyler}
\affiliation{Department of Physics and Astronomy, University of California, Los Angeles, CA 90095, USA}

\author[0000-0002-4290-6826]{Judah Van Zandt}
\affiliation{Department of Physics and Astronomy, University of California, Los Angeles, CA 90095, USA}


\author{Safaa Alhassan}
\affiliation{Citizen Scientist, Zooniverse c/o University of Oxford, Keble Road, Oxford OX1 3RH, UK}

\author[0009-0004-3918-7244]{Daval J. Amratlal}
\affiliation{Citizen Scientist, Zooniverse c/o University of Oxford, Keble Road, Oxford OX1 3RH, UK}

\author{Lais I. Antonel}
\affiliation{Citizen Scientist, Zooniverse c/o University of Oxford, Keble Road, Oxford OX1 3RH, UK}

\author{Simon L. S. Bentzen}
\affiliation{Citizen Scientist, Zooniverse c/o University of Oxford, Keble Road, Oxford OX1 3RH, UK}

\author{Milton K. D. Bosch}

\author{David Bundy}
\affiliation{Citizen Scientist, Zooniverse c/o University of Oxford, Keble Road, Oxford OX1 3RH, UK}

\author{Itayi Chitsiga}
\affiliation{Citizen Scientist, Zooniverse c/o University of Oxford, Keble Road, Oxford OX1 3RH, UK}

\author{Jérôme F. Delaunay}
\affiliation{Citizen Scientist, Zooniverse c/o University of Oxford, Keble Road, Oxford OX1 3RH, UK}

\author{Xavier Doisy}
\affiliation{Citizen Scientist, Zooniverse c/o University of Oxford, Keble Road, Oxford OX1 3RH, UK}

\author{Richard Ferstenou}
\affiliation{Citizen Scientist, Zooniverse c/o University of Oxford, Keble Road, Oxford OX1 3RH, UK}

\author{Mark Fynø}
\affiliation{Citizen Scientist, Zooniverse c/o University of Oxford, Keble Road, Oxford OX1 3RH, UK}

\author{James M. Geary}
\affiliation{Citizen Scientist, Zooniverse c/o University of Oxford, Keble Road, Oxford OX1 3RH, UK}

\author{Gerry Haynaly}
\affiliation{Citizen Scientist, Zooniverse c/o University of Oxford, Keble Road, Oxford OX1 3RH, UK}

\author{Pete Hermes}
\affiliation{Citizen Scientist, Zooniverse c/o University of Oxford, Keble Road, Oxford OX1 3RH, UK}

\author{Marc Huten}
\affiliation{Citizen Scientist, Zooniverse c/o University of Oxford, Keble Road, Oxford OX1 3RH, UK}

\author{Sam Lee}
\affiliation{Citizen Scientist, Zooniverse c/o University of Oxford, Keble Road, Oxford OX1 3RH, UK}

\author{Paul Metcalfe}
\affiliation{Citizen Scientist, Zooniverse c/o University of Oxford, Keble Road, Oxford OX1 3RH, UK}

\author{Garry J. Pennell}
\affiliation{Citizen Scientist, Zooniverse c/o University of Oxford, Keble Road, Oxford OX1 3RH, UK}

\author{Joanna Puszkarska}
\affiliation{Citizen Scientist, Zooniverse c/o University of Oxford, Keble Road, Oxford OX1 3RH, UK}

\author{Thomas Schäfer}
\affiliation{Citizen Scientist, Zooniverse c/o University of Oxford, Keble Road, Oxford OX1 3RH, UK}

\author[0000-0002-1825-7133]{Lisa Stiller}
\affiliation{Citizen Scientist, Zooniverse c/o University of Oxford, Keble Road, Oxford OX1 3RH, UK}

\author{Christopher Tanner}
\affiliation{Citizen Scientist, Zooniverse c/o University of Oxford, Keble Road, Oxford OX1 3RH, UK}

\author{Allan Tarr}
\affiliation{Citizen Scientist, Zooniverse c/o University of Oxford, Keble Road, Oxford OX1 3RH, UK}

\author{Andrew Wilkinson}
\affiliation{Citizen Scientist, Zooniverse c/o University of Oxford, Keble Road, Oxford OX1 3RH, UK}

\begin{abstract}
We report on the discovery and validation of a transiting long-period mini-Neptune orbiting a bright (V = 9.0 mag) G dwarf (TOI 4633; R = 1.05\,R$_{\odot}$, M = 1.10\,M$_{\odot}$). The planet was identified in data from the Transiting Exoplanet Survey Satellite by citizen scientists taking part in the Planet Hunters TESS project. Modelling of the transit events yields an orbital period of \textcolor{black}{271.9445 $\pm$ 0.0040 days and radius of 3.2 $\pm$ 0.20 R$\oplus$}. \textcolor{black}{The Earth-like orbital period and an incident flux of \insolationc\ places it in the optimistic habitable zone around the star.} Doppler spectroscopy of the system allowed us to place an upper mass limit on the transiting planet and revealed a non-transiting planet candidate in the system with a period of \textcolor{black}{\Pb[]} days. Furthermore, the combination of archival data dating back to 1905 with new high angular resolution imaging revealed a stellar companion orbiting the primary star with an orbital period of around 230 years and an eccentricity of about 0.9. \textcolor{black}{The long period of the transiting planet, combined with the high eccentricity and close approach of the companion star makes this a valuable system for testing the formation and stability of planets in binary systems.}
\end{abstract}

\keywords{planets and satellites: detection, planets and satellites: dynamical evolution and stability, (stars:) binaries: general}



\section{Introduction}
\label{sec:intro}

The advancement of space based \textcolor{black}{photometric} exoplanet missions, such as CoRoT \citep{2009AuvergneCorot}, \textit{Kepler} \citep{Borucki2010} and the Transiting Exoplanet Survey Satellite \citep[\textit{TESS}; ][]{ricker15} has significantly improved our understanding of extrasolar planetary systems, including our understanding of planet occurrence rates and system architectures. However, the detection of planets using the transit method is inherently biased towards short-period planets. This is in part due to the fact that the transit probability of a planet decreases with increased orbital distance from the host star, and in part due to the fact that automated transit detection pipelines typically require two or more transit events \textcolor{black}{in order to reach the signal-to-noise level required for detection and to achieve confidence that the signal is periodic.}

As a result, only 9.7\% of all confirmed transiting planets have orbital periods longer than 50 days, and 1.6\% have orbital periods longer than 200 days. Similarly, only around 2.5\% of known planets with a semi-major axis greater than 1 au were detected using the transit method, with over 70\% of them having been detected using Radial Velocity (RV) observations (NASA Exoplanet Archive).\footnote{\url{https://exoplanetarchive.ipac.caltech.edu}} While the RV method can yield planet properties such as the orbital period and \textcolor{black}{minimum mass measurements}, without the detection of a transit event the planet radius, and therefore the bulk density, cannot be constrained. \textcolor{black}{Furthermore, the detection of a transit event helps constrain the system inclination, thus enabling an absolute (instead of minimum) mass measurement.} Similarly, atmospheric characterisation via transmission spectroscopy is only possible for transiting planets. 

Transiting planets on long orbital periods, in particular, allow for new investigations into the formation, migration, and long-term stability of planetary systems. The comparison between planets with short and long orbital periods, for example, allows us to probe how equilibrium temperatures affect planet formation \citep[e.g., ][]{2018Lopez,2019Fernandes}. 

The long-term stability and evolution of planetary systems can also be affected by stellar binarity \citep{veras16, 2019Hamer}. As shown by \citet{2010Raghavan}, 54\%\,±\,2\% of solar-type stars are single, with the rest \textcolor{black}{existing} in pairs or higher order multiple systems. These companion stars can perturb planet orbits resulting in high eccentricity tidal migration which can produce hot Jupiters \citep{2012Naoz,2023Vick}, truncate protoplanetary disks and shorten disk lifespans \citep{2012Kraus,2019Manara,2020Winter,2022Zagaria}, and limit the formation of terrestrial planets \citep[when the binary separation is less than around 10 au; ][]{2007Quintana}. Due to the high fraction of stars that are part of binaries, a thorough understanding of how binary interactions affect planet formation, migration and long-term stability is important to constrain the underlying planet population in our Galaxy \citep{2021Moe}. 

In this paper we present the detection and validation of a transiting mini-Neptune (hereafter \target~c) orbiting a bright (Vmag = 9.0), nearby (d = 95 pc) \textcolor{black}{G Dwarf}, that was detected by citizen scientists. The period of 272 days makes this planet the second longest-period confirmed planet identified in the \tess\ data to date \citep[with the longest being TOI 4600\,c,][]{2023Mireles}, and only one of five confirmed \tess\ planets with orbital periods longer than 100 days \citep{2022Dalba,2023Heitzmann}. Furthermore, the long orbital period and incident flux of $\sim$1.6 F$\oplus$ places it in the habitable zone of its host star, making it only the fourth habitable zone planet identified in the TESS data to date, following TOI 700 d \citep{gilbert2020, Rodriguez2020}, TOI 700 e \citep{2023Gilbert}, and TOI 715 b \citep{2023Dransfield}.

RV monitoring revealed an additional, non-transiting, planet candidate with a 34 day period (hereafter \target~b). Furthermore, the combination of newly obtained speckle imaging and archival high-contrast imaging data dating back to 1905 revealed a bound stellar companion with a period of $\sim$ 230 years. As such, we present a bright multi-planet, multi-star system. 

The paper is structured as follows. In Section~\ref{sec:tess_discovery} we describe the discovery of \target~c in the \textit{TESS} data. In Section~\ref{sec:follw-up} we present the spectroscopic and imaging observations \textcolor{black}{and in Section~\ref{sec:RVdat_plc} we discuss activity indicators and statistically validate the transit signal. In Section~\ref{sec:stellar_system} we present the derivation of the parameters of both stars and the orbital properties of the binary system.} In Section~\ref{sec:planet_analysis} we discuss the planet parameters of \target~c and planet candidate \target~b and their long-term stability within the binary system. Finally, in Section~\ref{sec:results}, we place \target\ into the context of other long-period multi-planet systems as well as into the context of the population of confirmed planets in binaries.  


\section{Photometry and discovery of \target\ c} \label{sec:tess_discovery}

\target\ \citep[TIC\,307958020;][]{Stassun19} is located at high ecliptic latitude (near the \textcolor{black}{ecliptic pole}) and was observed by \tess\ nearly continually during years 2, 4 and 5 of the mission (Sectors 14--26, 40, 41, 47--53, 55, 56, 58, 59). \textcolor{black}{Following the identification of the first two transit events, we proposed for the target to be observed at the shortest \tess\ cadence. Therefore, from Sector 49 and onward, the data have been obtained at a 20-second cadence (proposal ID: DDT054; PI: Eisner). Prior to this sector all observations were obtained at a 2-minute cadence.}

The light curve of \target\ exhibits three transit events located in Sectors 20, 40 and 50. The former two transit events were identified by citizen scientists taking part in the Planet Hunters \textit{TESS} citizen science project \citep{eisner2020method}. The project, which is hosted by the Zooniverse platform \citep{lintott08,lintott11}, engages over 40,000 citizen scientists in the task of visually inspecting \tess\ data in the search for transit events. At any given time, Planet Hunters \textit{TESS} only ever displays the data from a single \tess\ sector. As such, the former two transit events were independently identified by 15 citizen scientists who were \textcolor{black}{randomly presented} the light curve of \target. The data consist of the Presearch Data Conditioning (PDC) 2-minute cadence observations, which are produced by the Science Processing Operations Center \citep[SPOC; ][]{Smith2012, Stumpe2012, jenkins16, Stumpe2012, Stumpe2014} pipeline. \textcolor{black}{Individual measurements that are flagged by the SPOC pipeline as being affected by various instrumental anomalies are not shown on Planet Hunters TESS. Unfortunately, the data around the time of the third transit event (in Sector 50) was identified as being affected by scattered light and as such this transit was not seen or identified by citizen scientists. For more details regarding the PHT pipeline and the identification of transit-like signals we refer the reader to \cite{eisner2020method}.}


Once the transit events were identified, we analyzed both the 2-minute cadence and 20-second cadence PDC SPOC light curves. As mentioned above, the 2-minute and 20-second cadence data around the time of the third transit event (BTJD$\sim$2680.6 days) were affected by scattered light.\oscar{
In order to recover the third transit, we performed a tailored correction of the Sector 50 light curve using cotrending basis vectors (CBVs). We followed a similar approach as the one presented by \citet{Barragan2022}, which makes use of the \texttt{lightkurve} package \citep{lightkurve2018}. In brief, we first created a light curve from the target pixel file (TPF) using the nominal \tess\ aperture and did not remove data that had been flagged as `bad' by the SPOC pipeline \textcolor{black}{(flag: bit 13, value 4096)}. The CBVs provided with the TPF were then used to correct the light curve using the built-in correction function \textcolor{black}{(\texttt{CBVCorrector}}\footnote{See \texttt{lightkurve} documentation \url{https://docs.lightkurve.org/tutorials/2-creating-light-curves/2-3-how-to-use-cbvcorrector.html}}}) in \texttt{lightkurve} and allowing for interpolation. This generates a light curve where the large scale trends are removed, including the trends around the times of the transit events.
Finally, we performed a crowding correction to account for extra flux from nearby stars that may be present in the Simple Aperture Photometry (SAP) mask. To do this we used the nominal crowding values given in the TPF to account and correct for the light curve contamination.\footnote{See \url{https://heasarc.gsfc.nasa.gov/docs/tess/UnderstandingCrowding.html} for more details on \tess\ crowding correction.} We note that this method did not correct for the light contribution from the close, bound companion star (see Section~\ref{sec:stellar_system}). \textcolor{black}{As a test, we also used this method to extract and detrend the light curves around the time of the other two transit events. We found there to be no significant difference between these and the SPOC light curves. As such, for these two former transits we use the SPOC data.}

\begin{figure*}
    \centering
    \includegraphics[width=1\textwidth]{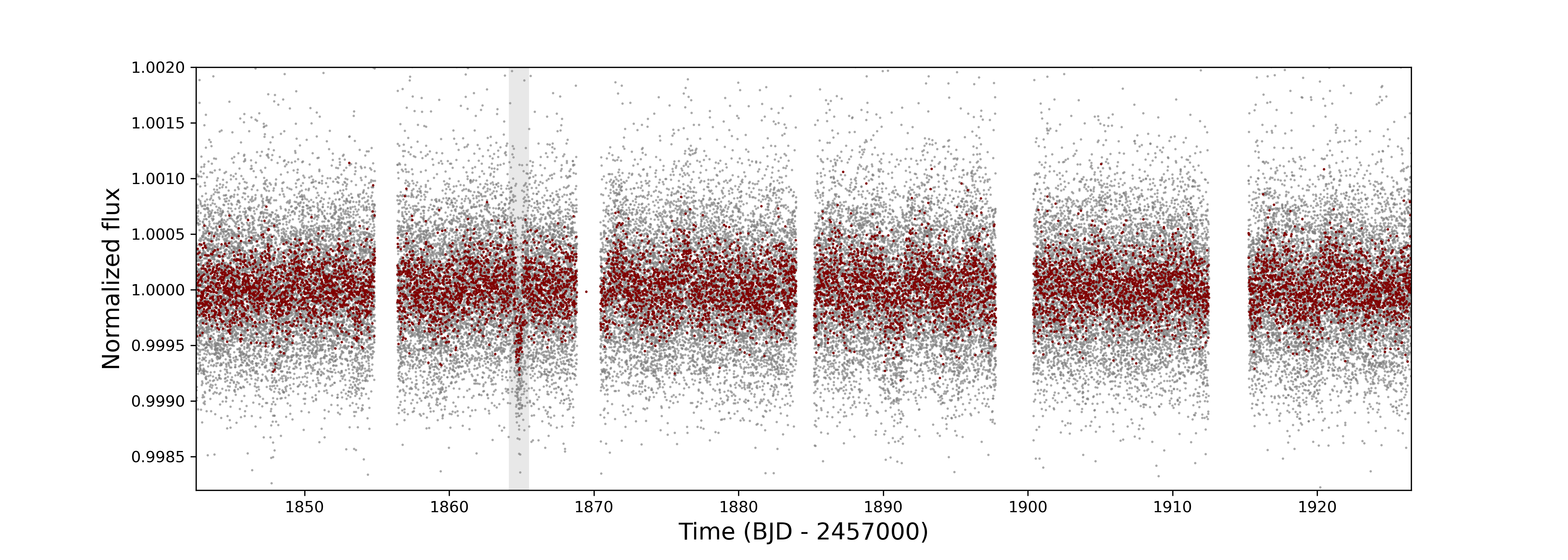}
    \caption{Normalized photometric data from \tess\ sectors 20 -- 22 to illustrate the stellar variability of the light curve. The first of the three transit events observed by \textit{TESS} to date is highlighted by the vertical grey column. The normalized 2-minute cadence flux binned to 10 minutes and unbinned are shown in maroon and grey, respectively.}
    \label{fig:full_LC}
\end{figure*}

\begin{figure}
    \centering
    \includegraphics[width=0.45\textwidth]{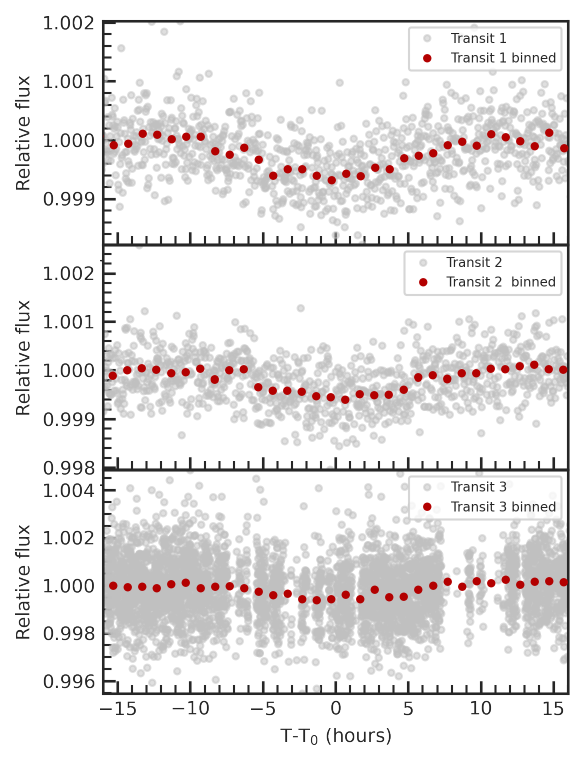}
    \caption{The three transits of \target\,c. Nominal \tess\ observations are shown in light grey while solid circles represent 60-minute binned data. }
    \label{fig:alltransits}
\end{figure}

The absence of further transit events in the \tess\ data allows us to confirm that the transiting object (\target~c) has an orbital period of $\sim$271.94 days. \textcolor{black}{ We confirm that all shorter aliases of this orbital period would result in at least one observable transit in the available \tess\ data.} A subset of the 2-minute cadence light curve is shown in Figure~\ref{fig:full_LC} and the three individual transits are shown in Figure~\ref{fig:alltransits}. 

\textcolor{black}{In addition to the \tess\ data, there is archival data from the All Sky Automated Survey for SuperNovae (ASAS-SN), consisting of 3482 observations obtained between 28 March 2012 and 10 November 2023, with 2481 observations in the \textit{g} filter, and 1001 in the \textit{V} filter \citep{2014Shappee, Kochanek2017}. However, we note that the \textit{g} filter data show a large change in the root mean square scatter before and after 2459330 HJD of around  105 mJy and 170 mJy, respectively. We, therefore, do not use the \textit{g} filter data for any further analysis. We find no evidence of any further transit signals in the ASAS-SN data}

\subsection{Excluding false positive scenarios}
\label{subsec:latte}

In order to help rule out instrumental and astrophysical false positive scenarios including background eclipsing binaries, systematic effects, and background events such as asteroids passing through the field of view we performed a number of standard diagnostic tests using the \tess\ data. We used the publicly available Lightcurve Analysis Tool for Transiting Exoplanets \citep[{\sc latte}; ][]{LATTE2020} for this analysis.\footnote{\url{http://latte-online.flatironinstitute.org/}}  For a full description of the diagnostic tests we refer the reader to \cite{LATTE2020}; however, in brief, the tests allowed us to ensure that:

\begin{itemize}
    \itemsep0em 
    \item the transit events do not coincide with the times of the periodic momentum dumps that introduce spurious signals into the data.
    \item the x and y centroid positions are smoothly varying with time in the vicinity of the transit events and \textcolor{black}{thus the transit events are unlikely to be caused by systematic effects or by background eclipsing binaries.}
    \item the light curves of the five nearest two-minute cadence \tess\ stars do not show similar signals at the same time \textcolor{black}{(projected distances between these stars and \target\ range from 2.7 to 28.7 arcminutes).}
    \item \textcolor{black}{the signal is on target by investigating the light curve extracted for each pixel surrounding the target in order to ensure that the transit events are not caused by spurious background signals on other pixels}.
    \item there are no spurious signals, such as sudden jumps or strong variations, in the background flux at the same time as the event.
    \item the transit shapes and depths when extracted with different aperture sizes are consistent.
    \item \textcolor{black}{the signal is on target by comparing the average in-transit and average out-of-transit flux, as well as the difference between them, which indicates the location of the change in flux (i.e., the location of the transit event).} 
\end{itemize}

We used these tests to show that the transit signals are unlikely to be systematic or astrophysical false positives. We uploaded \target\ c to the Exoplanet Follow-up Observing Program for \textit{TESS} (ExoFOP-TESS) site on 27 May 2020 as a community \textit{TESS} Object of Interest (cTOI). The planet was later promoted to a priority 1 (1 = highest priority, 5 = lowest priority) TOI candidate (TOI 4633.01) on 14 December 2022. 

While the SPOC did not recover the true period of \target~c, the first two transits were detected as a duo-transit in the two multi-sector searches conducted over sectors 14--23 and 14--26. Transit searches conducted with a noise-compensating matched filter over sectors 14--41, 14--50, 14--55, and 14--59 \citep{2002Jenkins, jenkins2010, 2020Jenkins} identified the first two transits at twice the actual orbital period. A limb-darkened transit model was fit to the transit signal in each case \citep{Li2019} and a suite of diagnostic tests were performed to assess the nature of the signal \citep{Twicken2018}. The signal passed all the diagnostic tests, including the difference image centroiding test, \textcolor{black}{which localized the source of the transit signal to within 2.3 $\pm$ 4.1 arcseconds of the presumed host star.} The \tess\ Science Office reviewed the Data Validation reports and issued an alert for TOI 4633 c on 19 November 2021 \citep{2021Guerrero}.


\section{Follow-up observations} \label{sec:follw-up}

In this section we outline the spectroscopic data, as well as the imaging data that revealed a close, bound stellar companion. The analysis of these data is discussed in Sections~\ref{sec:stellar_system} and \ref{sec:planet_analysis}. 



\subsection{Spectroscopic RV monitoring} \label{subsec:RV}

We acquired high-resolution (R\,$\sim$\,120,000) spectra with the HIgh Resolution Echelle Spectrometer (HIRES) mounted on the 10-m Keck telescope on Maunakea, Hawaii \citep[][Program numbers: 2022A\_N93, 2022B\_N025, and 2023A\_N085]{vogt1988hires}. The instrument has a wavelength coverage between $\sim$ 350 and $\sim$ 620 nm. The instrument passes the star's light through a heated iodine cell, allowing for precise wavelength calibration when determining relative RVs \citep{2010Howard}.  

We obtained 54 spectra using the iodine cell between 21 February 2022 and 22 July 2023 (mean per pixel S/N $\sim$ 167 at 550 nm). \textcolor{black}{Radial velocity measurements were determined following the approach of \citet{vogt1994} and then fit using the publicly available software package RadVel via comparison to a S/N = 260 template spectrum of G5 standard star HD 162232 taken on 20 June 2008 \citep{radvel}.} The RV analysis was also carried out using a template spectrum of \target\ taken on 24 November 2021 (per pixel S/N = 46), which yielded equivalent results within the uncertainties. The analysis of the HIRES spectra showed no evidence of a companion star, allowing us to place an upper limit of $\Delta$RV $\leq$ 10 km s$^{-1}$ on the relative motion of the two stars known to be in this system \citep{kolbl2015}. Furthermore, we note that we do not see any evidence for multiple stars in our radial velocity template or in the analysis of the measured radial velocity values alone. Due to the lack of the obvious signature of both stars in the spectra, we are unable to use a two-star template spectrum to extract the RVs.

Additional data were obtained with the SOPHIE high-resolution fiber-fed echelle spectrograph mounted on the 1.93-m Observatoire de Haute-Provence (OHP) telescope \citep{2008PerruchotOHP}. We used the high-resolution mode, which delivers a spectral resolution of R $\sim$ 75,000 across the wavelength range of 387--694 nm. We obtained 20 observations between 15 November 2021 and 5 March 2023 (mean per pixel S/N $\sim$ 33 at 555 nm). Standard stars that were observed at the same epochs using the same SOPHIE mode did not show significant instrumental drifts. The spectra were reduced using the standard SOPHIE RV reduction pipeline \citep{2009Bouchy}, including CCD charge transfer inefficiency correction \citep{2013Bouchy}. Following the method described, e.g., in \citet{2008Pollacco} and \citet{2008Hebrard}, we estimated and corrected for the sky background contamination (mainly due to the Moon) using the second SOPHIE fiber aperture, which is targeted 2 arcminutes away from the first one pointing toward the star. We estimated that only one of the 20 exposures was significantly polluted by sky background. \textcolor{black}{We note that the analysis of the SOPHIE spectra showed no evidence of a companion star.} The extracted HIRES and SOPHIE RV observations are listed in Table~1 and their implications for the system's architecture discussed in Section~\ref{sec:planet_analysis}.

The RV data allowed us to place an upper mass limit on planet~c and revealed an additional non-transiting planet candidate (with a period of $\sim$ 34 days; planet~b) in the system, as discussed in Section~\ref{sec:planet_analysis}. 

\begin{table*}[] \label{tab:rvs}
\begin{tabular}{ccccc|ccccc}
\hline
Time  & RV & RV error  & SNR & Instrument & Time & RV  & RV error & SNR & Instrument \\
(BJD - 2457000) & (m/s) &(m/s) &  & &  (BJD - 2457000) &  (m/s) &   (m/s) &  &\\
\hline
2534.3461	&	-47.40	&	7		&	34	&	OHP		&	2780.9910	&	14.50	&	1.60	&	216	&	HIRES	\\
2601.7097	&	-6.40	&	5.00	&	32	&	OHP		&	2782.4014	&	9.60	&	5.00	&	31	&	OHP  	\\
2606.7163	&	9.60	&	5.00	&	31	&	OHP		&	2785.9500	&	29.50	&	1.81	&	215	&	HIRES	\\
2622.7200	&	33.60	&	5.00	&	31	&	OHP		&	2786.9510	&	28.40	&	1.76	&	215	&	HIRES	\\
2623.6940	&	10.60	&	5.00	&	31	&	OHP		&	2789.8280	&	26.75	&	1.68	&	216	&	HIRES	\\
2631.6864	&	-9.40	&	5.00	&	31	&	OHP		&	2792.8320	&	13.20	&	1.50	&	214	&	HIRES	\\
2632.0840	&	-18.11	&	2.67	&	215	&	HIRES	&	2799.9740	&	-0.47	&	3.26	&	75	&	HIRES	\\
2648.6661	&	21.60	&	5.00	&	31	&	OHP		&	2806.8260	&	-15.09	&	3.04	&	75	&	HIRES	\\
2657.1140	&	22.26	&	2.23	&	214	&	HIRES	&	2809.8690	&	-12.29	&	2.97	&	75	&	HIRES	\\
2661.0570	&	1.01	&	2.49	&	214	&	HIRES	&	2812.7900	&	-13.26	&	2.76	&	76	&	HIRES	\\
2662.6285	&	-8.40	&	3.00	&	55	&	OHP		&	2822.8680	&	38.68	&	2.66	&	75	&	HIRES	\\
2672.0130	&	-16.19	&	2.20	&	216	&	HIRES	&	2824.4013	&	12.60	&	5.00	&	31	&	OHP   	\\
2681.0310	&	48.24	&	2.47	&	174	&	HIRES	&	2826.7810	&	14.72	&	2.53	&	76	&	HIRES	\\
2683.6198	&	14.60	&	5.00	&	30	&	OHP		&	2828.7990	&	16.66	&	3.02	&	76	&	HIRES	\\
2687.6359	&	8.60	&	5.00	&	31	&	OHP		&	2829.7400	&	4.81	&	3.26	&	76	&	HIRES	\\
2690.0110	&	9.61	&	2.11	&	216	&	HIRES	&	2831.8460	&	5.52	&	2.79	&	73	&	HIRES	\\
2695.0180	&	-11.66	&	2.05	&	214	&	HIRES	&	2833.7860	&	-4.98	&	2.74	&	75	&	HIRES	\\
2700.9730	&	-44.11	&	2.09	&	213	&	HIRES	&	2834.7620	&	-10.65	&	2.87	&	76	&	HIRES	\\
2710.0270	&	-11.04	&	1.76	&	215	&	HIRES	&	2835.7440	&	-8.97	&	2.69	&	77	&	HIRES	\\
2712.0200	&	3.27	&	2.27	&	163	&	HIRES	&	2838.8030	&	-8.72	&	2.87	&	75	&	HIRES	\\
2712.9410	&	18.14	&	1.87	&	215	&	HIRES	&	2840.7390	&	-10.70	&	3.22	&	76	&	HIRES	\\
2713.5387	&	6.60	&	5.00	&	31	&	OHP		&	2858.7850	&	37.97	&	3.17	&	74	&	HIRES	\\
2715.9930	&	-2.48	&	1.79	&	214	&	HIRES	&	2879.2694	&	-27.40	&	5.00	&	31	&	OHP	    \\
2725.5476	&	17.60	&	5.00	&	31	&	OHP		&	2892.3184	&	-9.40	&	5.00	&	33	&	OHP	    \\
2730.0720	&	5.38	&	1.82	&	214	&	HIRES	&	3009.6665	&	-5.40	&	6.00	&	31	&	OHP	    \\
2736.5730	&	-32.40	&	5.00	&	31	&	OHP		&	3043.0680	&	-20.97	&	2.65	&	143	&	HIRES	\\
2738.9790	&	-17.85	&	1.54	&	215	&	HIRES	&	3045.0940	&	-38.01	&	2.18	&	216	&	HIRES	\\
2741.9610	&	-19.38	&	1.51	&	215	&	HIRES	&	3047.1150	&	-36.49	&	2.07	&	202	&	HIRES	\\
2748.8440	&	-2.73	&	1.63	&	204	&	HIRES	&	3068.0640	&	-7.83	&	2.08	&	153	&	HIRES	\\
2749.8980	&	-0.83	&	1.48	&	216	&	HIRES	&	3070.9230	&	-31.26	&	2.01	&	214	&	HIRES	\\
2750.4574	&	12.60	&	4.00	&	42	&	OHP		&	3089.8800	&	-9.88	&	1.75	&	216	&	HIRES	\\
2768.9970	&	1.38	&	1.57	&	215	&	HIRES	&	3101.0680	&	15.11	&	1.51	&	210	&	HIRES	\\
2769.9960	&	-2.72	&	1.89	&	214	&	HIRES	&	3108.0060	&	-9.65	&	1.53	&	214	&	HIRES	\\
2773.4748	&	-11.40	&	6.00	&	31	&	OHP		&	3122.0480	&	5.74	&	1.63	&	171	&	HIRES	\\
2775.9770	&	-6.20	&	1.58	&	216	&	HIRES	&	3132.8510	&	6.65	&	1.51	&	216	&	HIRES	\\
2776.8030	&	-11.86	&	1.58	&	210	&	HIRES	&	3138.9630	&	5.20	&	1.86	&	196	&	HIRES	\\
2779.9810	&	3.22	&	1.69	&	214	&	HIRES	&	3148.0280	&	1.62	&	1.68	&	206	&	HIRES	\\
\end{tabular}
\caption{Spectroscopic data obtained with the OHP/SOPHIE (mean per pixel S/N $\sim$ 33 at 555 nm) and Keck/HIRES (mean per pixel S/N $\sim$ 167 at at 550 nm). \textcolor{black}{The reported RV values were systematically shifted to be centered around 0, as determined using a Keplerian fit}. The low resolution, iodine-free reconnaissance spectra obtained with Keck/HIRES are not included.}
\end{table*}

\subsection{Imaging observations and companion star} \label{subsec:imaging}

Archival data listed in the Washington Double Star Catalog (WDS) show that \target\ has a bound companion star. Hereafter, the stars will be referred to as star A and star B for the more and less massive star, respectively.  The 10 archival observations were obtained between February 1905 and June 2011 and their position angles (PA) and angular separations ($\rho$) are listed in Table~\ref{tab:imaging}. 

\begin{figure*}
    \centering
    \includegraphics[width=1\textwidth]{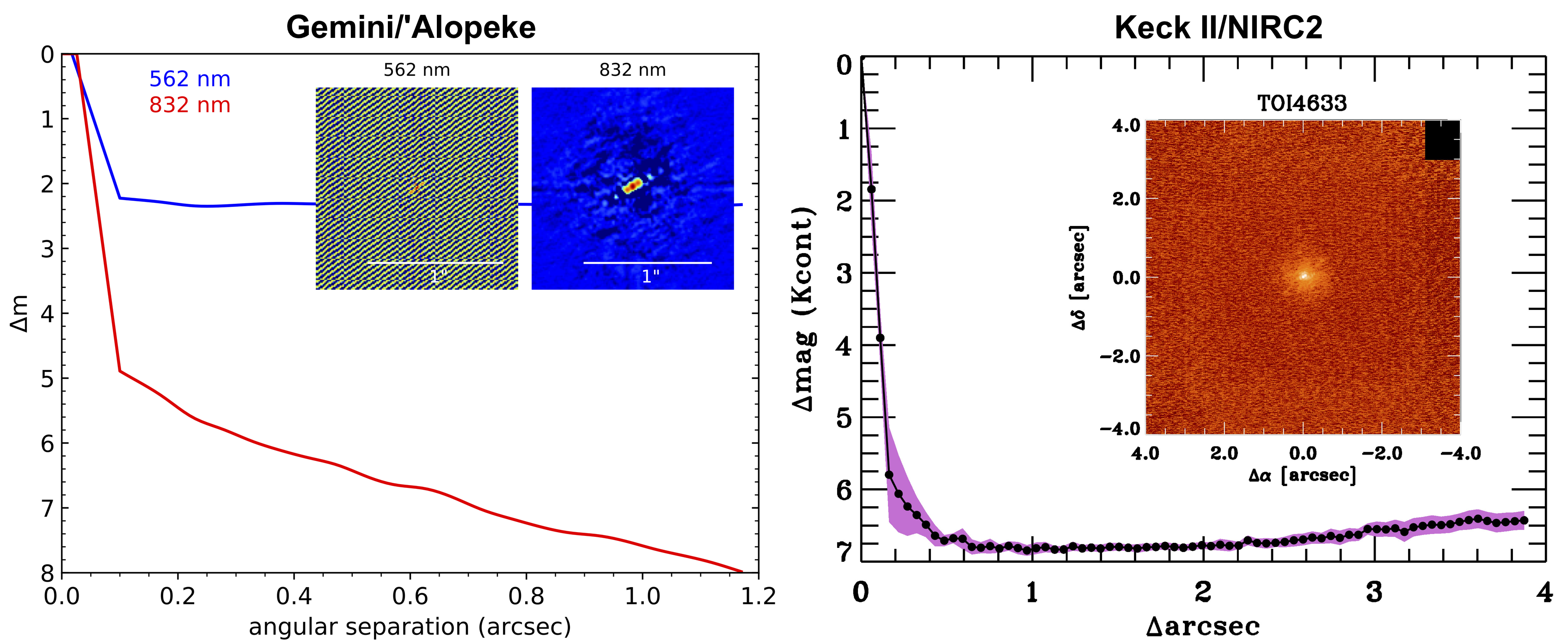}
    \caption{Contrast curves showing the 5 $\sigma$ detection sensitivity and speckle auto-correlation functions obtained using Gemini North/\okina Alopeke (left) and Keck II/NIRC2 (right).} 
    \label{fig:gemini}
\end{figure*}

In addition to archival data, we obtained high contrast imaging observations of \target\ to further constrain the orbit of the two stars. High contrast imaging was performed using the NESSI high-resolution speckle imaging instrument mounted on the 3.5-m WIYN telescope located at Kitt Peak National Observatory \citep{2018ScottNESSI} on 21 April 2022; the PHARO adaptive optics instrument mounted on 5.1-m Hale telescope at Palomar Observatory \citep{Hayward2001palomar} on 11 November 2021; the \okina Alopeke speckle instrument mounted on the 8.1-m Gemini North telescope on Maunakea \citep{Howell2011,Matson2019} on 9 May 2022; and the NIRC2 adaptive optics instrument mounted on the 10-m Keck II \citep{2000Wizinowich} located on Maunakea, Hawaii, on 26 April 2023. While the former two observations were unable to resolve the two stars, the \okina Alopeke and NIRC2 observations revealed the close companion at angular separations of 51 and 62 milliarcsecond, respectively. The observations showed that the two stars have a brightness ratio $F_B/F_A = 0.7$, as shown in Figure~\ref{fig:gemini}. The former two instruments were unable to resolve the two stars due to a lack of angular resolution to resolve a separation of $<$ 100 milliarcsecond. Similarly, neither Gaia DR2 nor Gaia DR3, which obtained observations of the target in 2015 and 2016, respectively, were able to resolve the two stars due to the small angular separation of the stars at the time of the observations. \textcolor{black}{Furthermore, due to the proximity of the two stars, the Gaia DR3 noise metric RUWE$\sim$1.15 is consistent with a single star solution, which highlights the need for high resolution imaging. Finally, we note that in addition to the archival data listed in Table~\ref{tab:imaging}, WDS records one additional observation obtained by \citet{2021Gili} in 2011.} \textcolor{black}{Their recorded values of position angle (37.6 $\deg$) and angular separation (0.233 arcseconds) are inconsistent with both the remainder of the archival and our newly obtained data. More specifically, there is no stable binary orbit that would be able to explain this measurement in combination with the archival and newly obtained measurements. Furthermore, the analysis method that was used to reduce these observations follows a unique procedure that is ambiguously defined and distinct from the processes used for any of the other imaging data, and thus we do not include it in any further analysis.}


\textcolor{black}{Given the available Gaia data, the proper motions of the two stars cannot be disentangled.} However, given the proper motion of the system (RA: -23.056, Dec: -67.0188 milliarcseconds/year) and the range of angular separations of the two stars reported between 1905 and 2023 (470 -- 50 milliarcseconds), we confirm that these two stars are bound and not a chance alignment along our line of sight.

\renewcommand{\arraystretch}{1.2}
\begin{table*}[]
\centering
\footnotesize
\begin{tabular}{ccccccccc}
\hline
\textbf{Date} & \textbf{PA}    & \textbf{$\rho$}     & \textbf{$\Delta$\ mag}  & \textbf{Aperture}    & \textbf{$\lambda$}    &  \textbf{Method$^a$} & \textbf{Reference} & \textbf{Notes} \\
\textbf{}     & \textbf{(deg)} & \textbf{(arcsec)}         &                       & \textbf{(m)}   & \textbf{(nm)}      & \textbf{}       & \textbf{}   & \textbf{}      \\ \hline
1905.170       & 124.7                & 0.42                         & 0.5                 & 0.9      &  --           & M               & \citet{1905Hussey}  &  WDS        \\
1922.876      & 124.1                 & 0.38                          & 0.4                 & 1.0     &  --            & M               & \citet{1927vanBiesbroeck}  &  WDS         \\
1959.440       & 124.9                & 0.4                          & 0.3                 & 0.4      &  --           & M               & \citet{1960Couteau}   &  WDS        \\
1963.790       & 124.8                & 0.47                         & 0.2                 & 0.4      &  --           & M               & \citet{1967Baize}    &  WDS        \\
1966.672      & 122.8                 & 0.42                          & 0.2                 & 0.7     &  --            & M               & \citet{1972Worley}     &  WDS       \\
1976.460       & 122.7                & 0.42                         & --                  & 0.5      &  --           & M               & \citet{1978Muller}   &  WDS        \\
1982.078      & 124                   & 0.3                           & 0.3                 & 0.7     &  --            & M               & \citet{1989Worley}     &  WDS       \\
1984.520       & 127.8                & 0.28                         & --                  & 0.6      &  --           & M               & \citet{1985Heintz}   &  WDS        \\
1993.100        & 119                 & 0.26                        & --                  & 0.5       &  --          & M               & \citet{1997Muller}   &  WDS         \\
\hline
\hline
2021.8616      & --                 & $<$\textcolor{black}{0.1}           & --             & 5.1     & 1000 -- 2500       & AO              & --    &  Palomar/PHARO    \\ 
2022.303      & --                  &  $<$\textcolor{black}{0.06}         & --             & 3.5     & 812 -- 852        & S               & --    &  WIYN/NESSI      \\ 
2022.7959     & 300.5 $\pm\ 0.5^*$           &  0.05  $\pm\ 0.01$         & 0.4            & 8.1     & 535 -- 589        & S               & --    &  Gemini/\okina Alopeke      \\ 
2022.7959     & 303.18 $\pm\ 1.29$           &  0.062  $\pm\ 0.01$        & 0.23          &     10   & 2256 -- 2285      & AO              & --    &  Keck II/NIRC2      \\ 
\hline
\end{tabular}
\caption{Archival and new imaging of \target. Notes: $^a$ M indicates that the observation were obtained using a micrometer on a refractor telescope; S indicates that the observations were obtained with the Speckle technique; and AO indicated the use of adaptive optics. $^*$180 degree ambiguity in the PA.}
\label{tab:imaging}
\end{table*}

\renewcommand{\arraystretch}{1}

\section{Activity indicators and statistical validation of the transit signals} 
\label{sec:RVdat_plc}

\textcolor{black}{In this section we discuss how the stellar activity affects our confidence in the planetary nature of planet candidate b. Furthermore, we present statistical validation of planet c.}

\subsection{Stellar activity indicators and planet candidate \target\ b}
\label{subsec:activity}

\begin{figure*}
    \includegraphics[width=1\textwidth]{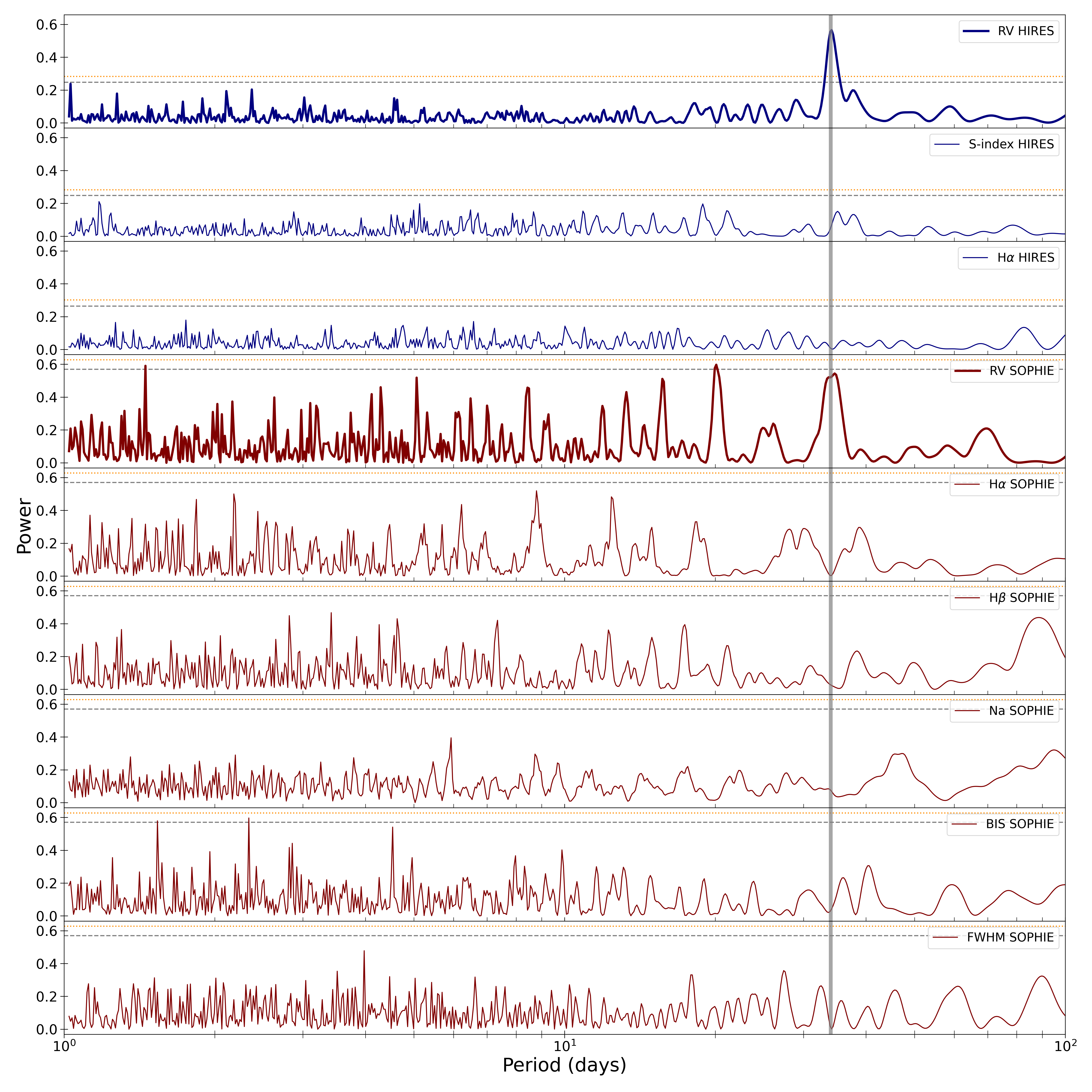}
    \caption{Generalized Lomb-Scargle Periodogram of the HIRES and SOPHIE RV data (navy and maroon bold panels, respectively). The non-bold panels show periodograms of various activity indicators for each instrument, while the vertical grey line indicates a period of 34 days. The horizontal dotted orange and dashed grey lines represent the 10\% and 30\% FAPs, respectively.} 
    \label{fig:activity_indicators}
\end{figure*}

The RV data exhibit a periodic \textcolor{black}{signal} with a period of $\sim$34 days. This can be seen in Figure~\ref{fig:activity_indicators}, which shows periodograms of the RV observations obtained with HIRES (navy) and SOPHIE (maroon), calculated using the PyAstronomy Generalized Lomb-Scargle Periodogram \citep{pya}. We carried out a number of tests in order to determine whether this periodic signal seen in the RV data is caused by a planetary body or by stellar activity. This is important as magnetic stellar activity manifests itself \textcolor{black}{by producing} brighter or darker regions on the surface of the star, such as faculae, starspots, and plages. This, in turn, affects the observed stellar spectra and can induce RV signals that mimic those from planets \citep[e.g.,][]{queloz2001no,figueira2010evidence,boisse2011disentangling,diaz2016harps}. 

For the HIRES data we investigate the chromospheric stellar activity using the S-index and its corresponding derivative $\log R'_{\rm HK}$ value, a parameter that provides a proxy for the level of magnetic activity on the surface of a star \citep[e.g.,][]{vaughan1978flux,1984Noyes,saar1998magnetic,santos2000coralie,cincunegui2007halpha,santos2010stellar}. In brief, $R'_{\rm HK}$ is the ratio of the emission in the cores of the Ca {\sc II} H and K lines (at 3933 \AA\ and 3968 \AA) to the total bolometric flux of the star, where the Ca {\sc II} H and K lines probe the temperature of the chromosphere of the star. As magnetic activity primarily heats up the chromosphere (resulting in stronger emission features), the ratio of the intensity of the emission feature in these lines to the bolometric flux probes the magnetic activity of the star. 

We used the H$\alpha$ Balmer line of hydrogen at 6563 \AA\ as an additional indicator of the chromospheric activity of the star. Using the HIRES data and following the methodology described by \citet{2011GomesdaSilva} we measured the ratio of the flux within $\pm$ 0.8 \AA\ of the H$\alpha$ line at 6562.808 \AA\ to the flux of the two wavelength regions of 6550.87 $\pm$ 5.375 \AA\ and 6580.31 $\pm$ 4.375 \AA. Prior to measuring the flux ratios, we derived a wavelength solution for the H$\alpha$ order of the spectrum and shifted \textcolor{black}{each spectrum} to the star's rest frame by cross correlating each spectra with the spectrum from the NSO solar atlas. Both the NSO solar atlas spectrum and the HIRES spectra were re-sampled to have wavelength steps of 0.003 \AA. The spectra were normalized with a third order polynomial (where the H$\alpha$ line $\pm$ 5 \AA\ was masked out). 

Similarly, we computed activity indicators using the SOPHIE data by measuring the ratio of the fluxes in the cores of the H$\alpha$ (6562.808~\AA), H$\beta$ (4861.363~\AA) and Na I D1 \& D2 (5895.92 \& 5889.95~\AA, respectively) \textcolor{black}{lines} to the flux in continuum regions around each of the lines. The same wavelength regions as described above were used for the H$\alpha$ index. For the H$\beta$ index we measured the flux within a 1.2~\AA\ window centered on the absorption line and divided by the two reference regions defined as 4855.0 $\pm$ 2.5~\AA\ and 4870.0 $\pm$ 2.5 \AA\ \citep{2022Klein}. The Na I index was defined as the flux within a 0.5 \AA\ window centered on each line divided by the two reference regions of 5805.0 $\pm$ 5~\AA\ and 6090.0 $\pm$ 10~\AA\ \citep{2011GomesdaSilva}. In addition to these stellar activity indicators, we made use of the fact that periodic changes in the shapes of absorption lines (e.g., FWHM, BIS) can be indicative of whether RV signals are caused by companions or by stellar activity. We note that different activity indicators are used for the HIRES and SOPHIE data due to the different spectral ranges that both of these instruments cover. 

We used the PyAstronomy Generalized Lomb-Scargle Periodogram \citep{pya} to search for periodicities in the chromospheric activity indicators, FWHM and BIS. As shown in Figure~\ref{fig:activity_indicators}, \textcolor{black}{where the dotted orange and dashed grey lines show the 10\% and 30\% False Alarm Probability (FAP), respectively, }there are no significant trends that coincide with the periodic signals seen in the RV data (indicated by the grey vertical line). We do note that the S-index shows a local power maximum at 37 days, which is near the radial velocity power maximum \textcolor{black}{at} 34 days. This power maximum is, however, not the largest peak in the S-index power \textcolor{black}{spectrum and has an FAP of 99.96\%.}

We also determine the degree of correlation between the RV time series and each of the activity indicators (independent for the HIRES and SOPHIE data) by computing Pearson's correlation coefficients \textit{r} and their \textit{p}-values. The test showed that there are no correlations between the RVs and any of the activity indicators with \textit{r} ranging from -0.28 to 0.14 and \textit{p}-values ranging from 0.23 to 0.29. \textcolor{black}{As all of the \textit{p}-values are significantly greater than 0.05, we conclude that there is no significant evidence for a correlation between the RV time series and the activity indicators for both the HIRES and SOPHIE data.}

\begin{figure*}
    \centering
    \includegraphics[width=0.9\textwidth]{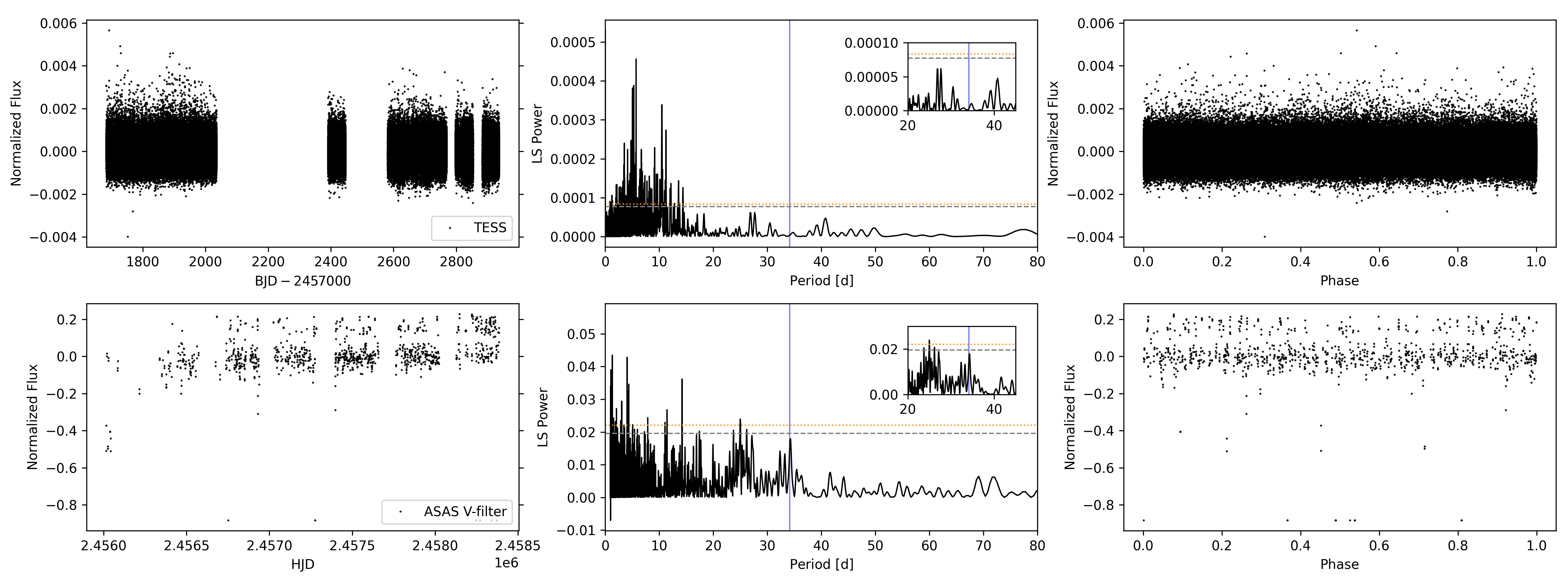}
    \caption{Normalized light curve (left), Generalized Lomb-Scargle periodogram (middle), and light curve phase folded at the period of planet candidate b (P$_{b}$ = \Pb) for the 2-minute cadence \tess\ data (top) and the \textit{V}-band ASAS-SN data. The vertical blue line in the middle panel indicates the period of planet candidate b, while the horizontal dotted orange and dashed grey lines represent the 10\% and 30\% FAPs, respectively.}
    \label{fig:periodogram_TESS_asas}
\end{figure*}

Finally, we ensured that the PyAstronomy Generalized Lomb-Scargle periodograms of the \tess\ \textcolor{black}{light curve and the ASAS-SN light curve (where the transit events in the \tess\ data are removed) show no significant periodic signals. The Lomb-Scargle periodograms of the \tess\ and ASAS-SN data are shown in Figure~\ref{fig:periodogram_TESS_asas}, alongside the light curves phase folded at the period of \target\ b (P = \Pb). This shows that there are no significant trends at this period with an FAP better than 30\% that could result from additional transit events, stellar rotation, or asteroseismic pulsations in either the \tess\ or the ASAS-SN data. This is further supported by the lack of a coherent signal in the phase folded data at the period of planet candidate b in the rightmost column of Figure~\ref{fig:periodogram_TESS_asas}.}

\textcolor{black}{In summary, based on our \textcolor{black}{analysis} of the stellar activity indicators we have been able to extract from the HIRES and SOPHIE spectra, there is no indication that the 34-day signal is caused by stellar activity, and we therefore consider the hypothesis that this signal is of planetary origin to be more likely. However, the limited precision of the HIRES activity indicators, and the limited number of SOPHIE observations, preclude a joint analysis of the RVs and activity indicators to disentangle the contributions of activity and planet(s) to the RVs. Consequently, we report the 34-day signal as a planet candidate, rather than a confirmed planet detection.}



\subsection{Statistical validation of planet c}

\textcolor{black}{The standard diagnostic tests presented in Section \ref{subsec:latte} allowed us to rule out a number of instrumental and astrophysical false positive scenarios that could mimic the transit-like signals seen in the \tess\ data. Furthermore, the spectroscopic follow-up observations presented in Sections~\ref{subsec:RV} allowed us to place an upper mass limit of 123 M$\oplus$ (99 \textcolor{black}{percentile of the credible interval}) on \target\ c, as discussed in Section~\ref{subsec:pyaneti}. As we do not have a 3-sigma mass measurement of planet c, we carry out a statistical analysis of the likelihood that the transit signals are caused by a planet as opposed to the range of alternative, astrophysical false positive scenarios.}

\textcolor{black}{We used the open source package \triceratops\ \citep{Giacalone2021}, which was specifically developed to aid in the vetting and validation of transit-like signals identified in the \tess\ data, to calculate the false positive probability (FPP) of the observed transit signals of \target\ c. In brief, \triceratops\ uses a Bayesian framework that incorporates prior knowledge of the target star, planet occurrence rates, and stellar multiplicity to calculate the probability that the transit signal is due to a transiting planet. It also makes use of the TRILEGAL \citep{Girardi2005} galactic model to simulate a population of stars around the line of sight of the target.}

\textcolor{black}{The resulting FPP quantifies the probability that the observed transit signal can be attributed to something other than a transiting planet. As inputs to the code we used the 2-minute cadence and 20-second cadence (where available) data (Section~\ref{sec:tess_discovery}), combined with the contrast curves obtained using \okina Alopeke and NIRC2 (Section~\ref{subsec:imaging}). We found the FPP to be 0.0003 $\pm$ 0.0005 and the nearby FPP \textcolor{black}{(NFFP)} to be 6.6 $\times$ 10 $^{-21}$ $\pm$ 4.6 $\times$ 10 $^{-21}$. Both of these are better than the commonly accepted validation threshold of FPP $<$ 0.015 and NFPP $<$ 10 $^{-3}$, as defined by \citet{Giacalone2021}, allowing us to conclude that \target\ c is a non-self-luminous object transiting one of the two stars in the binary. Furthermore, combined with the upper mass limit provided by the RV monitoring, we consider \target\ c to be a confirmed planet.}

\textcolor{black}{For the remainder of this paper, we consider the 34-day \textcolor{black}{RV-detected signal} to be a likely planet candidate (\target\ b), and the 272-day transit and RV-detected signal to be a confirmed planet (\target\ c).}

\section{Analysis of stellar system} \label{sec:stellar_system}

The combination of archival and new high resolution spectral imaging data revealed a bound companion star. Given the brightness ratio of the two stars of $F_B/F_A = 0.7$, we expect both stars to contribute to the spectra. However, there are no observable RV shifts from the companion star, detectable changes in the shape of the absorption lines, nor evidence for double lined spectra. As discussed in Section~\ref{subsec:RV}, HIRES is able to detect companion stars where $\Delta$RV $\geq$ 10 km s$^{-1}$. We assume for the remainder of this paper that the obtained HIRES and SOPHIE spectra are a composite of the light from both stars, where the relative RV shift between the two stars is less than 10 km s $^{-1}$. Similarly, we assume that the \tess\ light curve is a composite of the light from both stars. In this section we discuss the properties of the two stars and the stellar configuration. 

\subsection{Stellar parameter determination} \label{sec:stellar_params}

In order to determine stellar parameters for this system, we use precise multi-wavelength photometric measurements as well as high-resolution spectra. We note that while the multi-wavelength photometric fits account for the multiple stars in the system, the spectroscopic solutions only consider one star and are subject to error.



\subsubsection{Spectroscopic parameter determination}  \label{subsec:spec_params}

To extract stellar parameter values from the spectra, we used a moderate signal-to-noise (per pixel S/N = 46) iodine-free observation obtained as a reconnaissance observation using the HIRES spectrograph on the Keck I telescope \citep{vogt1994}. We measured the effective temperature (T$_\mathrm{eff}$), surface gravity (log $g$), iron abundance ( [Fe/H] ), and projected rotational velocity of the star using the tools available in the SpecMatch software package \citep{petigura2015}. We first corrected the observed wavelengths to be in the observer's rest frame by cross-correlating a solar model with the observed spectrum. Then, we fit for T$_\mathrm{eff}$, log $g$, [Fe/H], $v$sin$i$, and the instrumental \textcolor{black}{point spread function (PSF)} using the underlying Bayesian differential-evolution Markov Chain Monte Carlo (MCMC) machinery of ExoPy \citep{fulton2013}. At each step in the MCMC chains, a synthetic spectrum is created by interpolating the \citet{coelho2014} grid of stellar models for a set of T$_\mathrm{eff}$, log $g$, and [Fe/H] values and solar alpha abundance. We convolved this synthetic spectrum with a rotational plus macroturbulence broadening kernel using the prescriptions of \citet{valenti2005} and \citet{hirano2011}. Finally, we performed another convolution with a Gaussian kernel to account for the instrumental PSF, and compared the synthetic spectrum with the observed spectrum to assess the goodness of fit. The priors are uniform in T$_\mathrm{eff}$, log $g$, and [Fe/H], but we assign a Gaussian prior to the instrumental PSF that encompasses the typical variability in the PSF width caused by seeing changes and guiding errors. Five echelle orders of the spectrum were fit separately and the resulting posterior distributions were combined before taking the median values for each parameter. Parameter uncertainties were estimated as the scatter in spectroscopic parameters given by SpecMatch relative to the values for 352 stars in the \citet{valenti2005} sample and 76 stars in the \citet{huber2013} asteroseismic sample. Systematic trends in SpecMatch values as a function of T$_\mathrm{eff}$, log $g$, and [Fe/H] relative to these benchmark samples were fit for and removed in the final quoted parameter values.

Although the approach described above performs well for main sequence, single star parameter estimation, calibrations using an empirical spectral library can often result in more robust parameters. As such, we independently determined the stellar parameters of \target\ using SpecMatch-Emp, which follows a similar procedure as that described above but using an empirical library of stellar spectra taken with Keck/HIRES \citep{Yee2017}. While we find that our determinations of T$_\mathrm{eff}$, log $g$, and [Fe/H] are in excellent agreement between the two methods within their respective uncertainties, the determined values of the stellar radius are discrepant (R$_{\rm SpecMatch-Syn }$ = 1.50 $\pm$ 0.04 R$_\odot$ and R$_{\rm SpecMatch-Emp}$ = 1.11$\pm$0.18 R$_\odot$). In addition to the discrepancy between these two methods, both are affected by the assumption that the light originates from a single star. We, therefore, independently derived the stellar parameters using the spectral energy distribution (SED), as this is able to account for the light contributions of both stars in the system. 



\subsubsection{SED fitting} 
\label{sec:SED}

\begin{figure}[]
    \centering
    \includegraphics[width=1\columnwidth,trim=100 90 90 420,clip]{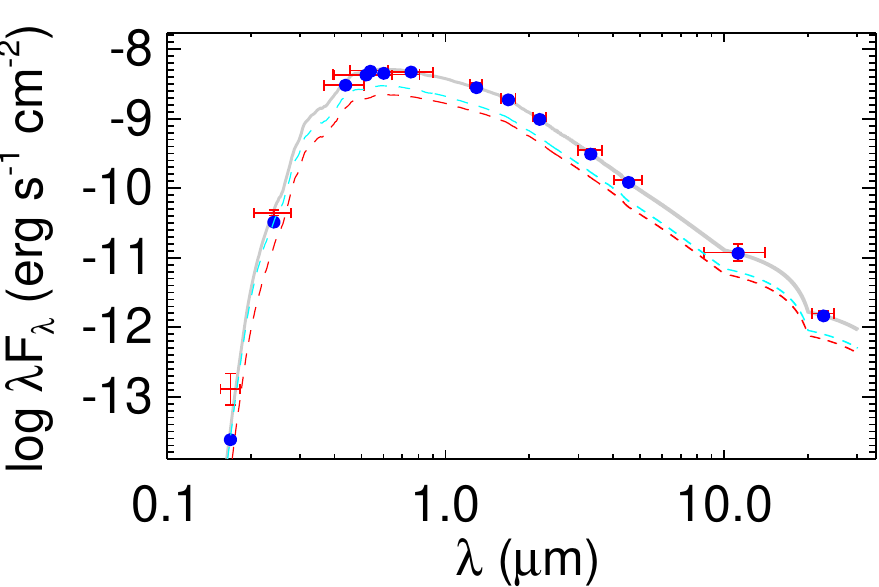}
    \caption{Spectral energy distribution of TIC~307958020. Red symbols represent the observed photometric measurements, where the horizontal bars represent the effective width of the passband. Blue symbols are the model fluxes from the best-fit Kurucz atmosphere model (grey) for the combined light of the binary star system. Cyan and red curves represent the Kurucz atmosphere models corresponding to the warmer (larger) and cooler (smaller) stellar components of the system, \textcolor{black}{respectively.}} 
    \label{fig:SED}
\end{figure}

As an independent determination of the stellar parameters, we performed an analysis of the broadband SED of the star together with the {\it Gaia\/} DR3 parallax \citep[with no systematic offset applied; see, e.g.,][]{StassunTorres:2021}, in order to determine an empirical measurement of the stellar radius, following the procedures described in \citet{Stassun:2016,Stassun:2017,Stassun:2018}. We pulled the $B_T V_T$ magnitudes from {\it Tycho-2}, the $JHK_S$ magnitudes from {\it 2MASS}, the W1--W4 magnitudes from {\it WISE}, the $G$, $G_{\rm BP}$, and $G_{\rm RP}$ magnitudes from {\it Gaia}, and the FUV and NUV magnitudes from {\it GALEX}. Together, the available photometry spans the full stellar SED over the wavelength range 0.2--22~$\mu$m (see Figure~\ref{fig:SED}).  

For an initial fit, we treated the SED as arising from a single star, with the fitted parameters being the effective temperature (\teff), surface gravity (\logg), and metallicity ([Fe/H]), as well as the extinction $A_V$, which we limited to the maximum line-of-sight value from the Galactic dust maps of \citet{Schlegel:1998}. For the initial single-star model fit, we adopted the \teff, \logg, and [Fe/H] from the spectroscopically determined values. The goodness of fit was, not surprisingly, only marginally good ($\chi_\nu^2 = 4.5$) and, given the stringent {\it Gaia\/} distance, implied an oversized star with $R_\star \approx 1.5$~R$_\odot$. 

Next, we performed a two-component fit using the methodology of \citet{Stassun:2016}, solving for the component \teff\ and $R_\star$ by requiring the flux-weighted average \teff\ to agree with the spectroscopic value used above while also requiring the flux ratio $F_{\rm B}/F_{\rm A}$ in the visible to match that determined from the high contrast imaging, and of course requiring agreement with the strict {\it Gaia\/} distance. 

The result, shown in Figure~\ref{fig:SED}, has $\chi_\nu^2 = 1.1$ and best-fit parameters of $A_V = 0.07 \pm 0.02$, $T_{\rm eff,A} = 5800 \pm 50$~K, $R_{\rm A} = 1.05 \pm 0.05$~R$_\odot$, $T_{\rm eff,B} = 5600 \pm 50$~K, and $R_{\rm B} = 0.98 \pm 0.05$~R$_\odot$. 
Based on the empirical relations of \citet{Torres:2010}, the hotter/larger star appears to have a mass $M_{\rm A} = 1.10 \pm 0.06$~M$_\odot$, and the cooler/smaller star has $M_{\rm B} = 1.05 \pm 0.06$~M$_\odot$, consistent with an interpretation of main-sequence stars. 

Finally, the $v$sin$i$ \textcolor{black}{measurement reported by APOGEE DR16 ($v$sin$i$ = 4.39 \kms) together with the brighter star's radius implies a maximum rotation} period of $12.1 \pm 1.3$~d. With the gyrochronology relations of \citet{Mamajek:2008}, this implies a minimum age of $1.3 \pm 0.3$~Gyr, again consistent with unevolved main-sequence stars. \textcolor{black}{As a comparison, the $v$sin$i$ derived using the SpecMatch software package from the HIRES data is 4.25 \kms, which is also consistent with the system containing main-sequence stars.} We note \textcolor{black}{that the $v$sin$i$ estimates are likely strongly affected by the contamination of the companion star. Finally, as mentioned in Section~\ref{subsec:activity}, the \tess\ and ASAS-SN data show no signs of measurable rotation and thus a more precise age of this system cannot be determined using gyrochronology.}

As the two-component SED analysis is able to account for the light contribution of both stars, we adopt these stellar parameters for the remainder of the paper. All parameters are listed in Table~\ref{tab:star}. 

\renewcommand{\arraystretch}{1.1}
\begin{table*}
\centering
  \caption{Stellar system parameters. \label{tab:star}}  
  \begin{tabular}{lcc}
  \hline
  \hline
  {\bf Identifiers} & Value & Source  \\
  \hline
  HU &  918     &    \citet{1905Hussey}   \\
  TOI &   4633      &       \\
  TIC  & 307958020 & \cite{Stassun19} \\
  Gaia DR3 & 1630906044157332224 &  \textit{Gaia} \textcolor{black}{DR3}$^{(\mathrm{a})}$ \\
  2MASS & J17072238+6228330 & 2MASS$^{(\mathrm{b})}$ \\
  \noalign{\smallskip}
  \hline
  {\bf Astrometry} & Value & Source  \\
  \hline  
  \noalign{\smallskip}
  $\alpha_{\rm J2000}$ & \textcolor{black}{17:07:22.396}  &   \textit{Gaia} \textcolor{black}{eDR3}$^{(\mathrm{a})}$ \\
  $\delta_{\rm J2000}$ & \textcolor{black}{62:28:33.011}  &  \textit{Gaia} \textcolor{black}{eDR3}$^{(\mathrm{a})}$ \\
  Distance (pc) & 95.20  $\pm$ 0.24 &  \cite{2018Bailer} \\
  $\pi$ (mas) &  \textcolor{black}{10.552  $\pm$ 0.014} & \textit{Gaia} \textcolor{black}{eDR3}$^{(\mathrm{a})}$ \\
  Spectral Type &  \textcolor{black}{early G Dwarf}  &  \\
  \noalign{\smallskip}
  \hline
  {\bf Photometry} & Magnitude & Source  \\
  \hline
  \noalign{\smallskip}
  B  &  $ 9.767 \pm 0.033  $ & \textit{Tycho-2}$^{(\mathrm{c})}$\\
  V  &  $ 9.017 \pm 0.002  $ & \textit{Tycho-2}$^{(\mathrm{c})}$\\
  J  &  $ 7.723 \pm 0.030 $ & 2MASS$^{(\mathrm{b})}$\\
  H  &  $ 7.448 \pm 0.044 $ & 2MASS$^{(\mathrm{b})}$\\
  K  &  $ 7.349 \pm 0.024 $ & 2MASS$^{(\mathrm{b})}$\\
  W1 &  $ 7.230  \pm 0.039 $ & WISE$^{(\mathrm{d})}$ \\
  W2 &  $ 7.326 \pm  0.020 $ & WISE$^{(\mathrm{d})}$\\
  W3 &  $ 7.305 \pm 0.016 $ & WISE$^{(\mathrm{d})}$\\
  \noalign{\smallskip}
  \hline
  {\bf SED derived properties}  & \textbf{Star A } & \textbf{Star B} \\
  \hline
  \noalign{\smallskip}
    Effective temperature $\mathrm{T_{eff}}$ (K)  & \textcolor{black}{$5800 \pm 50 $ }                & $5600 \pm 50 $ \\
    Stellar mass $M_{\star}$ ($M_\odot$)          & \textcolor{black}{$1.10 \pm 0.06 $ }             & $1.05 \pm 0.06 $ \\
    Stellar radius $R_{\star}$ ($R_\odot$)        & \textcolor{black}{$1.05 \pm 0.05 $ }            & $0.98 \pm 0.05 $ \\
    \noalign{\smallskip}
    \hline
    {\bf {\sc orbitize!} binary parameters}  & Prior$^e$ & Derived value \\
    \hline
    \noalign{\smallskip}
    Semi-major axis a$_{bin}$ (au) &   $\mathcal{U}[0,1000]$ & 48.6$_{- 3.5}^{ + 4.4}$  \\
    Eccentricity e$_{bin}$          &   $\mathcal{U}[0,1]$ & 0.91 $_{- 0.03}^{ + 0.03}$  \\
    Inclination i$_{bin}$ (deg)    &   $\mathcal{U}[0,180]$ & 90.1 $_{- 0.4}^{ + 0.4}$  \\
    $\omega$ (deg)      &     $\mathcal{U}[0,180]$ & 110.5$_{- 2.1 }^{ + 2.1}$  \\
    $\Omega$ (deg)      &     $\mathcal{U}[0,180]$ & 123.5$_{- 2.9}^{ + 3.3}$  \\ 
    Phase     &     $\mathcal{U}[0,1]$ & 0.42 $_{- 0.05}^{ + 0.05}$  \\ 
    Period (years)      &     derived & 231 $_{- 24}^{ + 32}$  \\ 
    Periastron (au)      &    derived & 4.5 $_{- 1.5}^{ + 2.1}$  \\ 
    \hline
  \end{tabular}
     \begin{tablenotes}\footnotesize
  \item \textit{Note} -- $^{(\mathrm{a})}$ \textit{Gaia} \textcolor{black}{early Data Release 3} \citep[eDR3; ][]{gaiaedr3}. $^{(\mathrm{b})}$ Two-micron All Sky Survey \citep[2MASS; ][]{2MASS2003}. $^{(\mathrm{c})}$ \textit{Tycho}-2 catalog \citep{Hog2000}.  $^{(\mathrm{d})}$ Wide-field Infrared Survey Explorer catalog \citep[WISE; ][]{Cutri2013}. $^{(\mathrm{e})}$ $\mathcal{U}[a,b]$ refers to uniform priors between $a$ and $b$. The uncertainties in the derived values represent the 68\% confidence interval.
\end{tablenotes}
\end{table*}


\subsection{Binary orbit modelling}
\label{subsec:binary_mod}

The orbital parameters of the binary system (\target\ AB) were determined by Bayesian parameter estimation using the open-source software `{\sc orbitize!}' \citep{2020Blunt}. The position angles and angular separations extracted from all of the available imaging data (listed in Table~\ref{tab:imaging}) were used as input data. Observations obtained prior to 2011 did not report uncertainties on their measurements. We, therefore, adopted large uncertainties for these archival position angles and angular separations of $\pm$~10 deg and $\pm$~50 milliarcseconds, respectively. As discussed in Section~\ref{subsec:RV}, the HIRES data allows us to place an upper limit on the relative motion of the two stars of 10 \kms. As such, RV values of 0 $\pm$ 5 \kms at the times of the HIRES observations are used as input RV measurements for the {\sc orbitize!} model.

We used the parallel-tempered Affine-invariant sampler {\sc ptemcee} \citep{mcmc2013ForemanMackey,2016Vousden} and adopted priors of 10.55 $\pm$ 0.013 milliarcseconds \citep{gaiaedr3} for the parallax and 1.10 $\pm$ 0.2~M$_{\odot}$ and 1.05 $\pm$ 0.2~M$_{\odot}$ for the masses of the primary and secondary stars, respectively (see section~\ref{sec:SED}). Due to the lack of more constraining RV measurements, values of $\omega$ and $\Omega$ that are separated by 180 $\deg$ are degenerate, as discussed in \citet{2020Blunt}. In order to account for this, we used uniform priors between 0 and 180 $\deg$ for both of these parameters. All priors are listed in Table~\ref{tab:star}. 

The sampler was run using 40 temperatures \textcolor{black}{(using an exponential ladder, with each temperature increasing by a factor of $\sqrt{2}$, so the highest temperature is $\sqrt{2}^{40}$)}, 1000 walkers per temperature and 50 million steps per walker. Convergence was assessed by visual inspection of the chains. \textcolor{black}{Due to the ambiguity in the two position angles ($\pm$ 180 $\deg$ ambiguity) derived from the speckle observations obtained in 2011 and 2022, we ran the {\sc orbitize!} model four times to account for each possible combination of position angles.} Convergence was not reached after 50 million runs when either of the two position angles obtained from speckle observations were rotated by 180 $\deg$ from the values listed in Table~\ref{tab:imaging}. As such, we will assume for the remainder of this paper that the position angles listed in this table are the most likely to be correct.

\begin{figure}
    \centering
    \includegraphics[width=1.02\columnwidth]{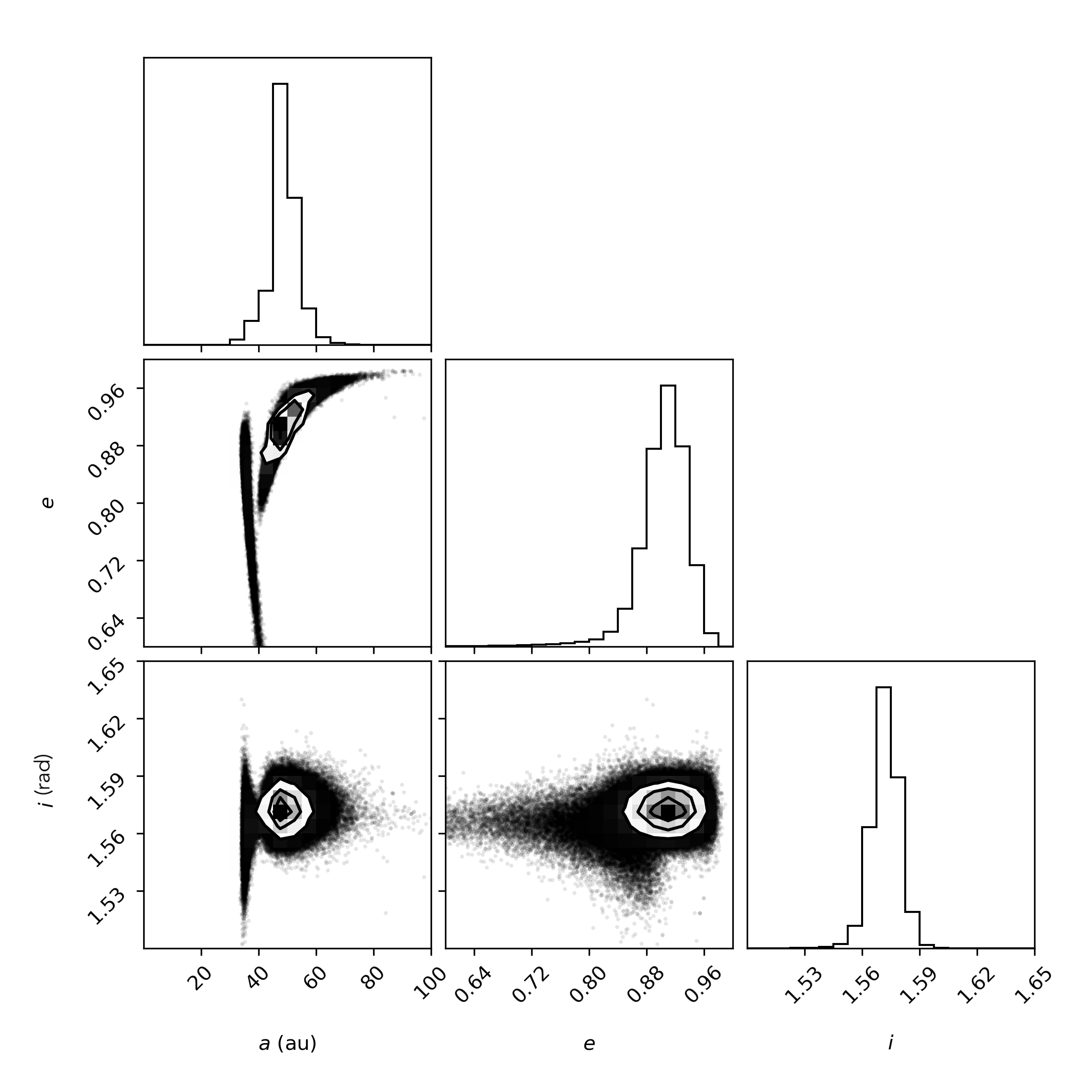}
    \caption{Corner plot of the posteriors of the {\sc orbitize!} results.} 
    \label{fig:orbitize_corner}
\end{figure}

\begin{figure}
    \centering
    \includegraphics[width=1.0\columnwidth]{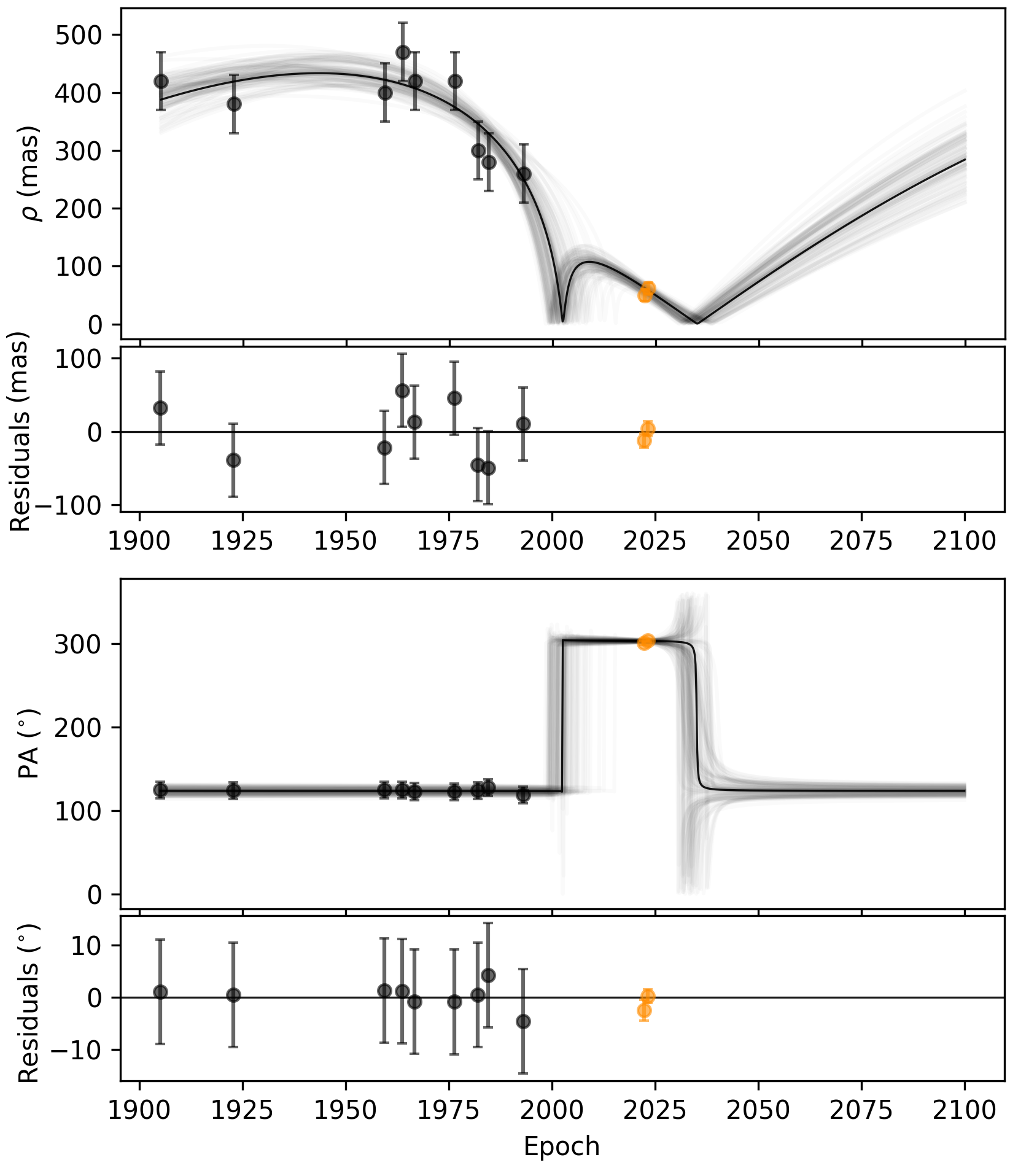}
    \caption{\textcolor{black}{100 orbits randomly drawn from the posterior distribution of the {\sc orbitize!} model (grey lines) for the angular separations (top panel) and position angle (bottom panel) of \target\ AB. The best fit models constructed from the median of the posteriors of each parameter are shown in black, and the residuals to these models are shown below each panel. Archival astrometric data, which have no uncertainties reported in the literature, are shown by the black points. We adopted large uncertainties for these archival position angles and angular separations of $\pm$~10 deg and $\pm$~50 milliarcseconds, respectively. All newly obtained data with reported uncertainties are shown in orange.}} 
    \label{fig:orbitize_model}
\end{figure}

\textcolor{black}{We found the binary semi-major axis, eccentricity and inclination to be constrained to a$_{\rm bin}$ = 48.6$_{- 3.5}^{ + 4.4}$  au, e$_{\rm bin}$ = 0.91 $_{- 0.03}^{ + 0.03}$, and i$_{\rm bin}$ = 90.1 $_{- 0.4}^{ + 0.4} \deg$ (see Figure~\ref{fig:orbitize_corner} for the posterior distributions of these three parameters). This corresponds to a stellar orbital period of 231 $_{- 24}^{ + 32}$ years. Figure~\ref{fig:orbitize_model} shows 100 model fits to the position angles and angular separation that were randomly sampled from the posteriors. The derived RV model indicates a semi-amplitude of 7.2 \kms, with a predicted RV shift of less than 0.2 \kms during the ground-based RV observational time base.}

\section{Analysis of planets} \label{sec:planet_analysis}

\textcolor{black}{In this Section we discuss the properties of the two observed planetary signals, where we make the assumption that both signals originate from planets that are orbiting the same star.} Given the current small angular separation of the two stars, we are unable to determine around which of the two stars either of the planets are orbiting. For the remainder of the paper we report all planet properties under the assumption that star A is the host star for both planets. We note that due to the similarities and large uncertainties in the derived properties of the two stars (M$_{A}$ = 1.10$\pm$ 0.06 M$_{\odot}$, R$_{A}$ = 1.05$\pm$0.05 R$_{\odot}$; M$_{B}$ = 1.05$\pm$0.06 M$_{\odot}$, R$_{B}$ = 0.98$\pm$0.05 R$_{\odot}$), the derived planet properties agree to within their uncertainties when derived with either of the two stars as the host.


\subsection{Joint transit and RV modelling} \label{subsec:pyaneti}

\begin{figure*}
    \centering
    \includegraphics[width=0.9\textwidth]{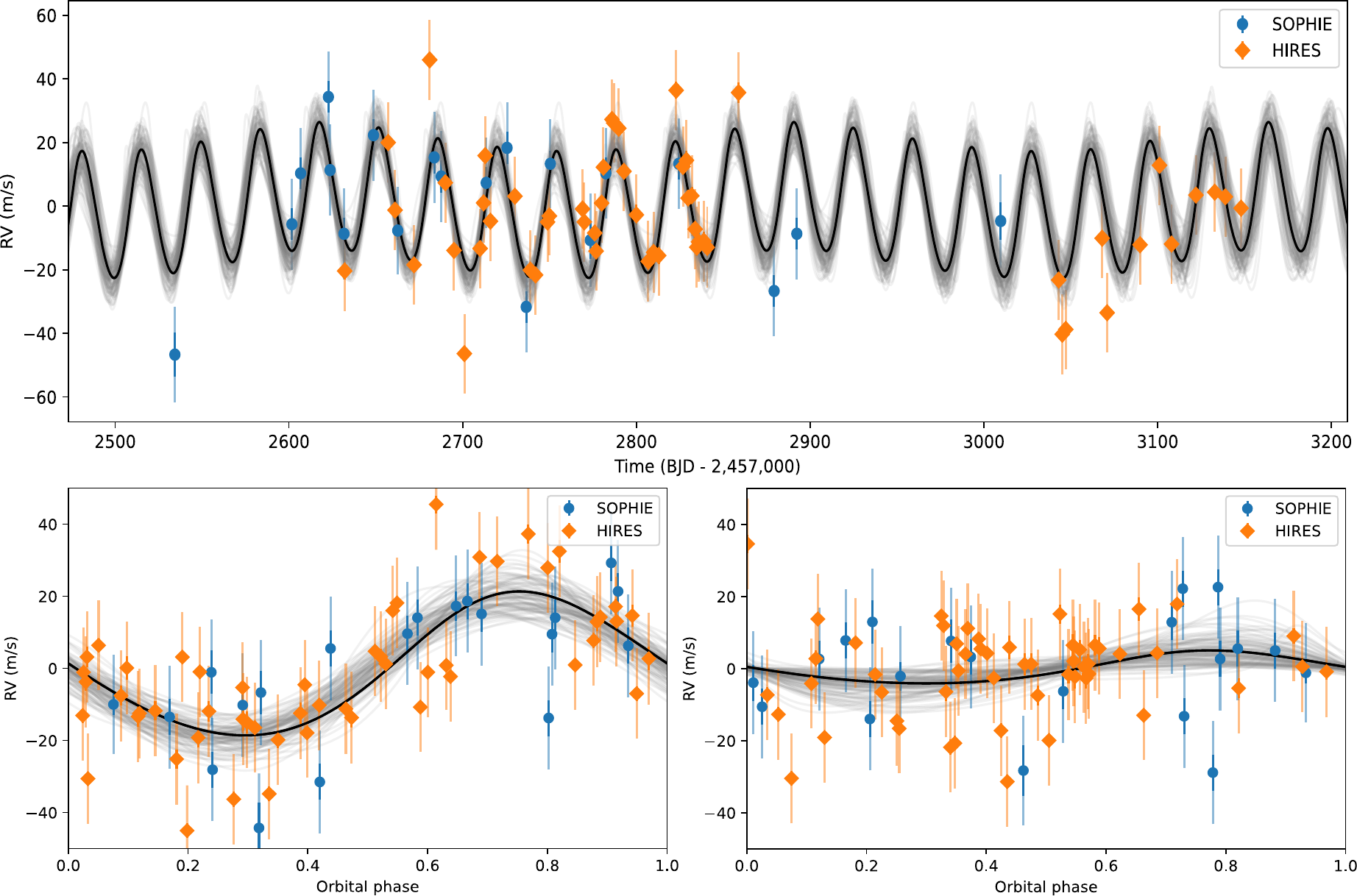}
    \caption{ RV time-series (top) and phase-folded plots for \target\,b (lower left) and \target\,c (lower right).
    SOPHIE (blue circles) and HIRES (orange diamonds) RV measurements are shown following the subtraction of the systemic velocities for each instrument.
    The light colored error bars show the uncertainties accounting for the jitter. Solid black lines show the inferred median model. Grey lines show models made with 100 random samples from the posterior distributions. 
    } 
    \label{fig:RV_modelling}
\end{figure*}

The transit and RV data were jointly modelled using \texttt{pyaneti} \citep[][]{pyaneti,pyaneti2022}. This \textcolor{black}{open-source software} creates marginalised posterior distributions for different orbital parameters by sampling the parameter space using \textcolor{black}{an} MCMC approach. The transits are modeled using the limb-darkened quadratic models by \citet{Mandel2002} while the RV data are fit with two Keplerian RV models. 
In order to account for an RV offset between the HIRES and SOPHIE data, we allow for a systematic velocity for each instrument and include a jitter term per instrument to account for imperfections in our transit and RV model.
\textcolor{black}{For each sector, we used either the 2-minute cadence or, where available, the 20-second cadence data for the transit model (see Section~\ref{sec:tess_discovery}) and both the HIRES and SOPHIE data for the RV fit.} 
All fitted parameters and priors used for the joint modelling are presented in Table~\ref{tab:parstarget}.  

\begin{figure}
    \centering
    \includegraphics[width=1\columnwidth]{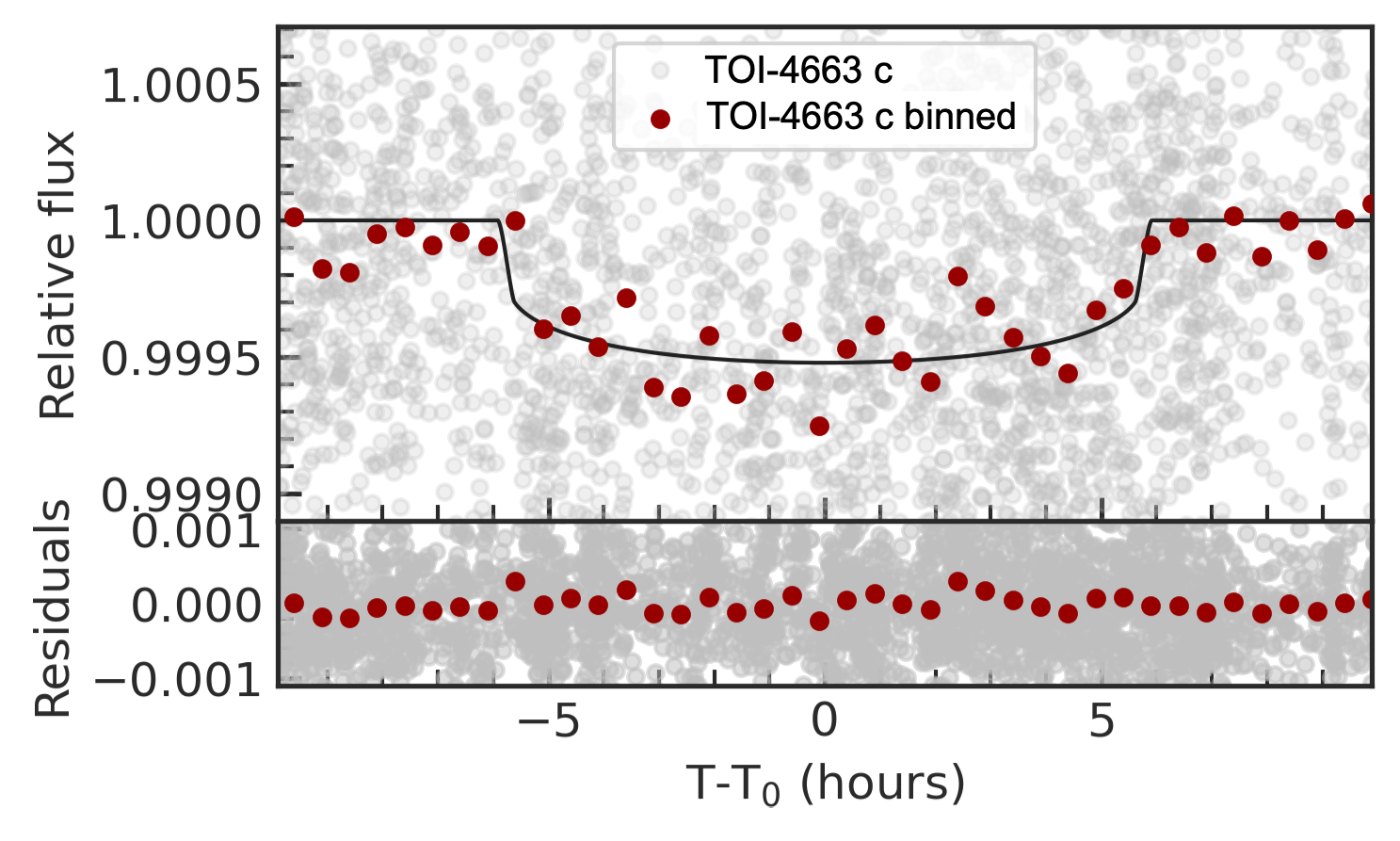}
    \caption{Phase-folded light curve for \target\,c. 2-minute and 20-second cadence \tess\ observations are shown in light grey. Solid color circles represent 60-min binned data. The inferred transit model is shown with a solid black line.} 
    \label{fig:transit_modelling}
\end{figure}

The parameter space was sampled using an MCMC approach with 250 \textcolor{black}{individual} chains and posterior distributions were generated using 5000 iterations of converged chains with a thin factor of 10. The inferred parameters extracted from the posteriors are shown in Table~\ref{tab:parstarget} while the inferred models are shown in Figures~\ref{fig:RV_modelling} and \ref{fig:transit_modelling}. \textcolor{black}{We note that we recover relatively large jitter terms in the RV modelling of \jHIRES and \jOHP for HIRES and SOPHIE, respectively.} This can be caused by a sub-optimal RV model or unknown systematics affecting the RV data. We suspect that the contamination from the second star also contributes to the observed scatter. \textcolor{black}{We see no evidence of a long-term RV trend.}

We checked for additional periodic signals in the data using the PyAstronomy Generalized Lomb-Scargle Periodogram on the residual RV data. We find a significant peak at a period of 182.8 days with a false alarm probability of $\sim$1.5\%. However, as this is half of the orbital period of the Earth, we consider this to most likely be a systematic effect and do not discuss this signal any further.

\renewcommand{\arraystretch}{1}
\begin{table*}
\centering
  \caption{System parameters. \label{tab:parstarget}}  
  \begin{tabular}{lccl}
  \hline
  Parameter & Prior$^{(\mathrm{a})}$ & Value$^{(\mathrm{b})}$ & Comments \\
  \hline
  \multicolumn{4}{l}{\textit{Model Parameters for \target\ b}} \\
  \noalign{\smallskip}
    Orbital period $P_{\mathrm{orb}}$ (days)  &  $\mathcal{U}[33,36]$ & \Pb[] \\
    Time of min. conjunction $T_0$ (BJD - 2457000)  & $\mathcal{U}[2787.32 , 2807.32]$ & \Tzerob[]  \\
    Parametrization $\sqrt{e} \sin \omega$  &  $\mathcal{U}[-1,1]$ & \esinb & The code ensures $e < 1$   \\
    Parametrization $\sqrt{e} \cos \omega$   &  $\mathcal{U}[-1,1]$ &\ecosb & The code ensures $e < 1$  \\
    Doppler semi-amplitude $K$ (\ms)$^{(\mathrm{c})}$ & $\mathcal{U}[0,30]$ & \kb[] & \\
    \noalign{\smallskip}
    \multicolumn{4}{l}{\textit{Model Parameters for \target\ c}} \\
    \noalign{\smallskip}
    Orbital period $P_{\mathrm{orb}}$ (days)  &  $\mathcal{U}[271.9 , 272.1 ]$ & \Pc[] &  \\
    Transit epoch $T_0$ (BJD - 2457000)  & $\mathcal{U}[ 1863.81 , 1865.81]$ & \Tzeroc[]  \\
    Parametrization $\sqrt{e} \sin \omega$  &  $\mathcal{U}[-1,1]$ &  \esinc & The code ensures $e < 1$   \\
    Parametrization $\sqrt{e} \cos \omega$   &  $\mathcal{U}[-1,1]$ & \ecosc & The code ensures $e < 1$  \\
    Observed scaled planet radius  $R_\mathrm{p}/R_{\star}$ &  $\mathcal{U}[0,0.1]$ & \rrc[]  \\
    Impact parameter $b$ &  $\mathcal{U}[0,1.1]$  & \bc[] \\
    Doppler semi-amplitude $K$ (\ms)$^{(\mathrm{c})}$  & $\mathcal{U}[0,30]$ & \kc[] &  11.5 ms$^{-1}$, 99\% percent upper limit   \\
    \multicolumn{4}{l}{\textit{Other Parameters}} \\
    \noalign{\smallskip}
    Stellar density $\rho_\star$ (${\rm g\,cm^{-3}}$) &  $\mathcal{N}[1.34,0.23]$ & \denstrc[] \\
    Parameterized limb-darkening coefficient $q_1$  & $\mathcal{U}[0,1]$ & \qone & $q_1$ parameter as in \citet{Kipping2013} \\
    Parameterized limb-darkening coefficient $q_2$ & $\mathcal{U}[0,1]$ & \qtwo & $q_2$ parameter as in \citet{Kipping2013} \\
    Offset velocity HIRES (\kms) & $\mathcal{U}[-0.50 , 0.50]$ & \HIRES[] \\
    Offset velocity SOPHIE (\kms) & $\mathcal{U}[-0.50 , 0.50]$ & \OHP[] \\
    Jitter HIRES (\ms) & $\mathcal{J}[1,100]$ & \jHIRES[] \\
    Jitter SOPHIE (\ms) & $\mathcal{J}[1,100]$ & \jOHP[] \\
    Jitter TESS (ppm) & $\mathcal{J}[1,100]$ & \jtr[] \\
    \hline
    \multicolumn{4}{l}{\textit{Derived parameters \target\ b}} \\
  \noalign{\smallskip}
    Planet minimum mass $M_{\rm p} \sin i$ ($M_{\oplus}$)$^{(\mathrm{c})}$  & $\cdots$ & \mpb[] &  \\
    Eccentricity $e$ & $\cdots$ & \eb[] &   \\
    Argument of periastron $w$ (deg) & $\cdots$ & \wb[] &   \\
    \multicolumn{4}{l}{\textit{Derived parameters \target\ c}}
    \\
  \noalign{\smallskip}
    Planet mass ($M_{\oplus}$)$^{(\mathrm{c})}$  & $\cdots$ & \mpc[] & 123 $M_{\oplus}$, 99\% percent upper limit    \\
    Observed planet radius ($R_{\oplus}$)  & $\cdots$ & \rpc[] \\
    Corrected planet radius ($R_{\oplus}$)  & $\cdots$ & \rpccorr[]  & Light contribution corrected \\
    Semi-major axis $a$ (au)  & $\cdots$ & \ac[] \\
    Eccentricity $e$ & $\cdots$ & \ec[] &  \\
    Argument of periastron $w$ (deg) & $\cdots$ & \wc[] &   \\
    Transit duration $\tau$ (hours) & $\cdots$ & \ttotc[] \\
    Orbit inclination $i$ (deg)  & $\cdots$ & \ic[] \\
    Insolation $F_{\rm p}$ ($F_{\oplus}$)   & $\cdots$ & \insolationc[] \\
    \hline
   \noalign{\smallskip}
  \end{tabular}
  \begin{tablenotes}\footnotesize
  \item \textit{Note} -- All parameters are calculated based on the assumption that both planets orbit star A. $^{(\mathrm{a})}$ $\mathcal{U}[a,b]$ refers to an uniform prior between $a$ and $b$, $\mathcal{N}[a,b]$ to a Gaussian prior with mean $a$ and standard deviation $b$, and $\mathcal{J}[a,b]$ to the modified Jeffrey's prior as defined by \citet[eq.~16]{Gregory2005}.
  $^{(\mathrm{b})}$  Inferred parameters and errors are defined as the median and 68.3 \% credible interval of the posterior distribution. $^{(\mathrm{c})}$ Due to the unknown effect of the second star on the observed semi-amplitude, these are lower limits.
\end{tablenotes}
\end{table*}

\subsubsection{Effects of companion star on planet radius and mass} \label{subsec:planet_radius}

Planet radii are calculated based on the observed transit depth and the measured radius of the host star. However, for multi-stellar systems, the relationship of the observed transit depth and the true planet radius depends on the brightness ratio of the brightness of the star which is being transited to the total brightness of all the stars in the system \citep{2017Furlan}. Assuming that the planet transits star A and given $F_{\rm B}/F_{\rm A} = 0.7$, we correct the observed planet radius of \target~c (R$_{\rm c, obs}$ = \rpc) by a factor of $\sqrt{\frac{F_{\rm A} + F_{\rm B}}{F_{\rm A}}} = \sqrt{1.7}$, resulting in a true planet radius of \rpccorr.  

\begin{figure}
    \centering
    \includegraphics[width=1\columnwidth]{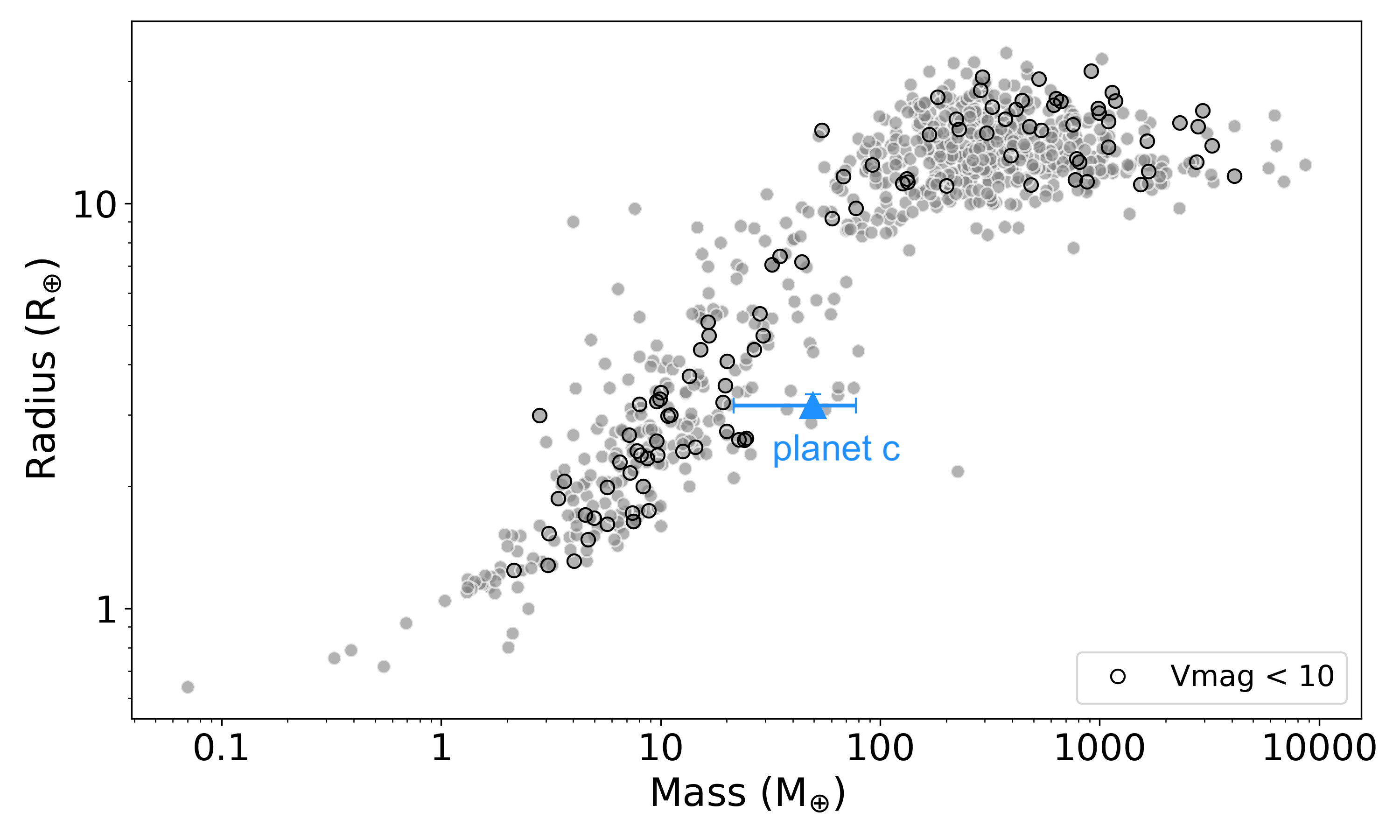}
    \caption{Planet radius versus planet mass for \textcolor{black}{all confirmed planets listed in the NASA exoplanet archive with a mass measurement with 30\% uncertainty or better.} Bright systems (m$_{\rm V}$ $<$ 10) are highlighted with a black outline. \target\ c is shown by the blue triangle, where uncertainties correspond to our radial velocity fit. Although the existence of a second star in this system means that the radial velocity amplitude is diluted and thus the true planet mass may be larger, the lack of planets with the radius of \target\ c at masses larger than 47 M$_\oplus$ suggests that the planet is unlikely to be significantly more massive than what is reported here.} 
    \label{fig:mass_rad}
\end{figure}

Similarly, the presence of the second star acts to reduce the observed amplitude of the RV shifts. The amount by which the observed signal is reduced depends on the relative strengths of the absorption lines (line depths) of both stars, as well as on their rotational velocities (line widths). As we are not able to disentangle the two spectra we cannot quantify the effect of the second star on the RV amplitude. As such, the observed RV amplitudes, and therefore the derived planet masses, are lower limits. However, given the radius of planet c and the known distribution of planet masses for a given planetary radius (as shown in Figure~\ref{fig:mass_rad}), we consider it unlikely for planet c to have a mass significantly greater than 47 M$_{\oplus}$.


\subsection{Search and recovery of transit signals}
\label{subsec:inj_recovery}

In order to search for transits of planet candidate b using the \tess\ data, and to search for additional transit signals, we searched the full \tess\ light curve using the Box Least Squares \citep[BLS; ][]{bls2002} algorithm. Before running the BLS search, we masked the transit signals of \target~c and used an iterative non-linear filter to subtract residual systematics on timescales $>1.7$ days \citep{Aigrain04}. We carried out the BLS search on an evenly sampled frequency grid ranging from 0.00125 to 1 d$^{-1}$ (1 to 800 days). The signal detection efficiency (SDE), defined as the ratio of the highest peak in the SNR periodogram relative to its standard deviation, was used to determine the significance of the recovery of the signal. The algorithms found no additional signals above an SDE of 7.6. 

Furthermore, we used an injection and recovery test to quantify the detectability of additional planets in the \tess\ data, following the methodology outlined in \citet{toi813}. In brief we injected transit signals generated using the {\sc batman} package \citep{Kreidberg15} into the PDC \tess\ light curve. The injected transit signals corresponded to planets with radii ranging from 1 to 12.5 R$_\oplus$ and periods ranging from 1 to 300~days, both sampled \textcolor{black}{randomly} from a log-uniform  distribution. The impact parameter and eccentricity were assumed to be zero for simplicity. We used the stellar parameters given in Table~\ref{tab:star} and adopted a quadratic limb-darkening law with $q_1$ and $q_2$ of 0.16 and 0.59, respectively, as taken from Table~15 in \cite{Claret2017}. 

\begin{figure}
    \centering
    \includegraphics[width=1.05\columnwidth]{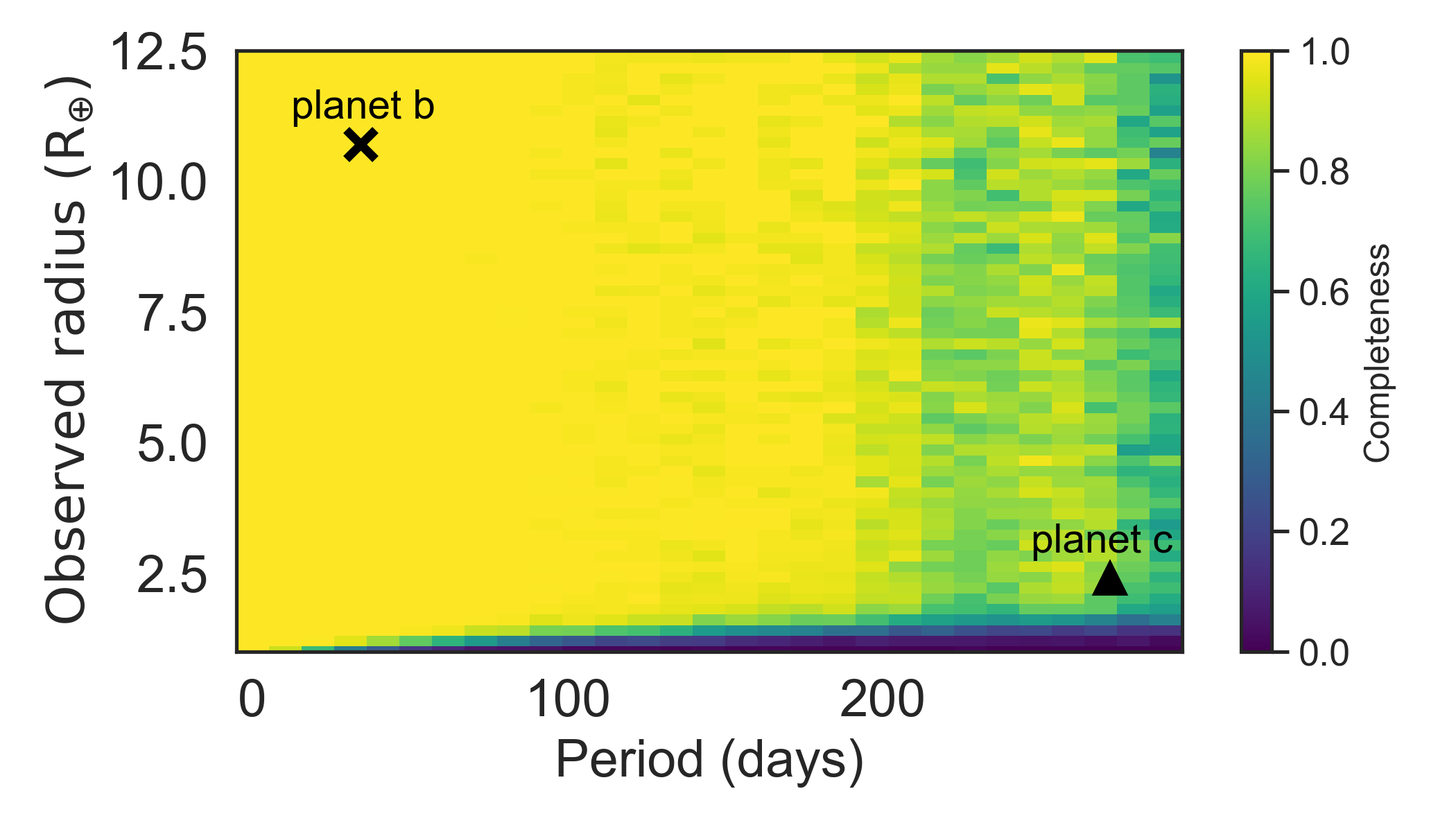}
    \caption{The recovery completeness of injected transit signals
into the light curve of \target\ as a function of the radius and orbital period. The radius of planet candidate b was estimated using the planet mass--radius scaling relations for volatile-rich planets by \citet{2020Otegi}. The signals were recovered using a BLS search. The figure clearly highlights that if planet candidate b were aligned such that it would transit star A, the transit events would have been detected by the BLS search algorithm.} 
    \label{fig:inj_recover}
\end{figure}

We simulated and injected transits for 500,000 planets and used the BLS methodology described above to try to recover each injected signal. For each simulation we identified the highest peak in the BLS periodogram. The signal was considered to be correctly identified when the corresponding period and orbital phase were within 1\% of the injected values. The fraction of recovered signals over a grid of period and radius bins was used to evaluate the completeness of the injection and recovery search. The radius and period bins have widths of 0.2 R$_\oplus$ and 10 days respectively, as shown in Figure~\ref{fig:inj_recover}. Using the planet mass--radius scaling relations for volatile-rich planets by \citet{2020Otegi} and the measured minimum planet mass of 109 M$_\oplus$, \textcolor{black}{planet candidate b has an estimated radius of $\sim$13.7 R$_\oplus$ corresponding to a minimum density of 0.11 g\,cm$^{-3}$ (assuming b=0).} Given the light contribution of the companion star which would act to dilute the transit, this would result in an observed radius of $\sim$10.7~R$_\oplus$ (indicated by the black cross). Figure~\ref{fig:inj_recover} shows that the properties of planet candidate b are such that if the planet were transiting we would be able to identify the transit events in the \tess\ data. This, in turn, allows us to constrain the inclination of planet candidate b to \textit{i} $<$ 88.5 deg.

\subsection{Dynamical stability}
\label{subsec:stability}

The orbital periods of planet candidate b and planet c are much shorter than the period of the binary orbit, indicating that this is a circumstellar, or ``s-type" system. \textcolor{black}{Furthermore, the large period ratio between the planets' orbits and the binary orbits indicate that this system is in a hierarchical configuration. Additionally, the significant eccentricity of the binary orbit and that the mutual inclination between the planets and the binary orbit is not known indicates that the stability of this system can be analyzed in light of the Eccentric Kozai-Lidov mechanism \cite[see e.g.,][]{Kozai,Lidov,Naoz16},
where the system is unstable if the mutual inclination between the binary and the planets lies between the ``Kozai Angles", that is $39.2^\circ \leq i_{mut} \leq 140.77^\circ$. Within these angles, the planets will undergo large oscillations in their eccentricities and inclinations. We explore the stability of orbital configurations with low mutual inclinations using the framework of} \citet{Quarles+20}, who numerically constrain the stability regime of Earth-mass planets in circumstellar orbits in binary systems with approximately 700 million N-body simulations. The grid map of \citet{Quarles+20} indicates that, \textcolor{black}{in the prograde orbit case,} without \target\ b, the transiting planet \target\ c remains stable against perturbation by the binary orbit if its mutual inclination with the binary orbit ($i_{\rm m,bin-c}$) is between $0$-$45^\circ$. 

However, as this is likely a two-planet system, planet-planet interactions should be considered. It has been shown that planet-planet interactions (and \textcolor{black}{general-relativistic} precession of the periapsis) can suppress eccentricity oscillations and destabilization from the influence of an outer perturber \citep[see e.g.,][]{Naoz16,Denham+19, Wei+21, Faridani+22}. In this system, however, the period ratio between the two planets is too large for planet-planet interactions to meaningfully stabilize \target\ c against the perturbations from the binary if the binary and the planets have meaningful mutual inclination \citep[as calculated using Equation 10 from][]{Denham+19}. Therefore, destabilizing planet-planet interactions, such as scattering, are considered.

\begin{figure*}
    \centering
    \includegraphics[width=1\textwidth]{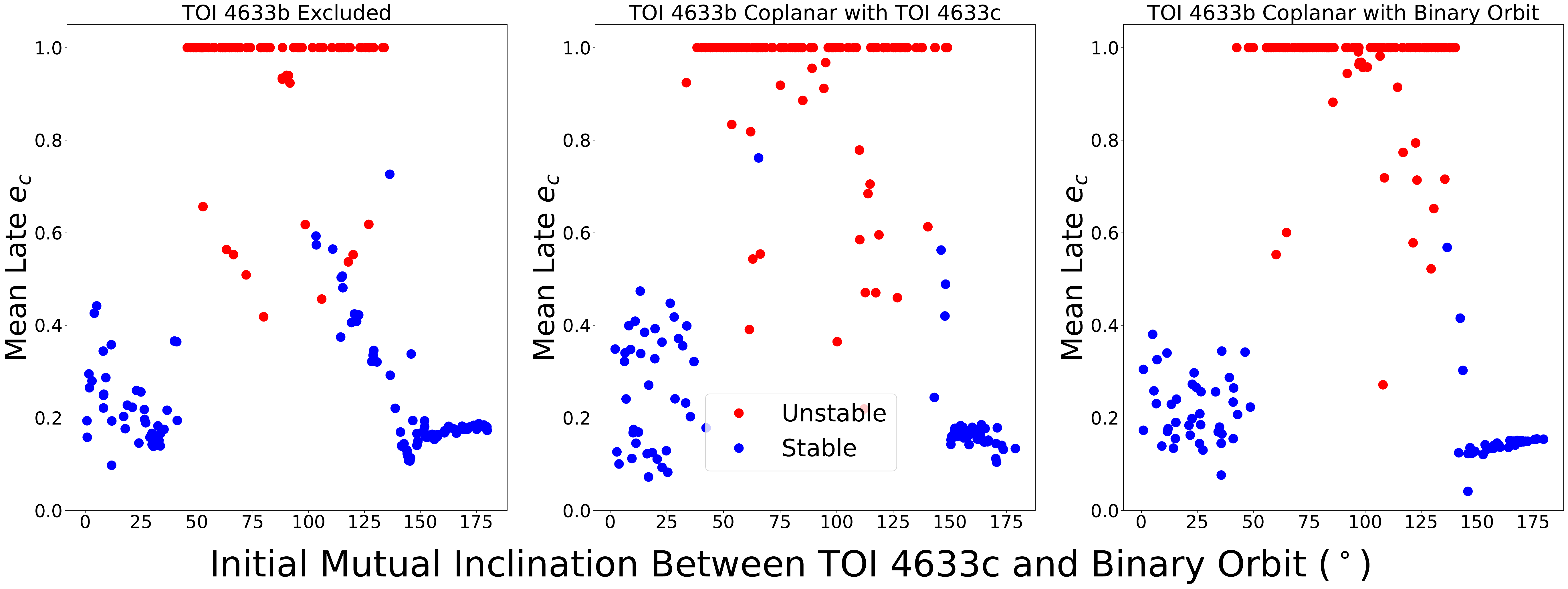}
    \caption{ Outcome of simulations organized by mass and initial mutual inclination between TOI 4633c and the binary orbit. We show 600 {\tt MERCURY} integrations run for $10^6$ years each across three scenarios indicated by the labels on the figures: \textbf{1)} without the 34-day planet, \textbf{2)} where TOI 4633b has initial inclination of $\sim 88^\circ$ narrowly avoiding transit and is coplanar with TOI 4633c, and \textbf{3)} where TOI 4633b is coplanar with the binary orbit. The x-axis is the initial mutual inclination set between TOI 4633c and the binary orbit, and the y-axis is the average eccentricity of TOI 4633c during the final 10\% of the integration time. Each simulation is represented by a point--blue for runs that remain stable and red for unstable. } 
    \label{fig:Faridani_Mercury_Sims}
\end{figure*}

Figure \ref{fig:Faridani_Mercury_Sims} shows the results of 600 N-body simulations of the \target\ system (with \target\ A hosting the planets) under three scenarios. The scenarios are (from left-to-right) 1) where \target\ b is not included, 2) where \target\ b has an initial inclination that narrowly avoids transit $(i\sim 88^\circ)$ but initially has a mutual inclination of $2^\circ$ or less with \target\ c, and 3) where \target\ b is initially coplanar with the binary orbit. 
For each scenario, the average eccentricity of \target\ c during the last $10\%$ of each simulation is plotted against \target\ c's initial mutual inclination with the binary orbit.
In each run, the longitude of ascending node and argument of periapsis of \target\ c are randomized, its eccentricity is initialized to $0.118$, and \target\ b (if present) has the same longitude of ascending node and argument of periapsis as \target\ c in the second scenario or the binary orbit in the third scenario. 
The runs were performed using the $N$-body code {\tt MERCURY} \citep[][]{mercury6}. This version of {\tt MERCURY} includes the first post-Newtonian term accounting for \textcolor{black}{general-relativistic precession}, (M.~Payne, private communication). 


The possible orbits of the planets in the system can be constrained by eliminating orbits that rapidly become unstable. The runs presented in Figure \ref{fig:Faridani_Mercury_Sims} only ran for $10^6$ years, a short time compared to the likely age of the system (estimated to be around 1.3 Gyr), meaning that unstable initial conditions can be excluded from consideration, and the orbits of the planets can be constrained to only orbits that remain stable. We find that without \target\ b, \target\ c is stable for a wide range of mutual inclinations with the binary--including many retrograde orbits, where mutual inclinations with the binary below $45^\circ$ or above $125^\circ$ (retrograde) remain stable. If \target\ b is included coplanar with \target\ c, \target\ c again remains stable if its mutual inclination with the binary is below $45^\circ$, but its range of stable retrograde orbits is lessened to above $145^\circ$. When \target\ b is initially coplanar with the binary, the stable regime of \target\ c expands slightly for both prograde orbits--stable up to $\sim 55^\circ$, but not for retrograde orbits where it is still only stable if its mutual inclination with the binary is above $145^\circ$. 

Between the three scenarios, we find that the presence of \target\ b is a lightly destabilizing presence in the system, reducing the number of stable retrograde orbital configurations.
Planet-planet interactions are not sufficient to suppress eccentricity excitation of \target\ c against perturbations from the binary if there is significant mutual inclination between the planets and the binary.
Moreover, because planet-planet interactions are not stabilizing \target\ c against perturbation, the influence of the masses of \target\ b and \target\ c within their allowed ranges are minor. This is because scattering is caused by perturbations by the binary rather than arising from the planet-planet interactions, so whether scattering occurs is not affected by varying the planet masses within their allowed uncertainties.

\textcolor{black}{We note that the choice of host star (whether the planets orbit \target\ A or \target\ B)  is not expected to significantly affect these results. Using the grids by \citet{Quarles+20}, we show that for mutual inclination ($i_{\rm m,bin-c}$) between 0 and $45^\circ$, planet c  would be stable orbiting around either of the two stars in the binary. As such, with the current available data, we cannot use stability arguments to determine around which of the two stars the planets are orbiting. Furthermore, as previously discussed, due to the similar masses and radii of the two stars in the binary, the derived planet properties are the same to within their uncertainties with either of the two stars as the planet host.}

\section{Discussion and conclusion} \label{sec:results}

The \target\ system consists of two G-type stars, a 272-day period transiting planet and a 34-day period non-transiting planet candidate. The system stands out from other transiting planet systems due to i) the long orbital period of planet c which places it in the habitable zone; ii) the brightness of the system; and iii) the similar-mass, bound stellar companion.

\subsection{Long period planet around a bright star}

\begin{figure}
    \centering
    \includegraphics[width=1.01\columnwidth]{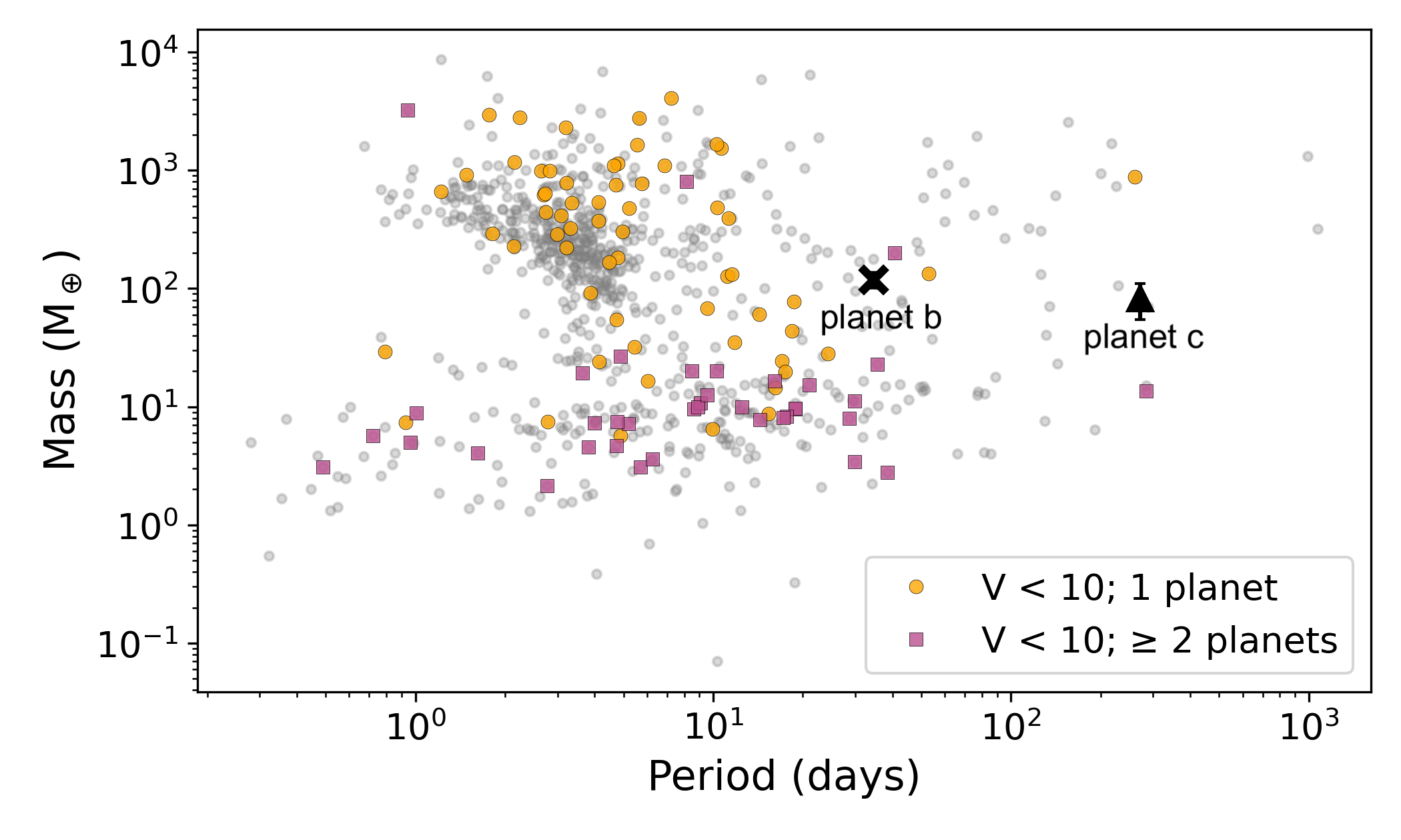}
    \caption{Planet mass versus planet orbital period for all systems with masses measured to 30\% or better uncertainty. We have highlighted planets around bright stars in color. Within the bright star sample, we distinguish between single-planet systems (orange circles) and systems with more than one planet (purple squares). \textcolor{black}{\target~c is the second longest period planet in a bright host star system (however, we do note that planet c does not have a mass measured to 30\% or better uncertainty)}. Furthermore, both planets in the system are fairly high mass and likely not rocky, which is unusual compared to the confirmed population of multi-planet systems.} 
    \label{fig:multis_context}
\end{figure}

To date, there is a distinct lack of confirmed transiting planets with long orbital periods, with only 175 planets with periods longer than 100 days, and 40 with periods longer than 250 days (only 5 of which are brighter than Vmag = 12).\footnote{NASA Exoplanet Archive} As such, with a period of $\sim$ 272 days and a host star brightness of Vmag = 9.0, planet c lies in an under-explored region of parameter space, as shown in Figure~\ref{fig:multis_context}. The figure shows the planet mass versus orbital period for all confirmed planets listed in the NASA exoplanet archive with a mass measurement with 30\% uncertainty or better as grey points, and highlights bright systems (m$_{\rm V}$ $<$ 10) in color. Bright systems with only one planet detected are shown as orange circles, while bright systems with more than one planet are shown as purple squares. Planet candidate b and c are shown by the black cross and triangle, respectively. The figure highlights a lack of confirmed planets with mass measurements to better that 30\% residing in bright (V$<$10) multi-planet systems. Furthermore, it shows that planet c is the second longest-period planet known around a bright star; \textcolor{black}{and that if both planets are orbiting around the same star, \target\ would be one of only a few bright, multi-planet systems.} The difficulty of detecting long-period planets is illustrated by the fact that the transit probability of a planet at the semi-major axis of \target\ c is 0.6\%. Overall, this highlights the importance of further characterisation of \target\ in order to help further our understanding of planetary system demographics. Particularly, the brightness of the host star makes \target\ c a prime candidate for further ground- and space-based characterisation. 

\subsection{Two planets, two stars}






The combination of new and archival high contrast imaging data dating back to 1905 showed that the system is comprised of two stars ($a_{\rm bin}$  = 48.6$_{- 4.4}^{ + 3.5}$ au, $e_{\rm bin}$ = 0.91$_{- 0.03}^{ + 0.03}$, $P_{\rm bin}$ = 231$_{- 32}^{ + 24}$ years). Due to the proximity of the two stars, we are unable to determine around which star the planets orbit, or \textcolor{black}{whether the two planets orbit the same star.}

\begin{figure}
    \centering
    \includegraphics[width=1.1\columnwidth]{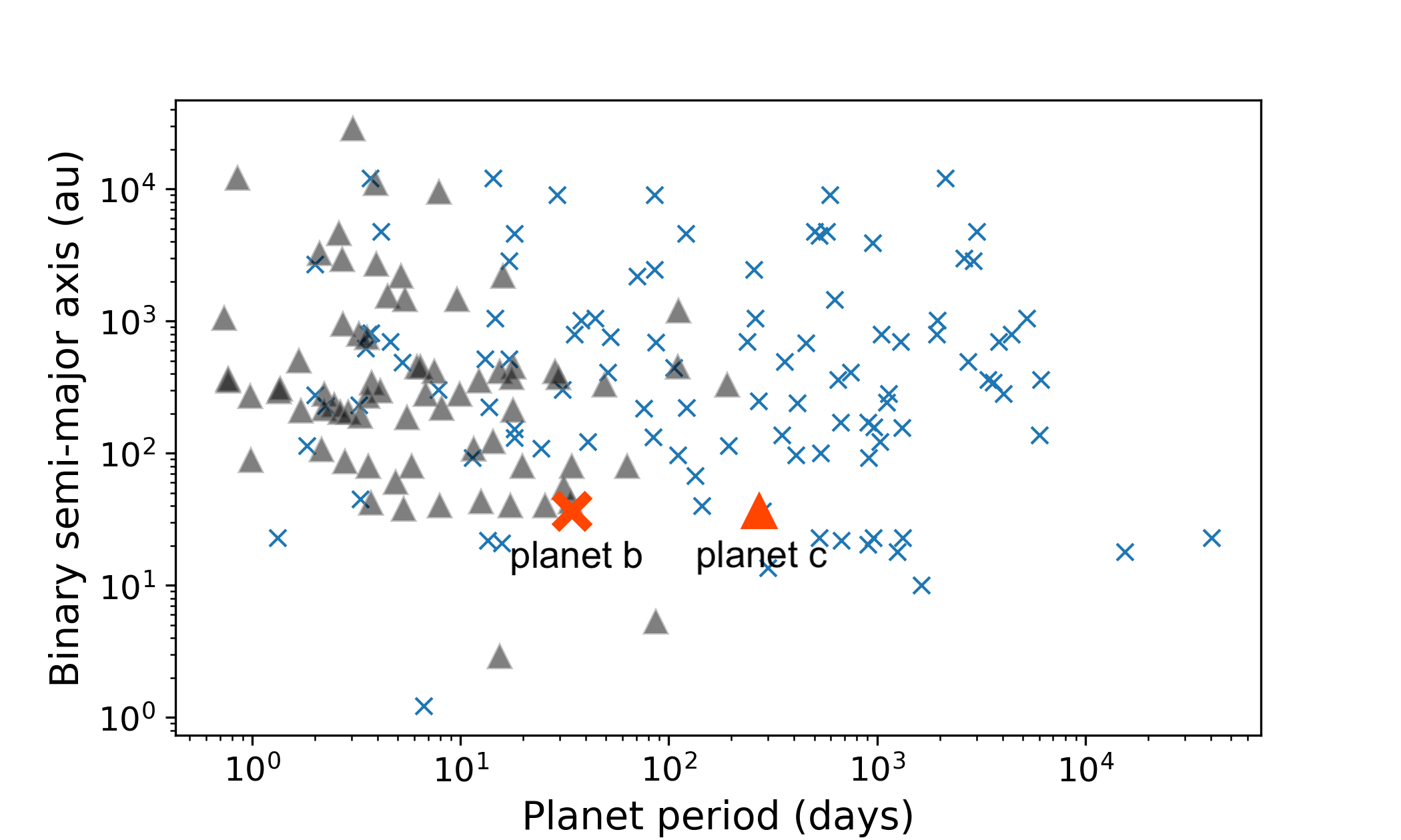}
    \caption{Binary semi-major axis versus planet period for planetary systems with more than one star. Transiting planets are shown as triangles, while radial velocity-detected planets are shown as crosses. \target\ planets are shown in red. \target\ has a smaller binary separation than most planet-hosting binary systems, and \target~c is the longest period transiting planet currently known in a binary system.} 
    \label{fig:binaries}
\end{figure}

Despite the large fraction of stars that reside in binaries, the known sample of confirmed planets in binaries remains limited. The catalogue of exoplanets in binary star systems \citep{2016Schwarz} lists 154 systems containing a total of 217 planets. Out of these, 27 are P-type (`circumbinary planets') and 190 are S-type (`circumstellar planets'). The properties of the S-type planets are shown in Figure~\ref{fig:binaries} on the binary semi-major axis versus planet orbital period plane. Planets listed as having been detected using the transit method (74 planets) and the radial velocity method (107 planets) are depicted by black triangles and blue crosses, respectively. The properties of \target\ b and c are shown in red. As highlighted by this figure, the stellar semi-major axis of \target\ AB, of $\sim$ 48.6 au, places \target\ in an under-sampled region of parameter space, with only 18 confirmed planets around a star with a binary semi-major axis less than that of \target\ AB, only two of which are transiting. Furthermore, it highlights that planet c has the longest orbital period of any confirmed transiting planet in an S-type binary. \textcolor{black}{In addition, there are currently only a handful of circumstellar systems where both stars are known to host a planet \citep[e.g., ][]{2016Teske, 2016Teskeb}. Thus, distinguishing around which star these planets orbit will, in the future, allow for a study of how differences in the host star properties (e.g., chemical abundances) can result in \textcolor{black}{different} planet properties.} 

Stellar multiplicity is widely believed to affect planet formation and evolution. However, the extent and details of the effect of a companion star remain topics of debate. For example, studies have suggested that giant planets on short orbital periods are preferentially found in systems with wide stellar companions \citep[e.g., ][]{2015Wang, 2016Ngo, 2018Ziegler, 2021Ziegler, 2021MoeKratter} due to the effect of the companion star triggering planet inward migration. Similarly, \citet{2021MoeKratter} combined a variety of RV and high-resolution imaging surveys to show that planet occurrence rates are suppressed as a function of binary separation (i.e., smaller binary semi-major axes result \textcolor{black}{in a lower occurrence of giant planets}). \textcolor{black}{\citet{2021MoeKratter} also show that at a binary separation of \target~AB$~\sim$~48.6 au, the observed occurrence rate of planets is around 50\% less compared to field stars. Other studies suggest that there is a suppression of transiting planets in S-type systems, due to the stellar companion disrupting the orbital coplanarity \citep[e.g.,][]{2015cWang}.}

\textcolor{black}{Recent studies have also been investigating whether binary orbital parameters (e.g., binary inclination and eccentricity) affect planet formation and observed planet properties. For example, \citet{2023Lester}, \citet{2022Behmard}, \citet{2022Christian}, and \citet{2022Dupuy} provide observational evidence that suggests that the orbital planes of the binary stars are preferentially aligned with the orbital planes of the planet(s).}



In addition to allowing for population studies that can inform theories of planet formation and migration, planets in binary systems are interesting due to the fact that the two stars formed simultaneously and thus are expected to have the same chemical abundance at the time of formation. As such, any observed differences in the chemical composition of the stars could be related to differences in the outcomes of planet formation and therefore may help provide constraints on how planet formation affects stellar properties \citep[e.g., ][]{2016Teske}. In turn, the identification of a correlation between certain chemical abundances and planet formation could help improve target selection of future space missions searching for planets, based on a single spectrum of the star. Alternatively, observed differences in chemical compositions between the two stars could indicate recent planet engulfment \citep{oh2018, 2023Behmard}. Even though the two stars in this binary are currently too close to one another to be able to separate their spectra and therefore to measure their individual chemical abundances, \textcolor{black}{our model of the orbit} indicates that in $\sim$ 30 years time the two stars will be separated by $\sim$ 150 milliarcseconds, which will allow us to observe the individual stars using instruments such as the Keck Planet Imager and Characterizer \citep{2020Delorme}. 

\subsection{\target\ c: a mini-Neptune in the habitable zone}

\begin{figure}
    \centering
    \includegraphics[width=1.1\columnwidth]{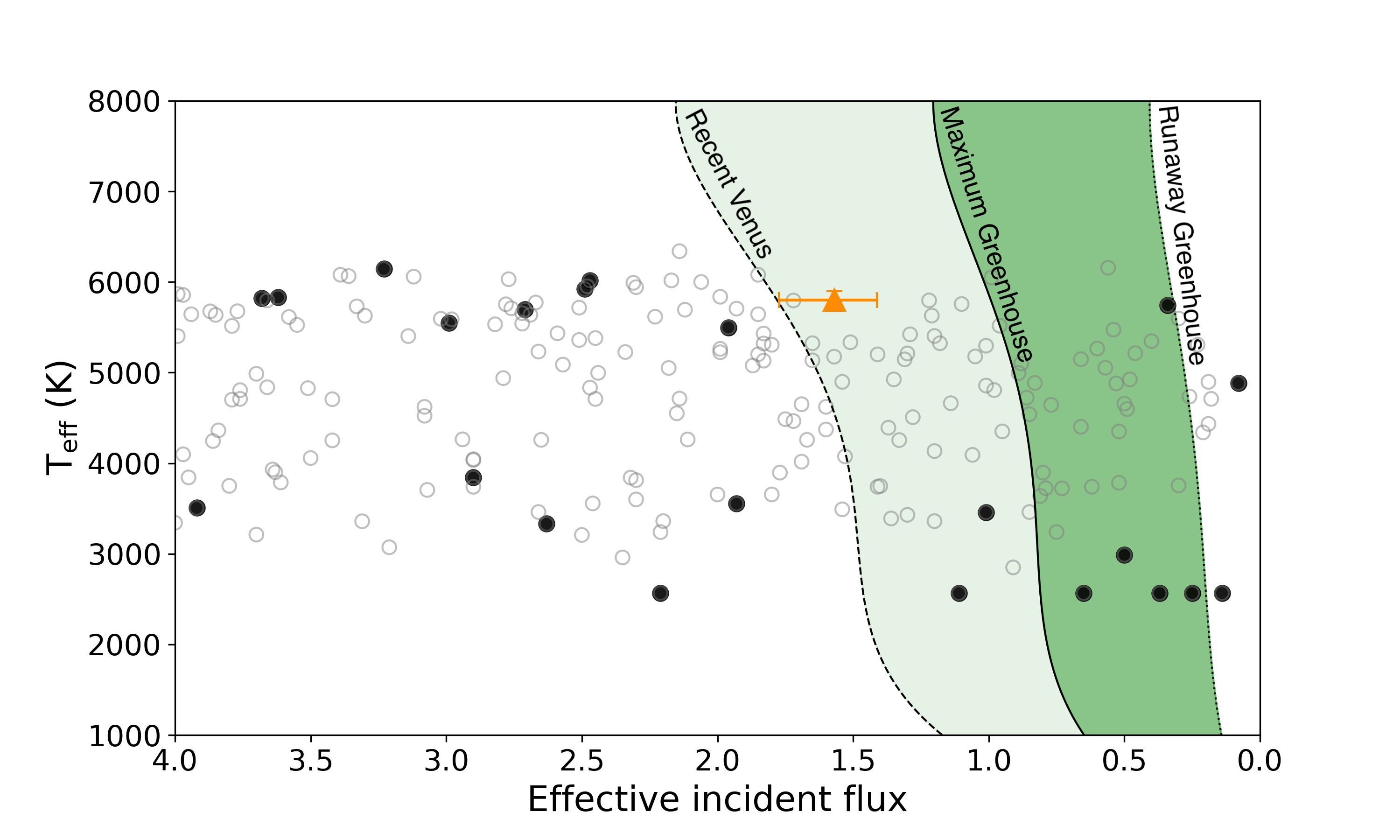}
    \caption{Stellar effective temperature versus effective incident flux on a planet. The boundaries of the habitable zone defined by \citet{2014Kopparapu} are illustrated as shaded regions. The black circles show all transiting exoplanets listed in the NASA exoplanet archive, where the filled circles show only planets that have a mass measurement to better than 30\% accuracy. The orange triangle shows the properties of \target\ c. There are no other transiting planets within the habitable zone that are brighter than 11th magnitude, and only four that are brighter than 13th magnitude. } 
    \label{fig:HZ}
\end{figure}

The search for exoplanets in the habitable zones of their host stars has been a focal point in the field of exoplanet research, where the habitable zone is defined as the region around a star where the surface temperature is favorable for liquid water to exist. We estimate the location of the habitable zone around \target\ following the prescriptions defined by \citet{2014Kopparapu}. The habitable zone boundaries of `Recent Venus', `Maximum Greenhouse', and `Runaway Greenhouse' conditions are shown in Figure~\ref{fig:HZ} \citep[for more details on how these boundaries are defined see ][]{2014Kopparapu}. Using this definition of the habitable zone we find that \target~c, with an insolation of \insolationc, lies inside the inner edge of the potentially habitable region of the star between the Recent Venus and Maximum Greenhouse boundaries. Using the Stefan-Boltzman law and assuming an albedo equivalent to that of Neptune (0.29), we show that the effective surface temperature of planet c is $\sim$ 290 K when only the energy contribution of star A is considered. When accounting for the energy contribution of stars A and B at the time of periastron (stellar separation $\sim$ 4.5 au), the effective surface temperature increases to $\sim$ 420 K. 

According to the Catalog of Habitable Zone Exoplanets by \citet{2023Hill}, \target\ is the brightest star known to host a transiting planet in the habitable zone, with no other transiting planets within the habitable zone that are brighter than 11th magnitude, five that are brighter than 13th magnitude, and 29 that are brighter than 15th magnitude. Even though planet c is at a distance from its host star where liquid water could potentially reside, the density of the planet \textcolor{black}{($>$0.11 g\,cm$^{-3}$)} suggests that it has a large and dense atmosphere that likely makes surface level liquid water impossible. However, as shown independently by \citet{2019Sucerquia} and \citet{2021Dobos}, the probability of a planet hosting a satellite increases with increased orbital period. As such, the brightness of the system and the long orbital period of planet c makes \target\ c a valuable target for further characterisation in the future and, in particular, for the search of planet satellites.


\subsection{Conclusion}



We present the discovery and validation of a transiting mini-Neptune (P$_c$ = \Pc, R$_c$ = \rpccorr, M$_c$ = \mpc) that was discovered by citizen scientists taking part in the Planet Hunters TESS citizen science project. The planet's long orbital period places it in the optimistic habitable zone of its host star, with an insolation of \insolationc. RV monitoring of the system with Keck/HIRES and OHP/SOPHIE revealed an additional 34-day periodic signal that is not seen in the photometric data. Due to a lack of strong evidence that this periodic signal is caused by stellar activity, we tentatively consider this signal to be a planet candidate (M$_b$sin$i$ $<$ 109 M$\oplus$). Furthermore, we used high contrast imaging observations spanning over 117 years (1905 - 2023) to constrain the orbit of the bound stellar companion star (a$_{bin}$ = 48.6$_{- 3.5}^{ + 4.4}$ \textcolor{black}{au}, e$_{bin}$ = 0.91$_{- 0.03}^{ + 0.03}$). 

Dynamical simulations were used to constrain the mutual inclination between the binary orbit and that of \target\ c. The N-body simulations showed that if the 34-day planet candidate is inclined such that it only narrowly avoids transiting, then the mutual inclination between planet c and the binary orbit has to be less than $\sim45^\circ$ to ensure that the system is stable for longer than $10^6$ years.

Overall, \target\ stands out due to its brightness, the long orbital period of the transiting planet \target\ c that places it in the optimistic habitable zone, and the presence of the bound, near equal mass, stellar companion. \target\ c is currently only the fourth habitable zone planet identified in the \tess\ data, and the only one transiting a G dwarf star. The system is valuable for studying planet formation and evolution in binaries, as well as for being an excellent target for detailed investigation of planets in habitable zones, particularly in multiple star systems. While the two stars in the system can not currently be resolved from the ground (due to their proximity), the binary orbit modelling showed that in around 30 years time the stars will be separated by more than 150 milliarcseconds on the sky, a distance large enough to be able to resolve with modern spectrographs such as the Keck Planet Imager and Characterizer. This will enable further characterisation of the two stars and the planets in the system.




\section*{Software and data availability} 

The \textit{TESS} data used within this article are hosted and made publicly available by the Mikulski Archive for Space Telescopes (MAST, \url{http://archive.stsci.edu/tess/}). All the {\it TESS} data used in this paper can be found in MAST: \dataset[10.17909/pepe-f853]{https://dx.doi.org/10.17909/t9-5z05-k040}. This work also used data from the NASA Exoplanet Archive \citep{nasaEA}, and ExoFOP \citep{exofop}.

This work made use of Astropy, a community-developed core Python package for Astronomy \citep{astropy2013}, matplotlib \citep{matplotlib}, pandas \citep{pandas}, NumPy \citep{numpy}, astroquery \citep{ginsburg2019astroquery}, sklearn \citep{pedregosa2011scikit}, emcee \citep{mcmc2013ForemanMackey}, {\sc ptemcee} \citep{2016Vousden}, {\sc triceratops} \citep{Giacalone2021}, {\sc latte} \citep{LATTE2020}, lightkurve \citep{lightkurve2018}, pyaneti \citep{pyaneti2022}, {\sc orbitize!} \citep{2020Blunt}, Mercury \citep{mercury6}, corner \citep{corner}.

\section*{Acknowledgments}

We would like to thank all of the planet hunters who have taken part in the Planet Hunters TESS citizen science project over the past six years. Without all of their dedicated work, many of whom have been part of the project \textit{ab initio}, this exciting system would not have been found. 

We acknowledge the use of public TESS data from pipelines at the TESS Science Office and at the TESS Science Processing Operations Center. Resources supporting this work were provided by the NASA High-End Computing (HEC) Program through the NASA Advanced Supercomputing (NAS) Division at Ames Research Center for the production of the SPOC data products.

We thank the Observatoire de Haute-Provence (CNRS) staff for its support. This work was supported by the ``Programme National de Plan\'etologie'' (PNP) of CNRS/INSU." Some of the observations in this paper made use of the high-resolution imaging instrument \okina Alopeke and were obtained under Gemini LLP Proposal Number: GN/S-2021A-LP-105. ‘Alopeke was funded by the NASA Exoplanet Exploration Program and built at the NASA Ames Research Center by Steve B. Howell, Nic Scott, Elliott P. Horch, and Emmett Quigley. \okina Alopeke was mounted on the Gemini North telescope of the international Gemini Observatory, a program of NSF’s OIR Lab, which is managed by the Association of Universities for Research in Astronomy (AURA) under a cooperative agreement with the National Science Foundation. on behalf of the Gemini partnership: the National Science Foundation (United States), National Research Council (Canada), Agencia Nacional de Investigación y Desarrollo (Chile), Ministerio de Ciencia, Tecnología e Innovación (Argentina), Ministério da Ciência, Tecnologia, Inovações e Comunicações (Brazil), and Korea Astronomy and Space Science Institute (Republic of Korea). 

N.E. thanks the LSSTC Data Science Fellowship Program, which is funded by LSSTC, NSF Cybertraining Grant number 1829740, the Brinson Foundation, and the Moore Foundation. C.A.C. acknowledges that this research was carried out at the Jet Propulsion Laboratory, California Institute of Technology, under a contract with the National Aeronautics and Space Administration (80NM0018D0004). N.S. acknowledges support from the National Science Foundation through the Graduate Research Fellowship Program under Grant 1842402. J.V.Z. acknowledges support from the Future Investigators in NASA Earth and Space Science and Technology (FINESST) grant 80NSSC22K1606. J.M.A.M. acknowledges support from the National Science Foundation Graduate Research Fellowship Program under Grant number DGE-1842400 and from NASA’s Interdisciplinary Consortia for Astrobiology Research (NNH19ZDA001N-ICAR) under award number 19-ICAR19\_2-0041. Xavier Delfosse acknowledges support  by the French National Research Agency in the framework of the Investissements d'Avenir program (ANR-15-IDEX-02), through the funding of the ``Origin of Life" project of the Grenoble-Alpes University.  F.K. acknowledges funding from the European Research Council (ERC) under the European Union's Horizon 2020 research and innovation programme (COBREX; grant agreement number 885593), and from the Initiative de Recherches Interdisciplinaires et Stratégiques (IRIS) of Université PSL ``Origines et Conditions d'Apparition de la Vie (OCAV)". E.M. acknowledges funding from FAPEMIG under project number APQ-02493-22 and a research productivity grant number 309829/2022-4 awarded by the CNPq, Brazil. CJ gratefully acknowledges support from the Netherlands Research School of Astronomy (NOVA) and from the Research Foundation Flanders (FWO) under grant agreement G0A2917N (BlackGEM).

Some of the data presented in this paper were obtained from the Mikulski Archive for Space Telescopes (MAST). The Space Telescope Science Institute (STScI) is operated by the Association of Universities for Research in Astronomy, Inc., under NASA contract NAS5-26555. Support for MAST for non-HST data is provided by the NASA Office of Space Science via grant NNX13AC07G and by other grants and contracts. This paper includes data collected with the \textit{TESS} mission, obtained from the MAST data archive at STScI. Funding for the \textit{TESS} mission is provided by the NASA Explorer Program. STScI is operated by the Association of Universities for Research in Astronomy, Inc., under NASA contract NAS 5–26555.

Finally, N.E. and S.G. wish to thank the Harry Potter franchise for providing us with the in-house nickname for this system of \textit{Percival}, inspired by Albus Percival Wulfric Brian Dumbledore's wisdom. 


%

\vspace{5mm}
\facilities{TESS, OHP/SOPHIE, Gemini North/\okina Alopeke, Keck/HIRES, Palomar/PHARO, WIYN/NESSI, Keck II/NIRC2}


\bibliography{references}{}

\begin{thebibliography}{}
\expandafter\ifx\csname natexlab\endcsname\relax\def\natexlab#1{#1}\fi
\providecommand{\url}[1]{\href{#1}{#1}}
\providecommand{\dodoi}[1]{doi:~\href{http://doi.org/#1}{\nolinkurl{#1}}}
\providecommand{\doeprint}[1]{\href{http://ascl.net/#1}{\nolinkurl{http://ascl.net/#1}}}
\providecommand{\doarXiv}[1]{\href{https://arxiv.org/abs/#1}{\nolinkurl{https://arxiv.org/abs/#1}}}

\bibitem[{{Aigrain} \& {Irwin}(2004)}]{Aigrain04}
{Aigrain}, S., \& {Irwin}, M. 2004, \mnras, 350, 331,
  \dodoi{10.1111/j.1365-2966.2004.07657.x}

\bibitem[{{Astropy Collaboration} {et~al.}(2013){Astropy Collaboration},
  {Robitaille}, {Tollerud}, {Greenfield}, {Droettboom}, {Bray}, {Aldcroft},
  {Davis}, {Ginsburg}, {Price-Whelan}, {Kerzendorf}, {Conley}, {Crighton},
  {Barbary}, {Muna}, {Ferguson}, {Grollier}, {Parikh}, {Nair}, {Unther},
  {Deil}, {Woillez}, {Conseil}, {Kramer}, {Turner}, {Singer}, {Fox}, {Weaver},
  {Zabalza}, {Edwards}, {Azalee Bostroem}, {Burke}, {Casey}, {Crawford},
  {Dencheva}, {Ely}, {Jenness}, {Labrie}, {Lim}, {Pierfederici}, {Pontzen},
  {Ptak}, {Refsdal}, {Servillat}, \& {Streicher}}]{astropy2013}
{Astropy Collaboration}, {Robitaille}, T.~P., {Tollerud}, E.~J., {et~al.} 2013,
  \aap, 558, A33, \dodoi{10.1051/0004-6361/201322068}

\bibitem[{{Auvergne} {et~al.}(2009){Auvergne}, {Bodin}, {Boisnard}, {Buey},
  {Chaintreuil}, {Epstein}, {Jouret}, {Lam-Trong}, {Levacher}, {Magnan},
  {Perez}, {Plasson}, {Plesseria}, {Peter}, {Steller}, {Tiph{\`e}ne}, {Baglin},
  {Agogu{\'e}}, {Appourchaux}, {Barbet}, {Beaufort}, {Bellenger}, {Berlin},
  {Bernardi}, {Blouin}, {Boumier}, {Bonneau}, {Briet}, {Butler}, {Cautain},
  {Chiavassa}, {Costes}, {Cuvilho}, {Cunha-Parro}, {de Oliveira Fialho},
  {Decaudin}, {Defise}, {Djalal}, {Docclo}, {Drummond}, {Dupuis}, {Exil},
  {Faur{\'e}}, {Gaboriaud}, {Gamet}, {Gavalda}, {Grolleau}, {Gueguen},
  {Guivarc'h}, {Guterman}, {Hasiba}, {Huntzinger}, {Hustaix}, {Imbert},
  {Jeanville}, {Johlander}, {Jorda}, {Journoud}, {Karioty}, {Kerjean},
  {Lafond}, {Lapeyrere}, {Landiech}, {Larqu{\'e}}, {Laudet}, {Le Merrer},
  {Leporati}, {Leruyet}, {Levieuge}, {Llebaria}, {Martin}, {Mazy}, {Mesnager},
  {Michel}, {Moalic}, {Monjoin}, {Naudet}, {Neukirchner}, {Nguyen-Kim},
  {Ollivier}, {Orcesi}, {Ottacher}, {Oulali}, {Parisot}, {Perruchot},
  {Piacentino}, {Pinheiro da Silva}, {Platzer}, {Pontet}, {Pradines},
  {Quentin}, {Rohbeck}, {Rolland}, {Rollenhagen}, {Romagnan}, {Russ}, {Samadi},
  {Schmidt}, {Schwartz}, {Sebbag}, {Smit}, {Sunter}, {Tello}, {Toulouse},
  {Ulmer}, {Vandermarcq}, {Vergnault}, {Wallner}, {Waultier}, \&
  {Zanatta}}]{2009AuvergneCorot}
{Auvergne}, M., {Bodin}, P., {Boisnard}, L., {et~al.} 2009, \aap, 506, 411,
  \dodoi{10.1051/0004-6361/200810860}

\bibitem[{{Bailer-Jones} {et~al.}(2018){Bailer-Jones}, {Rybizki}, {Fouesneau},
  {Mantelet}, \& {Andrae}}]{2018Bailer}
{Bailer-Jones}, C.~A.~L., {Rybizki}, J., {Fouesneau}, M., {Mantelet}, G., \&
  {Andrae}, R. 2018, \aj, 156, 58, \dodoi{10.3847/1538-3881/aacb21}

\bibitem[{{Baize}(1967)}]{1967Baize}
{Baize}, P. 1967, Journal des Observateurs, 50, 7

\bibitem[{{Barrag{\'a}n} {et~al.}(2022{\natexlab{a}}){Barrag{\'a}n}, {Aigrain},
  {Rajpaul}, \& {Zicher}}]{pyaneti2022}
{Barrag{\'a}n}, O., {Aigrain}, S., {Rajpaul}, V.~M., \& {Zicher}, N.
  2022{\natexlab{a}}, \mnras, 509, 866, \dodoi{10.1093/mnras/stab2889}

\bibitem[{Barrag\'an {et~al.}(2019)Barrag\'an, Gandolfi, \&
  Antoniciello}]{pyaneti}
Barrag\'an, O., Gandolfi, D., \& Antoniciello, G. 2019, \mnras, 482, 1017,
  \dodoi{10.1093/mnras/sty2472}

\bibitem[{{Barrag{\'a}n} {et~al.}(2022{\natexlab{b}}){Barrag{\'a}n},
  {Armstrong}, {Gandolfi}, {Carleo}, {Vidotto}, {Villarreal D'Angelo},
  {Oklop{\v{c}}i{\'c}}, {Isaacson}, {Oddo}, {Collins}, {Fridlund}, {Sousa},
  {Persson}, {Hellier}, {Howell}, {Howard}, {Redfield}, {Eisner}, {Georgieva},
  {Dragomir}, {Bayliss}, {Nielsen}, {Klein}, {Aigrain}, {Zhang}, {Teske},
  {Twicken}, {Jenkins}, {Esposito}, {Van Eylen}, {Rodler}, {Adibekyan},
  {Alarcon}, {Anderson}, {Akana Murphy}, {Barrado}, {Barros}, {Benneke},
  {Bouchy}, {Bryant}, {Butler}, {Burt}, {Cabrera}, {Casewell}, {Chaturvedi},
  {Cloutier}, {Cochran}, {Crane}, {Crossfield}, {Crouzet}, {Collins}, {Dai},
  {Deeg}, {Deline}, {Demangeon}, {Dumusque}, {Figueira}, {Furlan}, {Gnilka},
  {Goad}, {Goffo}, {Guti{\'e}rrez-Canales}, {Hadjigeorghiou}, {Hartman},
  {Hatzes}, {Harris}, {Henderson}, {Hirano}, {Hojjatpanah}, {Hoyer},
  {Kab{\'a}th}, {Korth}, {Lillo-Box}, {Luque}, {Marmier}, {Mo{\v{c}}nik},
  {Muresan}, {Murgas}, {Nagel}, {Osborne}, {Osborn}, {Osborn}, {Palle},
  {Raimbault}, {Ricker}, {Rubenzahl}, {Stockdale}, {Santos}, {Scott},
  {Schwarz}, {Shectman}, {Raimbault}, {Seager}, {S{\'e}gransan}, {Serrano},
  {Skarka}, {Smith}, {{\v{S}}ubjak}, {Tan}, {Udry}, {Watson}, {Wheatley},
  {West}, {Winn}, {Wang}, {Wolfgang}, \& {Ziegler}}]{Barragan2022}
{Barrag{\'a}n}, O., {Armstrong}, D.~J., {Gandolfi}, D., {et~al.}
  2022{\natexlab{b}}, \mnras, 514, 1606, \dodoi{10.1093/mnras/stac638}

\bibitem[{{Behmard} {et~al.}(2022){Behmard}, {Dai}, \& {Howard}}]{2022Behmard}
{Behmard}, A., {Dai}, F., \& {Howard}, A.~W. 2022, \aj, 163, 160,
  \dodoi{10.3847/1538-3881/ac53a7}

\bibitem[{{Behmard} {et~al.}(2023){Behmard}, {Sevilla}, \&
  {Fuller}}]{2023Behmard}
{Behmard}, A., {Sevilla}, J., \& {Fuller}, J. 2023, \mnras, 518, 5465,
  \dodoi{10.1093/mnras/stac3435}

\bibitem[{{Blunt} {et~al.}(2020){Blunt}, {Wang}, {Angelo}, {Ngo}, {Cody}, {De
  Rosa}, {Graham}, {Hirsch}, {Nagpal}, {Nielsen}, {Pearce}, {Rice}, \&
  {Tejada}}]{2020Blunt}
{Blunt}, S., {Wang}, J.~J., {Angelo}, I., {et~al.} 2020, \aj, 159, 89,
  \dodoi{10.3847/1538-3881/ab6663}

\bibitem[{{Boisse} {et~al.}(2011){Boisse}, {Bouchy}, {H{\'e}brard}, {Bonfils},
  {Santos}, \& {Vauclair}}]{boisse2011disentangling}
{Boisse}, I., {Bouchy}, F., {H{\'e}brard}, G., {et~al.} 2011, \aap, 528, A4,
  \dodoi{10.1051/0004-6361/201014354}

\bibitem[{{Borucki} {et~al.}(2010){Borucki}, {Koch}, {Basri}, {Batalha},
  {Brown}, {Caldwell}, {Caldwell}, {Christensen-Dalsgaard}, {Cochran},
  {DeVore}, {Dunham}, {Dupree}, {Gautier}, {Geary}, {Gilliland}, {Gould},
  {Howell}, {Jenkins}, {Kondo}, {Latham}, {Marcy}, {Meibom}, {Kjeldsen},
  {Lissauer}, {Monet}, {Morrison}, {Sasselov}, {Tarter}, {Boss}, {Brownlee},
  {Owen}, {Buzasi}, {Charbonneau}, {Doyle}, {Fortney}, {Ford}, {Holman},
  {Seager}, {Steffen}, {Welsh}, {Rowe}, {Anderson}, {Buchhave}, {Ciardi},
  {Walkowicz}, {Sherry}, {Horch}, {Isaacson}, {Everett}, {Fischer}, {Torres},
  {Johnson}, {Endl}, {MacQueen}, {Bryson}, {Dotson}, {Haas}, {Kolodziejczak},
  {Van Cleve}, {Chandrasekaran}, {Twicken}, {Quintana}, {Clarke}, {Allen},
  {Li}, {Wu}, {Tenenbaum}, {Verner}, {Bruhweiler}, {Barnes}, \&
  {Prsa}}]{Borucki2010}
{Borucki}, W.~J., {Koch}, D., {Basri}, G., {et~al.} 2010, Science, 327, 977,
  \dodoi{10.1126/science.1185402}

\bibitem[{{Bouchy} {et~al.}(2013){Bouchy}, {D{\'\i}az}, {H{\'e}brard},
  {Arnold}, {Boisse}, {Delfosse}, {Perruchot}, \& {Santerne}}]{2013Bouchy}
{Bouchy}, F., {D{\'\i}az}, R.~F., {H{\'e}brard}, G., {et~al.} 2013, \aap, 549,
  A49, \dodoi{10.1051/0004-6361/201219979}

\bibitem[{{Bouchy} {et~al.}(2009){Bouchy}, {H{\'e}brard}, {Udry}, {Delfosse},
  {Boisse}, {Desort}, {Bonfils}, {Eggenberger}, {Ehrenreich}, {Forveille},
  {Lagrange}, {Le Coroller}, {Lovis}, {Moutou}, {Pepe}, {Perrier}, {Pont},
  {Queloz}, {Santos}, {S{\'e}gransan}, \& {Vidal-Madjar}}]{2009Bouchy}
{Bouchy}, F., {H{\'e}brard}, G., {Udry}, S., {et~al.} 2009, \aap, 505, 853,
  \dodoi{10.1051/0004-6361/200912427}

\bibitem[{{Chambers}(1999)}]{mercury6}
{Chambers}, J.~E. 1999, \mnras, 304, 793,
  \dodoi{10.1046/j.1365-8711.1999.02379.x}

\bibitem[{{Christian} {et~al.}(2022){Christian}, {Vanderburg}, {Becker},
  {Yahalomi}, {Pearce}, {Zhou}, {Collins}, {Kraus}, {Stassun}, {de Beurs},
  {Ricker}, {Vanderspek}, {Latham}, {Winn}, {Seager}, {Jenkins}, {Abe},
  {Agabi}, {Amado}, {Baker}, {Barkaoui}, {Benkhaldoun}, {Benni}, {Berberian},
  {Berlind}, {Bieryla}, {Esparza-Borges}, {Bowen}, {Brown}, {Buchhave},
  {Burke}, {Buttu}, {Cadieux}, {Caldwell}, {Charbonneau}, {Chazov},
  {Chimaladinne}, {Collins}, {Combs}, {Conti}, {Crouzet}, {de Leon},
  {Deljookorani}, {Diamond}, {Doyon}, {Dragomir}, {Dransfield}, {Essack},
  {Evans}, {Fukui}, {Gan}, {Esquerdo}, {Gillon}, {Girardin}, {Guerra},
  {Guillot}, {K. Habich}, {Henriksen}, {Hoch}, {Isogai}, {Jehin}, {Jensen},
  {Johnson}, {Livingston}, {Kielkopf}, {Kim}, {Kawauchi}, {Krushinsky},
  {Kunzle}, {Laloum}, {Leger}, {Lewin}, {Mallia}, {Massey}, {Mori}, {McLeod},
  {M{\'e}karnia}, {Mireles}, {Mishevskiy}, {Tamura}, {Murgas}, {Narita},
  {Naves}, {Nelson}, {Osborn}, {Palle}, {Parviainen}, {Plavchan}, {Pozuelos},
  {Rabus}, {Relles}, {Rodr{\'\i}guez L{\'o}pez}, {Quinn}, {Schmider},
  {Schlieder}, {Schwarz}, {Shporer}, {Sibbald}, {Srdoc}, {Stibbards},
  {Stickler}, {Suarez}, {Stockdale}, {Tan}, {Terada}, {Triaud}, {Tronsgaard},
  {Waalkes}, {Wang}, {Watanabe}, {Wenceslas}, {Wingham}, {Wittrock}, \&
  {Ziegler}}]{2022Christian}
{Christian}, S., {Vanderburg}, A., {Becker}, J., {et~al.} 2022, \aj, 163, 207,
  \dodoi{10.3847/1538-3881/ac517f}

\bibitem[{{Cincunegui} {et~al.}(2007){Cincunegui}, {D{\'\i}az}, \&
  {Mauas}}]{cincunegui2007halpha}
{Cincunegui}, C., {D{\'\i}az}, R.~F., \& {Mauas}, P.~J.~D. 2007, \aap, 469,
  309, \dodoi{10.1051/0004-6361:20066503}

\bibitem[{{Claret}(2017)}]{Claret2017}
{Claret}, A. 2017, \aap, 600, A30, \dodoi{10.1051/0004-6361/201629705}

\bibitem[{{Coelho}(2014)}]{coelho2014}
{Coelho}, P.~R.~T. 2014, \mnras, 440, 1027, \dodoi{10.1093/mnras/stu365}

\bibitem[{{Couteau}(1960)}]{1960Couteau}
{Couteau}, P. 1960, Journal des Observateurs, 43, 1

\bibitem[{{Cutri} \& {et al.}(2013)}]{Cutri2013}
{Cutri}, R.~M., \& {et al.} 2013, VizieR Online Data Catalog, II/328

\bibitem[{{Cutri} {et~al.}(2003){Cutri}, {Skrutskie}, {van Dyk}, {Beichman},
  {Carpenter}, {Chester}, {Cambresy}, {Evans}, {Fowler}, {Gizis}, {Howard},
  {Huchra}, {Jarrett}, {Kopan}, {Kirkpatrick}, {Light}, {Marsh}, {McCallon},
  {Schneider}, {Stiening}, {Sykes}, {Weinberg}, {Wheaton}, {Wheelock}, \&
  {Zacarias}}]{2MASS2003}
{Cutri}, R.~M., {Skrutskie}, M.~F., {van Dyk}, S., {et~al.} 2003, VizieR Online
  Data Catalog, II/246

\bibitem[{{Czesla} {et~al.}(2019){Czesla}, {Schr{\"o}ter}, {Schneider},
  {Huber}, {Pfeifer}, {Andreasen}, \& {Zechmeister}}]{pya}
{Czesla}, S., {Schr{\"o}ter}, S., {Schneider}, C.~P., {et~al.} 2019, {PyA:
  Python astronomy-related packages}.
\newblock \doeprint{1906.010}

\bibitem[{{Dalba} {et~al.}(2022){Dalba}, {Kane}, {Dragomir}, {Villanueva},
  {Collins}, {Jacobs}, {LaCourse}, {Gagliano}, {Kristiansen}, {Omohundro},
  {Schwengeler}, {Terentev}, {Vanderburg}, {Fulton}, {Isaacson}, {Van Zandt},
  {Howard}, {Thorngren}, {Howell}, {Batalha}, {Chontos}, {Crossfield},
  {Dressing}, {Huber}, {Petigura}, {Robertson}, {Roy}, {Weiss}, {Behmard},
  {Beard}, {Brinkman}, {Giacalone}, {Hill}, {Lubin}, {Mayo}, {Mo{\v{c}}nik},
  {Akana Murphy}, {Polanski}, {Rice}, {Rosenthal}, {Rubenzahl}, {Scarsdale},
  {Turtelboom}, {Tyler}, {Benni}, {Boyce}, {Esposito}, {Girardin}, {Laloum},
  {Lewin}, {Mann}, {Marchis}, {Schwarz}, {Srdoc}, {Steuer}, {Sivarani}, {Unni},
  {Eisner}, {Fetherolf}, {Li}, {Yao}, {Pepper}, {Ricker}, {Vanderspek},
  {Latham}, {Seager}, {Winn}, {Jenkins}, {Burke}, {Eastman}, {Lund},
  {Rodriguez}, {Rowden}, {Ting}, \& {Villase{\~n}or}}]{2022Dalba}
{Dalba}, P.~A., {Kane}, S.~R., {Dragomir}, D., {et~al.} 2022, \aj, 163, 61,
  \dodoi{10.3847/1538-3881/ac415b}

\bibitem[{{Delorme} {et~al.}(2020){Delorme}, {Jovanovic}, {Echeverri}, {Mawet},
  {Wallace}, {Bartos}, {Cetre}, {Wizinowich}, {Ragland}, {Wang}, {Ruffio},
  {Lilley}, {Wethrell}, {Doppmann}, {Fitzgerald}, {Ruane}, {Schofield},
  {Calvin}, {Morris}, {Pezzato}, {Llop Sayson}, {Magnone}, {Johnson}, {Shohn},
  {Bond}, {Chun}, {Suominen}, \& {Skemer}}]{2020Delorme}
{Delorme}, J.-R., {Jovanovic}, N., {Echeverri}, D., {et~al.} 2020, in Society
  of Photo-Optical Instrumentation Engineers (SPIE) Conference Series, Vol.
  11447, Society of Photo-Optical Instrumentation Engineers (SPIE) Conference
  Series, 114471P, \dodoi{10.1117/12.2562836}

\bibitem[{{Denham} {et~al.}(2019){Denham}, {Naoz}, {Hoang}, {Stephan}, \&
  {Farr}}]{Denham+19}
{Denham}, P., {Naoz}, S., {Hoang}, B.-M., {Stephan}, A.~P., \& {Farr}, W.~M.
  2019, \mnras, 482, 4146, \dodoi{10.1093/mnras/sty2830}

\bibitem[{{D{\'\i}az} {et~al.}(2016){D{\'\i}az}, {S{\'e}gransan}, {Udry},
  {Lovis}, {Pepe}, {Dumusque}, {Marmier}, {Alonso}, {Benz}, {Bouchy},
  {Coffinet}, {Collier Cameron}, {Deleuil}, {Figueira}, {Gillon}, {Lo Curto},
  {Mayor}, {Mordasini}, {Motalebi}, {Moutou}, {Pollacco}, {Pompei}, {Queloz},
  {Santos}, \& {Wyttenbach}}]{diaz2016harps}
{D{\'\i}az}, R.~F., {S{\'e}gransan}, D., {Udry}, S., {et~al.} 2016, \aap, 585,
  A134, \dodoi{10.1051/0004-6361/201526729}

\bibitem[{{Dobos} {et~al.}(2021){Dobos}, {Charnoz}, {P{\'a}l}, {Roque-Bernard},
  \& {Szab{\'o}}}]{2021Dobos}
{Dobos}, V., {Charnoz}, S., {P{\'a}l}, A., {Roque-Bernard}, A., \& {Szab{\'o}},
  G.~M. 2021, \pasp, 133, 094401, \dodoi{10.1088/1538-3873/abfe04}

\bibitem[{{Dransfield} {et~al.}(2023){Dransfield}, {Timmermans}, {Triaud},
  {D{\'e}vora-Pajares}, {Aganze}, {Barkaoui}, {Burgasser}, {Collins},
  {Cointepas}, {Ducrot}, {G{\"u}nther}, {Howell}, {Murray}, {Niraula},
  {Rackham}, {Sebastian}, {Stassun}, {Z{\'u}{\~n}iga-Fern{\'a}ndez},
  {Almenara}, {Bonfils}, {Bouchy}, {Burke}, {Charbonneau}, {Christiansen},
  {Delrez}, {Gan}, {Garc{\'\i}a}, {Gillon}, {Chew}, {Hesse}, {Hooton}, {Isopi},
  {Jehin}, {Jenkins}, {Latham}, {Mallia}, {Murgas}, {Pedersen}, {Pozuelos},
  {Queloz}, {Rodriguez}, {Schanche}, {Seager}, {Srdoc}, {Stockdale}, {Twicken},
  {Vanderspek}, {Wells}, {Winn}, {de Wit}, \& {Zapparata}}]{2023Dransfield}
{Dransfield}, G., {Timmermans}, M., {Triaud}, A. H.~M.~J., {et~al.} 2023,
  \mnras, \dodoi{10.1093/mnras/stad1439}

\bibitem[{{Dupuy} {et~al.}(2022){Dupuy}, {Kraus}, {Kratter}, {Rizzuto}, {Mann},
  {Huber}, \& {Ireland}}]{2022Dupuy}
{Dupuy}, T.~J., {Kraus}, A.~L., {Kratter}, K.~M., {et~al.} 2022, \mnras, 512,
  648, \dodoi{10.1093/mnras/stac306}

\bibitem[{{Eisner} {et~al.}(2020{\natexlab{a}}){Eisner}, {Lintott}, \&
  {Aigrain}}]{LATTE2020}
{Eisner}, N., {Lintott}, C., \& {Aigrain}, S. 2020{\natexlab{a}}, The Journal
  of Open Source Software, 5, 2101, \dodoi{10.21105/joss.02101}

\bibitem[{{Eisner} {et~al.}(2020{\natexlab{b}}){Eisner}, {Barrag{\'a}n},
  {Aigrain}, {Lintott}, {Miller}, {Zicher}, {Boyajian}, {Brice{\~n}o},
  {Bryant}, {Christiansen}, {Feinstein}, {Flor-Torres}, {Fridlund}, {Gandolfi},
  {Gilbert}, {Guerrero}, {Jenkins}, {Jones}, {Kristiansen}, {Vanderburg},
  {Law}, {L{\'o}pez-S{\'a}nchez}, {Mann}, {Safron}, {Schwamb}, {Stassun},
  {Osborn}, {Wang}, {Zic}, {Ziegler}, {Barnet}, {Bean}, {Bundy}, {Chetnik},
  {Dawson}, {Garstone}, {Stenner}, {Huten}, {Larish}, {Melanson}, {Mitchell},
  {Moore}, {Peltsch}, {Rogers}, {Schuster}, {Smith}, {Simister}, {Tanner},
  {Terentev}, \& {Tsymbal}}]{toi813}
{Eisner}, N.~L., {Barrag{\'a}n}, O., {Aigrain}, S., {et~al.}
  2020{\natexlab{b}}, \mnras, 494, 750, \dodoi{10.1093/mnras/staa138}

\bibitem[{{Eisner} {et~al.}(2021){Eisner}, {Barrag{\'a}n}, {Lintott},
  {Aigrain}, {Nicholson}, {Boyajian}, {Howell}, {Johnston}, {Lakeland},
  {Miller}, {McMaster}, {Parviainen}, {Safron}, {Schwamb}, {Trouille},
  {Vaughan}, {Zicher}, {Allen}, {Allen}, {Bouslog}, {Johnson}, {Simon},
  {Wolfenbarger}, {Baeten}, {Bundy}, \& {Hoffman}}]{eisner2020method}
{Eisner}, N.~L., {Barrag{\'a}n}, O., {Lintott}, C., {et~al.} 2021, \mnras, 501,
  4669, \dodoi{10.1093/mnras/staa3739}

\bibitem[{{ExoFOP}(2019)}]{exofop}
{ExoFOP}. 2019, Exoplanet Follow-up Observing Program - TESS,  IPAC,
  \dodoi{10.26134/EXOFOP3}

\bibitem[{{Faridani} {et~al.}(2022){Faridani}, {Naoz}, {Wei}, \&
  {Farr}}]{Faridani+22}
{Faridani}, T.~H., {Naoz}, S., {Wei}, L., \& {Farr}, W.~M. 2022, \apj, 932, 78,
  \dodoi{10.3847/1538-4357/ac6e38}

\bibitem[{{Fernandes} {et~al.}(2019){Fernandes}, {Mulders}, {Pascucci},
  {Mordasini}, \& {Emsenhuber}}]{2019Fernandes}
{Fernandes}, R.~B., {Mulders}, G.~D., {Pascucci}, I., {Mordasini}, C., \&
  {Emsenhuber}, A. 2019, \apj, 874, 81, \dodoi{10.3847/1538-4357/ab0300}

\bibitem[{{Figueira} {et~al.}(2010){Figueira}, {Marmier}, {Bonfils}, {di
  Folco}, {Udry}, {Santos}, {Lovis}, {M{\'e}gevand}, {Melo}, {Pepe}, {Queloz},
  {S{\'e}gransan}, {Triaud}, \& {Viana Almeida}}]{figueira2010evidence}
{Figueira}, P., {Marmier}, M., {Bonfils}, X., {et~al.} 2010, \aap, 513, L8,
  \dodoi{10.1051/0004-6361/201014323}

\bibitem[{Foreman-Mackey(2016)}]{corner}
Foreman-Mackey, D. 2016, The Journal of Open Source Software, 1, 24,
  \dodoi{10.21105/joss.00024}

\bibitem[{{Foreman-Mackey} {et~al.}(2013){Foreman-Mackey}, {Hogg}, {Lang}, \&
  {Goodman}}]{mcmc2013ForemanMackey}
{Foreman-Mackey}, D., {Hogg}, D.~W., {Lang}, D., \& {Goodman}, J. 2013, \pasp,
  125, 306, \dodoi{10.1086/670067}

\bibitem[{{Fulton} {et~al.}(2018){Fulton}, {Petigura}, {Blunt}, \&
  {Sinukoff}}]{radvel}
{Fulton}, B.~J., {Petigura}, E.~A., {Blunt}, S., \& {Sinukoff}, E. 2018, \pasp,
  130, 044504, \dodoi{10.1088/1538-3873/aaaaa8}

\bibitem[{{Fulton} {et~al.}(2013){Fulton}, {Howard}, {Winn}, {Albrecht},
  {Marcy}, {Crepp}, {Bakos}, {Johnson}, {Hartman}, {Isaacson}, {Knutson}, \&
  {Zhao}}]{fulton2013}
{Fulton}, B.~J., {Howard}, A.~W., {Winn}, J.~N., {et~al.} 2013, \apj, 772, 80,
  \dodoi{10.1088/0004-637X/772/2/80}

\bibitem[{{Furlan} {et~al.}(2017){Furlan}, {Ciardi}, {Everett}, {Saylors},
  {Teske}, {Horch}, {Howell}, {van Belle}, {Hirsch}, {Gautier}, {Adams},
  {Barrado}, {Cartier}, {Dressing}, {Dupree}, {Gilliland}, {Lillo-Box},
  {Lucas}, \& {Wang}}]{2017Furlan}
{Furlan}, E., {Ciardi}, D.~R., {Everett}, M.~E., {et~al.} 2017, \aj, 153, 71,
  \dodoi{10.3847/1538-3881/153/2/71}

\bibitem[{{Gaia Collaboration} {et~al.}(2021){Gaia Collaboration}, {Brown},
  {Vallenari}, {Prusti}, {de Bruijne}, {Babusiaux}, {Biermann}, {Creevey},
  {Evans}, {Eyer}, {Hutton}, {Jansen}, {Jordi}, {Klioner}, {Lammers},
  {Lindegren}, {Luri}, {Mignard}, {Panem}, {Pourbaix}, {Randich}, {Sartoretti},
  {Soubiran}, {Walton}, {Arenou}, {Bailer-Jones}, {Bastian}, {Cropper},
  {Drimmel}, {Katz}, {Lattanzi}, {van Leeuwen}, {Bakker}, {Cacciari},
  {Casta{\~n}eda}, {De Angeli}, {Ducourant}, {Fabricius}, {Fouesneau},
  {Fr{\'e}mat}, {Guerra}, {Guerrier}, {Guiraud}, {Jean-Antoine Piccolo},
  {Masana}, {Messineo}, {Mowlavi}, {Nicolas}, {Nienartowicz}, {Pailler},
  {Panuzzo}, {Riclet}, {Roux}, {Seabroke}, {Sordo}, {Tanga}, {Th{\'e}venin},
  {Gracia-Abril}, {Portell}, {Teyssier}, {Altmann}, {Andrae}, {Bellas-Velidis},
  {Benson}, {Berthier}, {Blomme}, {Brugaletta}, {Burgess}, {Busso}, {Carry},
  {Cellino}, {Cheek}, {Clementini}, {Damerdji}, {Davidson}, {Delchambre},
  {Dell'Oro}, {Fern{\'a}ndez-Hern{\'a}ndez}, {Galluccio}, {Garc{\'\i}a-Lario},
  {Garcia-Reinaldos}, {Gonz{\'a}lez-N{\'u}{\~n}ez}, {Gosset}, {Haigron},
  {Halbwachs}, {Hambly}, {Harrison}, {Hatzidimitriou}, {Heiter},
  {Hern{\'a}ndez}, {Hestroffer}, {Hodgkin}, {Holl}, {Jan{\ss}en}, {Jevardat de
  Fombelle}, {Jordan}, {Krone-Martins}, {Lanzafame}, {L{\"o}ffler}, {Lorca},
  {Manteiga}, {Marchal}, {Marrese}, {Moitinho}, {Mora}, {Muinonen}, {Osborne},
  {Pancino}, {Pauwels}, {Petit}, {Recio-Blanco}, {Richards}, {Riello},
  {Rimoldini}, {Robin}, {Roegiers}, {Rybizki}, {Sarro}, {Siopis}, {Smith},
  {Sozzetti}, {Ulla}, {Utrilla}, {van Leeuwen}, {van Reeven}, {Abbas}, {Abreu
  Aramburu}, {Accart}, {Aerts}, {Aguado}, {Ajaj}, {Altavilla}, {{\'A}lvarez},
  {{\'A}lvarez Cid-Fuentes}, {Alves}, {Anderson}, {Anglada Varela}, {Antoja},
  {Audard}, {Baines}, {Baker}, {Balaguer-N{\'u}{\~n}ez}, {Balbinot}, {Balog},
  {Barache}, {Barbato}, {Barros}, {Barstow}, {Bartolom{\'e}}, {Bassilana},
  {Bauchet}, {Baudesson-Stella}, {Becciani}, {Bellazzini}, {Bernet}, {Bertone},
  {Bianchi}, {Blanco-Cuaresma}, {Boch}, {Bombrun}, {Bossini}, {Bouquillon},
  {Bragaglia}, {Bramante}, {Breedt}, {Bressan}, {Brouillet}, {Bucciarelli},
  {Burlacu}, {Busonero}, {Butkevich}, {Buzzi}, {Caffau}, {Cancelliere},
  {C{\'a}novas}, {Cantat-Gaudin}, {Carballo}, {Carlucci}, {Carnerero},
  {Carrasco}, {Casamiquela}, {Castellani}, {Castro-Ginard}, {Castro Sampol},
  {Chaoul}, {Charlot}, {Chemin}, {Chiavassa}, {Cioni}, {Comoretto}, {Cooper},
  {Cornez}, {Cowell}, {Crifo}, {Crosta}, {Crowley}, {Dafonte}, {Dapergolas},
  {David}, {David}, {de Laverny}, {De Luise}, {De March}, {De Ridder}, {de
  Souza}, {de Teodoro}, {de Torres}, {del Peloso}, {del Pozo}, {Delbo},
  {Delgado}, {Delgado}, {Delisle}, {Di Matteo}, {Diakite}, {Diener},
  {Distefano}, {Dolding}, {Eappachen}, {Edvardsson}, {Enke}, {Esquej}, {Fabre},
  {Fabrizio}, {Faigler}, {Fedorets}, {Fernique}, {Fienga}, {Figueras},
  {Fouron}, {Fragkoudi}, {Fraile}, {Franke}, {Gai}, {Garabato},
  {Garcia-Gutierrez}, {Garc{\'\i}a-Torres}, {Garofalo}, {Gavras}, {Gerlach},
  {Geyer}, {Giacobbe}, {Gilmore}, {Girona}, {Giuffrida}, {Gomel}, {Gomez},
  {Gonzalez-Santamaria}, {Gonz{\'a}lez-Vidal}, {Granvik},
  {Guti{\'e}rrez-S{\'a}nchez}, {Guy}, {Hauser}, {Haywood}, {Helmi}, {Hidalgo},
  {Hilger}, {H{\l}adczuk}, {Hobbs}, {Holland}, {Huckle}, {Jasniewicz},
  {Jonker}, {Juaristi Campillo}, {Julbe}, {Karbevska}, {Kervella}, {Khanna},
  {Kochoska}, {Kontizas}, {Kordopatis}, {Korn}, {Kostrzewa-Rutkowska},
  {Kruszy{\'n}ska}, {Lambert}, {Lanza}, {Lasne}, {Le Campion}, {Le Fustec},
  {Lebreton}, {Lebzelter}, {Leccia}, {Leclerc}, {Lecoeur-Taibi}, {Liao},
  {Licata}, {Lindstr{\o}m}, {Lister}, {Livanou}, {Lobel}, {Madrero Pardo},
  {Managau}, {Mann}, {Marchant}, {Marconi}, {Marcos Santos}, {Marinoni},
  {Marocco}, {Marshall}, {Martin Polo}, {Mart{\'\i}n-Fleitas}, {Masip},
  {Massari}, {Mastrobuono-Battisti}, {Mazeh}, {McMillan}, {Messina},
  {Michalik}, {Millar}, {Mints}, {Molina}, {Molinaro}, {Moln{\'a}r},
  {Montegriffo}, {Mor}, {Morbidelli}, {Morel}, {Morris}, {Mulone}, {Munoz},
  {Muraveva}, {Murphy}, {Musella}, {Noval}, {Ord{\'e}novic}, {Orr{\`u}},
  {Osinde}, {Pagani}, {Pagano}, {Palaversa}, {Palicio}, {Panahi}, {Pawlak},
  {Pe{\~n}alosa Esteller}, {Penttil{\"a}}, {Piersimoni}, {Pineau}, {Plachy},
  {Plum}, {Poggio}, {Poretti}, {Poujoulet}, {Pr{\v{s}}a}, {Pulone}, {Racero},
  {Ragaini}, {Rainer}, {Raiteri}, {Rambaux}, {Ramos}, {Ramos-Lerate}, {Re
  Fiorentin}, {Regibo}, {Reyl{\'e}}, {Ripepi}, {Riva}, {Rixon}, {Robichon},
  {Robin}, {Roelens}, {Rohrbasser}, {Romero-G{\'o}mez}, {Rowell}, {Royer},
  {Rybicki}, {Sadowski}, {Sagrist{\`a} Sell{\'e}s}, {Sahlmann}, {Salgado},
  {Salguero}, {Samaras}, {Sanchez Gimenez}, {Sanna}, {Santove{\~n}a},
  {Sarasso}, {Schultheis}, {Sciacca}, {Segol}, {Segovia}, {S{\'e}gransan},
  {Semeux}, {Shahaf}, {Siddiqui}, {Siebert}, {Siltala}, {Slezak}, {Smart},
  {Solano}, {Solitro}, {Souami}, {Souchay}, {Spagna}, {Spoto}, {Steele},
  {Steidelm{\"u}ller}, {Stephenson}, {S{\"u}veges}, {Szabados}, {Szegedi-Elek},
  {Taris}, {Tauran}, {Taylor}, {Teixeira}, {Thuillot}, {Tonello}, {Torra},
  {Torra}, {Turon}, {Unger}, {Vaillant}, {van Dillen}, {Vanel}, {Vecchiato},
  {Viala}, {Vicente}, {Voutsinas}, {Weiler}, {Wevers}, {Wyrzykowski}, {Yoldas},
  {Yvard}, {Zhao}, {Zorec}, {Zucker}, {Zurbach}, \& {Zwitter}}]{gaiaedr3}
{Gaia Collaboration}, {Brown}, A.~G.~A., {Vallenari}, A., {et~al.} 2021, \aap,
  649, A1, \dodoi{10.1051/0004-6361/202039657}

\bibitem[{{Giacalone} {et~al.}(2021){Giacalone}, {Dressing}, {Jensen},
  {Collins}, {Ricker}, {Vanderspek}, {Seager}, {Winn}, {Jenkins}, {Barclay},
  {Barkaoui}, {Cadieux}, {Charbonneau}, {Collins}, {Conti}, {Doyon}, {Evans},
  {Ghachoui}, {Gillon}, {Guerrero}, {Hart}, {Jehin}, {Kielkopf}, {McLean},
  {Murgas}, {Palle}, {Parviainen}, {Pozuelos}, {Relles}, {Shporer}, {Socia},
  {Stockdale}, {Tan}, {Torres}, {Twicken}, {Waalkes}, \&
  {Waite}}]{Giacalone2021}
{Giacalone}, S., {Dressing}, C.~D., {Jensen}, E. L.~N., {et~al.} 2021, \aj,
  161, 24, \dodoi{10.3847/1538-3881/abc6af}

\bibitem[{{Gilbert} {et~al.}(2020){Gilbert}, {Barclay}, {Schlieder},
  {Quintana}, {Hord}, {Kostov}, {Lopez}, {Rowe}, {Hoffman}, {Walkowicz},
  {Silverstein}, {Rodriguez}, {Vanderburg}, {Suissa}, {Airapetian}, {Clement},
  {Raymond}, {Mann}, {Kruse}, {Lissauer}, {Col{\'o}n}, {Kopparapu},
  {Kreidberg}, {Zieba}, {Collins}, {Quinn}, {Howell}, {Ziegler}, {Vrijmoet},
  {Adams}, {Arney}, {Boyd}, {Brande}, {Burke}, {Cacciapuoti}, {Chance},
  {Christiansen}, {Covone}, {Daylan}, {Dineen}, {Dressing}, {Essack},
  {Fauchez}, {Galgano}, {Howe}, {Kaltenegger}, {Kane}, {Lam}, {Lee}, {Lewis},
  {Logsdon}, {Mandell}, {Monsue}, {Mullally}, {Mullally}, {Paudel},
  {Pidhorodetska}, {Plavchan}, {Reyes}, {Rinehart}, {Rojas-Ayala}, {Smith},
  {Stassun}, {Tenenbaum}, {Vega}, {Villanueva}, {Wolf}, {Youngblood}, {Ricker},
  {Vanderspek}, {Latham}, {Seager}, {Winn}, {Jenkins}, {Bakos}, {Brice{\~n}o},
  {Ciardi}, {Cloutier}, {Conti}, {Couperus}, {Di Sora}, {Eisner}, {Everett},
  {Gan}, {Hartman}, {Henry}, {Isopi}, {Jao}, {Jensen}, {Law}, {Mallia},
  {Matson}, {Shappee}, {Le Wood}, \& {Winters}}]{gilbert2020}
{Gilbert}, E.~A., {Barclay}, T., {Schlieder}, J.~E., {et~al.} 2020, \aj, 160,
  116, \dodoi{10.3847/1538-3881/aba4b2}

\bibitem[{{Gilbert} {et~al.}(2023){Gilbert}, {Vanderburg}, {Rodriguez}, {Hord},
  {Clement}, {Barclay}, {Quintana}, {Schlieder}, {Kane}, {Jenkins}, {Twicken},
  {Kunimoto}, {Vanderspek}, {Arney}, {Charbonneau}, {G{\"u}nther}, {Huang},
  {Isopi}, {Kostov}, {Kristiansen}, {Latham}, {Mallia}, {Mamajek}, {Mireles},
  {Quinn}, {Ricker}, {Schulte}, {Seager}, {Suissa}, {Winn}, {Youngblood}, \&
  {Zapparata}}]{2023Gilbert}
{Gilbert}, E.~A., {Vanderburg}, A., {Rodriguez}, J.~E., {et~al.} 2023, \apjl,
  944, L35, \dodoi{10.3847/2041-8213/acb599}

\bibitem[{{Gili} {et~al.}(2021){Gili}, {Prieur}, {Rivet}, {Vakili}, {Scardia},
  {Pansecchi}, {Argyle}, {Ling}, {Piccotti}, {Aristidi}, {Koechlin}, {Bonneau},
  {Maccarini}, \& {Serot}}]{2021Gili}
{Gili}, R., {Prieur}, J.-L., {Rivet}, J.-P., {et~al.} 2021, Astronomische
  Nachrichten, 342, 865, \dodoi{10.1002/asna.202113985}

\bibitem[{Ginsburg {et~al.}(2019)Ginsburg, Sip{\H{o}}cz, Brasseur,
  Cowperthwaite, Craig, Deil, Groener, Guillochon, Guzman, Liedtke,
  {et~al.}}]{ginsburg2019astroquery}
Ginsburg, A., Sip{\H{o}}cz, B.~M., Brasseur, C., {et~al.} 2019, The
  Astronomical Journal, 157, 98

\bibitem[{{Girardi} {et~al.}(2005){Girardi}, {Groenewegen}, {Hatziminaoglou},
  \& {da Costa}}]{Girardi2005}
{Girardi}, L., {Groenewegen}, M.~A.~T., {Hatziminaoglou}, E., \& {da Costa}, L.
  2005, \aap, 436, 895, \dodoi{10.1051/0004-6361:20042352}

\bibitem[{{Gomes da Silva} {et~al.}(2011){Gomes da Silva}, {Santos}, {Bonfils},
  {Delfosse}, {Forveille}, \& {Udry}}]{2011GomesdaSilva}
{Gomes da Silva}, J., {Santos}, N.~C., {Bonfils}, X., {et~al.} 2011, \aap, 534,
  A30, \dodoi{10.1051/0004-6361/201116971}

\bibitem[{{Gregory}(2005)}]{Gregory2005}
{Gregory}, P.~C. 2005, \apj, 631, 1198, \dodoi{10.1086/432594}

\bibitem[{{Guerrero} {et~al.}(2021){Guerrero}, {Seager}, {Huang}, {Vanderburg},
  {Garcia Soto}, {Mireles}, {Hesse}, {Fong}, {Glidden}, {Shporer}, {Latham},
  {Collins}, {Quinn}, {Burt}, {Dragomir}, {Crossfield}, {Vanderspek},
  {Fausnaugh}, {Burke}, {Ricker}, {Daylan}, {Essack}, {G{\"u}nther}, {Osborn},
  {Pepper}, {Rowden}, {Sha}, {Villanueva}, {Yahalomi}, {Yu}, {Ballard},
  {Batalha}, {Berardo}, {Chontos}, {Dittmann}, {Esquerdo}, {Mikal-Evans},
  {Jayaraman}, {Krishnamurthy}, {Louie}, {Mehrle}, {Niraula}, {Rackham},
  {Rodriguez}, {Rowden}, {Sousa-Silva}, {Watanabe}, {Wong}, {Zhan},
  {Zivanovic}, {Christiansen}, {Ciardi}, {Swain}, {Lund}, {Mullally},
  {Fleming}, {Rodriguez}, {Boyd}, {Quintana}, {Barclay}, {Col{\'o}n},
  {Rinehart}, {Schlieder}, {Clampin}, {Jenkins}, {Twicken}, {Caldwell},
  {Coughlin}, {Henze}, {Lissauer}, {Morris}, {Rose}, {Smith}, {Tenenbaum},
  {Ting}, {Wohler}, {Bakos}, {Bean}, {Berta-Thompson}, {Bieryla}, {Bouma},
  {Buchhave}, {Butler}, {Charbonneau}, {Doty}, {Ge}, {Holman}, {Howard},
  {Kaltenegger}, {Kane}, {Kjeldsen}, {Kreidberg}, {Lin}, {Minsky}, {Narita},
  {Paegert}, {P{\'a}l}, {Palle}, {Sasselov}, {Spencer}, {Sozzetti}, {Stassun},
  {Torres}, {Udry}, \& {Winn}}]{2021Guerrero}
{Guerrero}, N.~M., {Seager}, S., {Huang}, C.~X., {et~al.} 2021, \apjs, 254, 39,
  \dodoi{10.3847/1538-4365/abefe1}

\bibitem[{{Hamer} \& {Schlaufman}(2019)}]{2019Hamer}
{Hamer}, J.~H., \& {Schlaufman}, K.~C. 2019, \aj, 158, 190,
  \dodoi{10.3847/1538-3881/ab3c56}

\bibitem[{{Hayward} {et~al.}(2001){Hayward}, {Brandl}, {Pirger}, {Blacken},
  {Gull}, {Schoenwald}, \& {Houck}}]{Hayward2001palomar}
{Hayward}, T.~L., {Brandl}, B., {Pirger}, B., {et~al.} 2001, \pasp, 113, 105,
  \dodoi{10.1086/317969}

\bibitem[{{H{\'e}brard} {et~al.}(2008){H{\'e}brard}, {Bouchy}, {Pont},
  {Loeillet}, {Rabus}, {Bonfils}, {Moutou}, {Boisse}, {Delfosse}, {Desort},
  {Eggenberger}, {Ehrenreich}, {Forveille}, {Lagrange}, {Lovis}, {Mayor},
  {Pepe}, {Perrier}, {Queloz}, {Santos}, {S{\'e}gransan}, {Udry}, \&
  {Vidal-Madjar}}]{2008Hebrard}
{H{\'e}brard}, G., {Bouchy}, F., {Pont}, F., {et~al.} 2008, \aap, 488, 763,
  \dodoi{10.1051/0004-6361:200810056}

\bibitem[{{Heintz}(1985)}]{1985Heintz}
{Heintz}, W.~D. 1985, \apjs, 58, 439, \dodoi{10.1086/191048}

\bibitem[{{Heitzmann} {et~al.}(2023){Heitzmann}, {Zhou}, {Quinn}, {Huang},
  {Dong}, {Bouma}, {Dawson}, {Marsden}, {Wright}, {Petit}, {Collins},
  {Barkaoui}, {Wittenmyer}, {Gillen}, {Brahm}, {Hobson}, {Hellier}, {Ziegler},
  {Brice{\~n}o}, {Law}, {Mann}, {Howell}, {Gnilka}, {Littlefield}, {Latham},
  {Lissauer}, {Newton}, {Krolikowski}, {Kerr}, {Rampalli}, {Douglas}, {Eisner},
  {Guedj}, {Sun}, {Smit}, {Huten}, {Eschweiler}, {Abe}, {Guillot}, {Ricker},
  {Vanderspek}, {Seager}, {Jenkins}, {Ting}, {Winn}, {Ciardi}, {Vanderburg},
  {Burke}, {Rodriguez}, \& {Daylan}}]{2023Heitzmann}
{Heitzmann}, A., {Zhou}, G., {Quinn}, S.~N., {et~al.} 2023, \aj, 165, 121,
  \dodoi{10.3847/1538-3881/acb5a2}

\bibitem[{{Hill} {et~al.}(2023){Hill}, {Bott}, {Dalba}, {Fetherolf}, {Kane},
  {Kopparapu}, {Li}, \& {Ostberg}}]{2023Hill}
{Hill}, M.~L., {Bott}, K., {Dalba}, P.~A., {et~al.} 2023, \aj, 165, 34,
  \dodoi{10.3847/1538-3881/aca1c0}

\bibitem[{{Hirano} {et~al.}(2011){Hirano}, {Suto}, {Winn}, {Taruya}, {Narita},
  {Albrecht}, \& {Sato}}]{hirano2011}
{Hirano}, T., {Suto}, Y., {Winn}, J.~N., {et~al.} 2011, \apj, 742, 69,
  \dodoi{10.1088/0004-637X/742/2/69}

\bibitem[{{H{\o}g} {et~al.}(2000){H{\o}g}, {Fabricius}, {Makarov}, {Urban},
  {Corbin}, {Wycoff}, {Bastian}, {Schwekendiek}, \& {Wicenec}}]{Hog2000}
{H{\o}g}, E., {Fabricius}, C., {Makarov}, V.~V., {et~al.} 2000, \aap, 355, L27

\bibitem[{{Howard} {et~al.}(2010){Howard}, {Johnson}, {Marcy}, {Fischer},
  {Wright}, {Bernat}, {Henry}, {Peek}, {Isaacson}, {Apps}, {Endl}, {Cochran},
  {Valenti}, {Anderson}, \& {Piskunov}}]{2010Howard}
{Howard}, A.~W., {Johnson}, J.~A., {Marcy}, G.~W., {et~al.} 2010, \apj, 721,
  1467, \dodoi{10.1088/0004-637X/721/2/1467}

\bibitem[{{Howell} {et~al.}(2011){Howell}, {Everett}, {Sherry}, {Horch}, \&
  {Ciardi}}]{Howell2011}
{Howell}, S.~B., {Everett}, M.~E., {Sherry}, W., {Horch}, E., \& {Ciardi},
  D.~R. 2011, \aj, 142, 19, \dodoi{10.1088/0004-6256/142/1/19}

\bibitem[{{Huber} {et~al.}(2013){Huber}, {Chaplin}, {Christensen-Dalsgaard},
  {Gilliland}, {Kjeldsen}, {Buchhave}, {Fischer}, {Lissauer}, {Rowe},
  {Sanchis-Ojeda}, {Basu}, {Handberg}, {Hekker}, {Howard}, {Isaacson},
  {Karoff}, {Latham}, {Lund}, {Lundkvist}, {Marcy}, {Miglio}, {Silva Aguirre},
  {Stello}, {Arentoft}, {Barclay}, {Bedding}, {Burke}, {Christiansen},
  {Elsworth}, {Haas}, {Kawaler}, {Metcalfe}, {Mullally}, \&
  {Thompson}}]{huber2013}
{Huber}, D., {Chaplin}, W.~J., {Christensen-Dalsgaard}, J., {et~al.} 2013,
  \apj, 767, 127, \dodoi{10.1088/0004-637X/767/2/127}

\bibitem[{{Hunter}(2007)}]{matplotlib}
{Hunter}, J.~D. 2007, Computing in Science Engineering, 9, 90

\bibitem[{{Hussey}(1905)}]{1905Hussey}
{Hussey}, W.~J. 1905, Lick Observatory Bulletin, 74, 95,
  \dodoi{10.5479/ADS/bib/1905LicOB.3.95H}

\bibitem[{{Jenkins}(2002)}]{2002Jenkins}
{Jenkins}, J.~M. 2002, \apj, 575, 493, \dodoi{10.1086/341136}

\bibitem[{{Jenkins} {et~al.}(2020){Jenkins}, {Tenenbaum}, {Seader}, {Burke},
  {McCauliff}, {Smith}, {Twicken}, \& {Chandrasekaran}}]{2020Jenkins}
{Jenkins}, J.~M., {Tenenbaum}, P., {Seader}, S., {et~al.} 2020, {Kepler Data
  Processing Handbook: Transiting Planet Search}, Kepler Science Document
  KSCI-19081-003

\bibitem[{{Jenkins} {et~al.}(2010){Jenkins}, {Chandrasekaran}, {McCauliff},
  {Caldwell}, {Tenenbaum}, {Li}, {Klaus}, {Cote}, \& {Middour}}]{jenkins2010}
{Jenkins}, J.~M., {Chandrasekaran}, H., {McCauliff}, S.~D., {et~al.} 2010, in
  Society of Photo-Optical Instrumentation Engineers (SPIE) Conference Series,
  Vol. 7740, Software and Cyberinfrastructure for Astronomy, ed. N.~M.
  {Radziwill} \& A.~{Bridger}, 77400D, \dodoi{10.1117/12.856764}

\bibitem[{{Jenkins} {et~al.}(2016){Jenkins}, {Twicken}, {McCauliff},
  {Campbell}, {Sanderfer}, {Lung}, {Mansouri-Samani}, {Girouard}, {Tenenbaum},
  {Klaus}, {Smith}, {Caldwell}, {Chacon}, {Henze}, {Heiges}, {Latham},
  {Morgan}, {Swade}, {Rinehart}, \& {Vanderspek}}]{jenkins16}
{Jenkins}, J.~M., {Twicken}, J.~D., {McCauliff}, S., {et~al.} 2016, in Society
  of Photo-Optical Instrumentation Engineers (SPIE) Conference Series, Vol.
  9913, \procspie, 99133E, \dodoi{10.1117/12.2233418}

\bibitem[{{Kipping}(2013)}]{Kipping2013}
{Kipping}, D.~M. 2013, \mnras, 435, 2152, \dodoi{10.1093/mnras/stt1435}

\bibitem[{{Klein} {et~al.}(2022){Klein}, {Zicher}, {Kavanagh}, {Nielsen},
  {Aigrain}, {Vidotto}, {Barrag{\'a}n}, {Strugarek}, {Nicholson}, {Donati}, \&
  {Bouvier}}]{2022Klein}
{Klein}, B., {Zicher}, N., {Kavanagh}, R.~D., {et~al.} 2022, \mnras, 512, 5067,
  \dodoi{10.1093/mnras/stac761}

\bibitem[{{Kochanek} {et~al.}(2017){Kochanek}, {Shappee}, {Stanek}, {Holoien},
  {Thompson}, {Prieto}, {Dong}, {Shields}, {Will}, {Britt}, {Perzanowski}, \&
  {Pojma{\'n}ski}}]{Kochanek2017}
{Kochanek}, C.~S., {Shappee}, B.~J., {Stanek}, K.~Z., {et~al.} 2017, \pasp,
  129, 104502, \dodoi{10.1088/1538-3873/aa80d9}

\bibitem[{{Kolbl} {et~al.}(2015){Kolbl}, {Marcy}, {Isaacson}, \&
  {Howard}}]{kolbl2015}
{Kolbl}, R., {Marcy}, G.~W., {Isaacson}, H., \& {Howard}, A.~W. 2015, \aj, 149,
  18, \dodoi{10.1088/0004-6256/149/1/18}

\bibitem[{{Kopparapu} {et~al.}(2014){Kopparapu}, {Ramirez}, {SchottelKotte},
  {Kasting}, {Domagal-Goldman}, \& {Eymet}}]{2014Kopparapu}
{Kopparapu}, R.~K., {Ramirez}, R.~M., {SchottelKotte}, J., {et~al.} 2014,
  \apjl, 787, L29, \dodoi{10.1088/2041-8205/787/2/L29}

\bibitem[{{Kov{\'a}cs} {et~al.}(2002){Kov{\'a}cs}, {Zucker}, \&
  {Mazeh}}]{bls2002}
{Kov{\'a}cs}, G., {Zucker}, S., \& {Mazeh}, T. 2002, \aap, 391, 369,
  \dodoi{10.1051/0004-6361:20020802}

\bibitem[{{Kozai}(1962)}]{Kozai}
{Kozai}, Y. 1962, \aj, 67, 591, \dodoi{10.1086/108790}

\bibitem[{{Kraus} {et~al.}(2012){Kraus}, {Ireland}, {Hillenbrand}, \&
  {Martinache}}]{2012Kraus}
{Kraus}, A.~L., {Ireland}, M.~J., {Hillenbrand}, L.~A., \& {Martinache}, F.
  2012, \apj, 745, 19, \dodoi{10.1088/0004-637X/745/1/19}

\bibitem[{{Kreidberg}(2015)}]{Kreidberg15}
{Kreidberg}, L. 2015, \pasp, 127, 1161, \dodoi{10.1086/683602}

\bibitem[{{Lester} {et~al.}(2023){Lester}, {Howell}, {Matson}, {Furlan},
  {Gnilka}, {Littlefield}, {Ciardi}, {Everett}, {Fajardo-Acosta}, \&
  {Clark}}]{2023Lester}
{Lester}, K.~V., {Howell}, S.~B., {Matson}, R.~A., {et~al.} 2023, \aj, 166,
  166, \dodoi{10.3847/1538-3881/acf563}

\bibitem[{{Li} {et~al.}(2019){Li}, {Tenenbaum}, {Twicken}, {Burke}, {Jenkins},
  {Quintana}, {Rowe}, \& {Seader}}]{Li2019}
{Li}, J., {Tenenbaum}, P., {Twicken}, J.~D., {et~al.} 2019, \pasp, 131, 024506,
  \dodoi{10.1088/1538-3873/aaf44d}

\bibitem[{{Lidov}(1962)}]{Lidov}
{Lidov}, M.~L. 1962, \planss, 9, 719, \dodoi{10.1016/0032-0633(62)90129-0}

\bibitem[{{Lightkurve Collaboration} {et~al.}(2018){Lightkurve Collaboration},
  {Cardoso}, {Hedges}, {Gully-Santiago}, {Saunders}, {Cody}, {Barclay}, {Hall},
  {Sagear}, {Turtelboom}, {Zhang}, {Tzanidakis}, {Mighell}, {Coughlin}, {Bell},
  {Berta-Thompson}, {Williams}, {Dotson}, \& {Barentsen}}]{lightkurve2018}
{Lightkurve Collaboration}, {Cardoso}, J.~V.~d.~M., {Hedges}, C., {et~al.}
  2018, {Lightkurve: Kepler and TESS time series analysis in Python},
  Astrophysics Source Code Library.
\newblock \doeprint{1812.013}

\bibitem[{{Lintott} {et~al.}(2011){Lintott}, {Schawinski}, {Bamford}, {Slosar},
  {Land}, {Thomas}, {Edmondson}, {Masters}, {Nichol}, \& {Raddick}}]{lintott11}
{Lintott}, C., {Schawinski}, K., {Bamford}, S., {et~al.} 2011, VizieR Online
  Data Catalog, J/MNRAS/410/166

\bibitem[{{Lintott} {et~al.}(2008){Lintott}, {Schawinski}, {Slosar}, {Land},
  {Bamford}, {Thomas}, {Raddick}, {Nichol}, {Szalay}, \&
  {Andreescu}}]{lintott08}
{Lintott}, C.~J., {Schawinski}, K., {Slosar}, A., {et~al.} 2008, \mnras, 389,
  1179, \dodoi{10.1111/j.1365-2966.2008.13689.x}

\bibitem[{{Lopez} \& {Rice}(2018)}]{2018Lopez}
{Lopez}, E.~D., \& {Rice}, K. 2018, \mnras, 479, 5303,
  \dodoi{10.1093/mnras/sty1707}

\bibitem[{{Mamajek} \& {Hillenbrand}(2008)}]{Mamajek:2008}
{Mamajek}, E.~E., \& {Hillenbrand}, L.~A. 2008, \apj, 687, 1264,
  \dodoi{10.1086/591785}

\bibitem[{{Manara} {et~al.}(2019){Manara}, {Tazzari}, {Long}, {Herczeg},
  {Lodato}, {Rota}, {Cazzoletti}, {van der Plas}, {Pinilla}, {Dipierro},
  {Edwards}, {Harsono}, {Johnstone}, {Liu}, {Menard}, {Nisini}, {Ragusa},
  {Boehler}, \& {Cabrit}}]{2019Manara}
{Manara}, C.~F., {Tazzari}, M., {Long}, F., {et~al.} 2019, \aap, 628, A95,
  \dodoi{10.1051/0004-6361/201935964}

\bibitem[{{Mandel} \& {Agol}(2002)}]{Mandel2002}
{Mandel}, K., \& {Agol}, E. 2002, \apjl, 580, L171, \dodoi{10.1086/345520}

\bibitem[{{Matson} {et~al.}(2019){Matson}, {Howell}, \& {Ciardi}}]{Matson2019}
{Matson}, R.~A., {Howell}, S.~B., \& {Ciardi}, D.~R. 2019, \aj, 157, 211,
  \dodoi{10.3847/1538-3881/ab1755}

\bibitem[{McKinney {et~al.}(2010)}]{pandas}
McKinney, W., {et~al.} 2010, in Proceedings of the 9th Python in Science
  Conference, Vol. 445, Austin, TX, 51--56

\bibitem[{{Mireles} {et~al.}(2023){Mireles}, {Dragomir}, {Osborn}, {Hesse},
  {Collins}, {Villanueva}, {Bieryla}, {Ciardi}, {Stassun}, {Harris},
  {Lissauer}, {Schwarz}, {Srdoc}, {Barkaoui}, {Riffeser}, {McLeod}, {Pepper},
  {Grieves}, {Passegger}, {Ulmer-Moll}, {Rodriguez}, {Feliz}, {Quinn}, {Boyle},
  {Fausnaugh}, {Kunimoto}, {Rowden}, {Vanderburg}, {Wohler}, {Jenkins},
  {Latham}, {Ricker}, {Seager}, \& {Winn}}]{2023Mireles}
{Mireles}, I., {Dragomir}, D., {Osborn}, H.~P., {et~al.} 2023, \apjl, 954, L15,
  \dodoi{10.3847/2041-8213/aceb69}

\bibitem[{{Moe} \& {Kratter}(2021{\natexlab{a}})}]{2021Moe}
{Moe}, M., \& {Kratter}, K.~M. 2021{\natexlab{a}}, \mnras, 507, 3593,
  \dodoi{10.1093/mnras/stab2328}

\bibitem[{{Moe} \& {Kratter}(2021{\natexlab{b}})}]{2021MoeKratter}
---. 2021{\natexlab{b}}, \mnras, 507, 3593, \dodoi{10.1093/mnras/stab2328}

\bibitem[{{Muller}(1978)}]{1978Muller}
{Muller}, P. 1978, \aaps, 33, 275

\bibitem[{{Muller}(1997)}]{1997Muller}
---. 1997, \aaps, 126, 273, \dodoi{10.1051/aas:1997387}

\bibitem[{{Naoz}(2016)}]{Naoz16}
{Naoz}, S. 2016, \araa, 54, 441, \dodoi{10.1146/annurev-astro-081915-023315}

\bibitem[{{Naoz} {et~al.}(2012){Naoz}, {Farr}, \& {Rasio}}]{2012Naoz}
{Naoz}, S., {Farr}, W.~M., \& {Rasio}, F.~A. 2012, \apjl, 754, L36,
  \dodoi{10.1088/2041-8205/754/2/L36}

\bibitem[{{NASA Exoplanet Archive}(2019)}]{nasaEA}
{NASA Exoplanet Archive}. 2019, Composite Planet Data Table,  IPAC,
  \dodoi{10.26133/NEA2}

\bibitem[{{Ngo} {et~al.}(2016){Ngo}, {Knutson}, {Hinkley}, {Bryan}, {Crepp},
  {Batygin}, {Crossfield}, {Hansen}, {Howard}, {Johnson}, {Mawet}, {Morton},
  {Muirhead}, \& {Wang}}]{2016Ngo}
{Ngo}, H., {Knutson}, H.~A., {Hinkley}, S., {et~al.} 2016, \apj, 827, 8,
  \dodoi{10.3847/0004-637X/827/1/8}

\bibitem[{{Noyes} {et~al.}(1984){Noyes}, {Hartmann}, {Baliunas}, {Duncan}, \&
  {Vaughan}}]{1984Noyes}
{Noyes}, R.~W., {Hartmann}, L.~W., {Baliunas}, S.~L., {Duncan}, D.~K., \&
  {Vaughan}, A.~H. 1984, \apj, 279, 763, \dodoi{10.1086/161945}

\bibitem[{{Oh} {et~al.}(2018){Oh}, {Price-Whelan}, {Brewer}, {Hogg}, {Spergel},
  \& {Myles}}]{oh2018}
{Oh}, S., {Price-Whelan}, A.~M., {Brewer}, J.~M., {et~al.} 2018, \apj, 854,
  138, \dodoi{10.3847/1538-4357/aaab4d}

\bibitem[{{Otegi} {et~al.}(2020){Otegi}, {Bouchy}, \& {Helled}}]{2020Otegi}
{Otegi}, J.~F., {Bouchy}, F., \& {Helled}, R. 2020, \aap, 634, A43,
  \dodoi{10.1051/0004-6361/201936482}

\bibitem[{Pedregosa {et~al.}(2011)Pedregosa, Varoquaux, Gramfort, Michel,
  Thirion, Grisel, Blondel, Prettenhofer, Weiss, Dubourg,
  {et~al.}}]{pedregosa2011scikit}
Pedregosa, F., Varoquaux, G., Gramfort, A., {et~al.} 2011, the Journal of
  machine Learning research, 12, 2825

\bibitem[{{Perruchot} {et~al.}(2008){Perruchot}, {Kohler}, {Bouchy}, {Richaud},
  {Richaud}, {Moreaux}, {Merzougui}, {Sottile}, {Hill}, {Knispel}, {Regal},
  {Meunier}, {Ilovaisky}, {Le Coroller}, {Gillet}, {Schmitt}, {Pepe}, {Fleury},
  {Sosnowska}, {Vors}, {M{\'e}gevand}, {Blanc}, {Carol}, {Point}, {Laloge}, \&
  {Brunel}}]{2008PerruchotOHP}
{Perruchot}, S., {Kohler}, D., {Bouchy}, F., {et~al.} 2008, in Society of
  Photo-Optical Instrumentation Engineers (SPIE) Conference Series, Vol. 7014,
  Ground-based and Airborne Instrumentation for Astronomy II, ed. I.~S.
  {McLean} \& M.~M. {Casali}, 70140J, \dodoi{10.1117/12.787379}

\bibitem[{{Petigura}(2015)}]{petigura2015}
{Petigura}, E.~A. 2015, PhD thesis, University of California, Berkeley

\bibitem[{{Pollacco} {et~al.}(2008){Pollacco}, {Skillen}, {Collier Cameron},
  {Loeillet}, {Stempels}, {Bouchy}, {Gibson}, {Hebb}, {H{\'e}brard}, {Joshi},
  {McDonald}, {Smalley}, {Smith}, {Street}, {Udry}, {West}, {Wilson},
  {Wheatley}, {Aigrain}, {Alsubai}, {Benn}, {Bruce}, {Christian}, {Clarkson},
  {Enoch}, {Evans}, {Fitzsimmons}, {Haswell}, {Hellier}, {Hickey}, {Hodgkin},
  {Horne}, {Hrudkov{\'a}}, {Irwin}, {Kane}, {Keenan}, {Lister}, {Maxted},
  {Mayor}, {Moutou}, {Norton}, {Osborne}, {Parley}, {Pont}, {Queloz}, {Ryans},
  \& {Simpson}}]{2008Pollacco}
{Pollacco}, D., {Skillen}, I., {Collier Cameron}, A., {et~al.} 2008, \mnras,
  385, 1576, \dodoi{10.1111/j.1365-2966.2008.12939.x}

\bibitem[{{Quarles} {et~al.}(2020){Quarles}, {Li}, {Kostov}, \&
  {Haghighipour}}]{Quarles+20}
{Quarles}, B., {Li}, G., {Kostov}, V., \& {Haghighipour}, N. 2020, \aj, 159,
  80, \dodoi{10.3847/1538-3881/ab64fa}

\bibitem[{{Queloz} {et~al.}(2001){Queloz}, {Henry}, {Sivan}, {Baliunas},
  {Beuzit}, {Donahue}, {Mayor}, {Naef}, {Perrier}, \& {Udry}}]{queloz2001no}
{Queloz}, D., {Henry}, G.~W., {Sivan}, J.~P., {et~al.} 2001, \aap, 379, 279,
  \dodoi{10.1051/0004-6361:20011308}

\bibitem[{{Quintana} {et~al.}(2007){Quintana}, {Adams}, {Lissauer}, \&
  {Chambers}}]{2007Quintana}
{Quintana}, E.~V., {Adams}, F.~C., {Lissauer}, J.~J., \& {Chambers}, J.~E.
  2007, \apj, 660, 807, \dodoi{10.1086/512542}

\bibitem[{{Raghavan} {et~al.}(2010){Raghavan}, {McAlister}, {Henry}, {Latham},
  {Marcy}, {Mason}, {Gies}, {White}, \& {ten Brummelaar}}]{2010Raghavan}
{Raghavan}, D., {McAlister}, H.~A., {Henry}, T.~J., {et~al.} 2010, \apjs, 190,
  1, \dodoi{10.1088/0067-0049/190/1/1}

\bibitem[{{Ricker} {et~al.}(2015){Ricker}, {Winn}, {Vanderspek}, {Latham},
  {Bakos}, {Bean}, {Berta-Thompson}, {Brown}, {Buchhave}, \&
  {Butler}}]{ricker15}
{Ricker}, G.~R., {Winn}, J.~N., {Vanderspek}, R., {et~al.} 2015, Journal of
  Astronomical Telescopes, Instruments, and Systems, 1, 014003,
  \dodoi{10.1117/1.JATIS.1.1.014003}

\bibitem[{{Rodriguez} {et~al.}(2020){Rodriguez}, {Vanderburg}, {Zieba},
  {Kreidberg}, {Morley}, {Eastman}, {Kane}, {Spencer}, {Quinn}, {Cloutier},
  {Huang}, {Collins}, {Mann}, {Gilbert}, {Schlieder}, {Quintana}, {Barclay},
  {Suissa}, {Kopparapu}, {Dressing}, {Ricker}, {Vanderspek}, {Latham},
  {Seager}, {Winn}, {Jenkins}, {Berta-Thompson}, {Boyd}, {Charbonneau},
  {Caldwell}, {Chiang}, {Christiansen}, {Ciardi}, {Col{\'o}n}, {Doty}, {Gan},
  {Guerrero}, {G{\"u}nther}, {Lee}, {Levine}, {Lopez}, {Muirhead}, {Newton},
  {Rose}, {Twicken}, \& {Villase{\~n}or}}]{Rodriguez2020}
{Rodriguez}, J.~E., {Vanderburg}, A., {Zieba}, S., {et~al.} 2020, \aj, 160,
  117, \dodoi{10.3847/1538-3881/aba4b3}

\bibitem[{{Saar} {et~al.}(1998){Saar}, {Butler}, \& {Marcy}}]{saar1998magnetic}
{Saar}, S.~H., {Butler}, R.~P., \& {Marcy}, G.~W. 1998, \apjl, 498, L153,
  \dodoi{10.1086/311325}

\bibitem[{{Santos} {et~al.}(2010){Santos}, {Gomes da Silva}, {Lovis}, \&
  {Melo}}]{santos2010stellar}
{Santos}, N.~C., {Gomes da Silva}, J., {Lovis}, C., \& {Melo}, C. 2010, \aap,
  511, A54, \dodoi{10.1051/0004-6361/200913433}

\bibitem[{{Santos} {et~al.}(2000){Santos}, {Mayor}, {Naef}, {Pepe}, {Queloz},
  {Udry}, \& {Blecha}}]{santos2000coralie}
{Santos}, N.~C., {Mayor}, M., {Naef}, D., {et~al.} 2000, \aap, 361, 265

\bibitem[{{Schlegel} {et~al.}(1998){Schlegel}, {Finkbeiner}, \&
  {Davis}}]{Schlegel:1998}
{Schlegel}, D.~J., {Finkbeiner}, D.~P., \& {Davis}, M. 1998, \apj, 500, 525,
  \dodoi{10.1086/305772}

\bibitem[{{Schwarz} {et~al.}(2016){Schwarz}, {Funk}, {Zechner}, \&
  {Bazs{\'o}}}]{2016Schwarz}
{Schwarz}, R., {Funk}, B., {Zechner}, R., \& {Bazs{\'o}}, {\'A}. 2016, \mnras,
  460, 3598, \dodoi{10.1093/mnras/stw1218}

\bibitem[{{Scott} {et~al.}(2018){Scott}, {Howell}, {Horch}, \&
  {Everett}}]{2018ScottNESSI}
{Scott}, N.~J., {Howell}, S.~B., {Horch}, E.~P., \& {Everett}, M.~E. 2018,
  \pasp, 130, 054502, \dodoi{10.1088/1538-3873/aab484}

\bibitem[{{Shappee} {et~al.}(2014){Shappee}, {Prieto}, {Grupe}, {Kochanek},
  {Stanek}, {De Rosa}, {Mathur}, {Zu}, {Peterson}, {Pogge}, {Komossa}, {Im},
  {Jencson}, {Holoien}, {Basu}, {Beacom}, {Szczygie{\l}}, {Brimacombe},
  {Adams}, {Campillay}, {Choi}, {Contreras}, {Dietrich}, {Dubberley},
  {Elphick}, {Foale}, {Giustini}, {Gonzalez}, {Hawkins}, {Howell}, {Hsiao},
  {Koss}, {Leighly}, {Morrell}, {Mudd}, {Mullins}, {Nugent}, {Parrent},
  {Phillips}, {Pojmanski}, {Rosing}, {Ross}, {Sand}, {Terndrup}, {Valenti},
  {Walker}, \& {Yoon}}]{2014Shappee}
{Shappee}, B.~J., {Prieto}, J.~L., {Grupe}, D., {et~al.} 2014, \apj, 788, 48,
  \dodoi{10.1088/0004-637X/788/1/48}

\bibitem[{{Smith} {et~al.}(2012){Smith}, {Stumpe}, {Van Cleve}, {Jenkins},
  {Barclay}, {Fanelli}, {Girouard}, {Kolodziejczak}, {McCauliff}, {Morris}, \&
  {Twicken}}]{Smith2012}
{Smith}, J.~C., {Stumpe}, M.~C., {Van Cleve}, J.~E., {et~al.} 2012, \pasp, 124,
  1000, \dodoi{10.1086/667697}

\bibitem[{{Stassun} {et~al.}(2017){Stassun}, {Collins}, \&
  {Gaudi}}]{Stassun:2017}
{Stassun}, K.~G., {Collins}, K.~A., \& {Gaudi}, B.~S. 2017, \aj, 153, 136,
  \dodoi{10.3847/1538-3881/aa5df3}

\bibitem[{{Stassun} {et~al.}(2018){Stassun}, {Corsaro}, {Pepper}, \&
  {Gaudi}}]{Stassun:2018}
{Stassun}, K.~G., {Corsaro}, E., {Pepper}, J.~A., \& {Gaudi}, B.~S. 2018, \aj,
  155, 22, \dodoi{10.3847/1538-3881/aa998a}

\bibitem[{{Stassun} \& {Torres}(2016)}]{Stassun:2016}
{Stassun}, K.~G., \& {Torres}, G. 2016, \aj, 152, 180,
  \dodoi{10.3847/0004-6256/152/6/180}

\bibitem[{{Stassun} \& {Torres}(2021)}]{StassunTorres:2021}
---. 2021, \apjl, 907, L33, \dodoi{10.3847/2041-8213/abdaad}

\bibitem[{{Stassun} {et~al.}(2019){Stassun}, {Oelkers}, {Paegert}, {Torres},
  {Pepper}, {De Lee}, {Collins}, {Latham}, {Muirhead}, {Chittidi},
  {Rojas-Ayala}, {Fleming}, {Rose}, {Tenenbaum}, {Ting}, {Kane}, {Barclay},
  {Bean}, {Brassuer}, {Charbonneau}, {Ge}, {Lissauer}, {Mann}, {McLean},
  {Mullally}, {Narita}, {Plavchan}, {Ricker}, {Sasselov}, {Seager}, {Sharma},
  {Shiao}, {Sozzetti}, {Stello}, {Vanderspek}, {Wallace}, \&
  {Winn}}]{Stassun19}
{Stassun}, K.~G., {Oelkers}, R.~J., {Paegert}, M., {et~al.} 2019, \aj, 158,
  138, \dodoi{10.3847/1538-3881/ab3467}

\bibitem[{{Stumpe} {et~al.}(2014){Stumpe}, {Smith}, {Catanzarite}, {Van Cleve},
  {Jenkins}, {Twicken}, \& {Girouard}}]{Stumpe2014}
{Stumpe}, M.~C., {Smith}, J.~C., {Catanzarite}, J.~H., {et~al.} 2014, \pasp,
  126, 100, \dodoi{10.1086/674989}

\bibitem[{{Stumpe} {et~al.}(2012){Stumpe}, {Smith}, {Van Cleve}, {Twicken},
  {Barclay}, {Fanelli}, {Girouard}, {Jenkins}, {Kolodziejczak}, {McCauliff}, \&
  {Morris}}]{Stumpe2012}
{Stumpe}, M.~C., {Smith}, J.~C., {Van Cleve}, J.~E., {et~al.} 2012, \pasp, 124,
  985, \dodoi{10.1086/667698}

\bibitem[{{Sucerquia} {et~al.}(2019){Sucerquia}, {Alvarado-Montes}, {Zuluaga},
  {Cuello}, \& {Giuppone}}]{2019Sucerquia}
{Sucerquia}, M., {Alvarado-Montes}, J.~A., {Zuluaga}, J.~I., {Cuello}, N., \&
  {Giuppone}, C. 2019, \mnras, 489, 2313, \dodoi{10.1093/mnras/stz2110}

\bibitem[{{Teske} {et~al.}(2016{\natexlab{a}}){Teske}, {Khanal}, \&
  {Ram{\'\i}rez}}]{2016Teske}
{Teske}, J.~K., {Khanal}, S., \& {Ram{\'\i}rez}, I. 2016{\natexlab{a}}, \apj,
  819, 19, \dodoi{10.3847/0004-637X/819/1/19}

\bibitem[{{Teske} {et~al.}(2016{\natexlab{b}}){Teske}, {Shectman}, {Vogt},
  {D{\'\i}az}, {Butler}, {Crane}, {Thompson}, \& {Arriagada}}]{2016Teskeb}
{Teske}, J.~K., {Shectman}, S.~A., {Vogt}, S.~S., {et~al.} 2016{\natexlab{b}},
  \aj, 152, 167, \dodoi{10.3847/0004-6256/152/6/167}

\bibitem[{{Torres} {et~al.}(2010){Torres}, {Andersen}, \&
  {Gim{\'e}nez}}]{Torres:2010}
{Torres}, G., {Andersen}, J., \& {Gim{\'e}nez}, A. 2010, \aapr, 18, 67,
  \dodoi{10.1007/s00159-009-0025-1}

\bibitem[{{Twicken} {et~al.}(2018){Twicken}, {Catanzarite}, {Clarke},
  {Girouard}, {Jenkins}, {Klaus}, {Li}, {McCauliff}, {Seader}, {Tenenbaum},
  {Wohler}, {Bryson}, {Burke}, {Caldwell}, {Haas}, {Henze}, \&
  {Sanderfer}}]{Twicken2018}
{Twicken}, J.~D., {Catanzarite}, J.~H., {Clarke}, B.~D., {et~al.} 2018, \pasp,
  130, 064502, \dodoi{10.1088/1538-3873/aab694}

\bibitem[{{Valenti} \& {Fischer}(2005)}]{valenti2005}
{Valenti}, J.~A., \& {Fischer}, D.~A. 2005, \apjs, 159, 141,
  \dodoi{10.1086/430500}

\bibitem[{{van Biesbroeck}(1927)}]{1927vanBiesbroeck}
{van Biesbroeck}, G. 1927, Publications of the Yerkes Observatory, 5, 1.vii

\bibitem[{Vaughan {et~al.}(1978)Vaughan, Preston, \& Wilson}]{vaughan1978flux}
Vaughan, A.~H., Preston, G.~W., \& Wilson, O.~C. 1978, \pasp, 90, 267

\bibitem[{{Veras}(2016)}]{veras16}
{Veras}, D. 2016, Royal Society Open Science, 3, 150571,
  \dodoi{10.1098/rsos.150571}

\bibitem[{{Vick} {et~al.}(2023){Vick}, {Su}, \& {Lai}}]{2023Vick}
{Vick}, M., {Su}, Y., \& {Lai}, D. 2023, \apjl, 943, L13,
  \dodoi{10.3847/2041-8213/acaea6}

\bibitem[{Vogt \& Donald~Penrod(1988)}]{vogt1988hires}
Vogt, S.~S., \& Donald~Penrod, G. 1988, in Instrumentation for Ground-Based
  Optical Astronomy (Springer), 68--103

\bibitem[{{Vogt} {et~al.}(1994){Vogt}, {Allen}, {Bigelow}, {Bresee}, {Brown},
  {Cantrall}, {Conrad}, {Couture}, {Delaney}, {Epps}, {Hilyard}, {Hilyard},
  {Horn}, {Jern}, {Kanto}, {Keane}, {Kibrick}, {Lewis}, {Osborne},
  {Pardeilhan}, {Pfister}, {Ricketts}, {Robinson}, {Stover}, {Tucker}, {Ward},
  \& {Wei}}]{vogt1994}
{Vogt}, S.~S., {Allen}, S.~L., {Bigelow}, B.~C., {et~al.} 1994, in Society of
  Photo-Optical Instrumentation Engineers (SPIE) Conference Series, Vol. 2198,
  Instrumentation in Astronomy VIII, ed. D.~L. {Crawford} \& E.~R. {Craine},
  362, \dodoi{10.1117/12.176725}

\bibitem[{{Vousden} {et~al.}(2016){Vousden}, {Farr}, \& {Mandel}}]{2016Vousden}
{Vousden}, W.~D., {Farr}, W.~M., \& {Mandel}, I. 2016, \mnras, 455, 1919,
  \dodoi{10.1093/mnras/stv2422}

\bibitem[{Walt {et~al.}(2011)Walt, Colbert, \& Varoquaux}]{numpy}
Walt, S. v.~d., Colbert, S.~C., \& Varoquaux, G. 2011, Computing in science \&
  engineering, 13, 22

\bibitem[{{Wang} {et~al.}(2015{\natexlab{a}}){Wang}, {Fischer}, {Xie}, \&
  {Ciardi}}]{2015Wang}
{Wang}, J., {Fischer}, D.~A., {Xie}, J.-W., \& {Ciardi}, D.~R.
  2015{\natexlab{a}}, \apj, 813, 130, \dodoi{10.1088/0004-637X/813/2/130}

\bibitem[{{Wang} {et~al.}(2015{\natexlab{b}}){Wang}, {Fischer}, {Xie}, \&
  {Ciardi}}]{2015cWang}
---. 2015{\natexlab{b}}, \apj, 813, 130, \dodoi{10.1088/0004-637X/813/2/130}

\bibitem[{{Wei} {et~al.}(2021){Wei}, {Naoz}, {Faridani}, \& {Farr}}]{Wei+21}
{Wei}, L., {Naoz}, S., {Faridani}, T., \& {Farr}, W.~M. 2021, \apj, 923, 118,
  \dodoi{10.3847/1538-4357/ac2c70}

\bibitem[{{Winter} {et~al.}(2020){Winter}, {Kruijssen}, {Longmore}, \&
  {Chevance}}]{2020Winter}
{Winter}, A.~J., {Kruijssen}, J.~M.~D., {Longmore}, S.~N., \& {Chevance}, M.
  2020, \nat, 586, 528, \dodoi{10.1038/s41586-020-2800-0}

\bibitem[{{Wizinowich} {et~al.}(2000){Wizinowich}, {Acton}, {Shelton},
  {Stomski}, {Gathright}, {Ho}, {Lupton}, {Tsubota}, {Lai}, {Max}, {Brase},
  {An}, {Avicola}, {Olivier}, {Gavel}, {Macintosh}, {Ghez}, \&
  {Larkin}}]{2000Wizinowich}
{Wizinowich}, P., {Acton}, D.~S., {Shelton}, C., {et~al.} 2000, \pasp, 112,
  315, \dodoi{10.1086/316543}

\bibitem[{{Worley}(1972)}]{1972Worley}
{Worley}, C.~E. 1972, Publications of the U.S. Naval Observatory Second Series,
  22, 29

\bibitem[{{Worley}(1989)}]{1989Worley}
---. 1989, Publications of the U.S. Naval Observatory Second Series, 25, 1

\bibitem[{{Yee} {et~al.}(2017){Yee}, {Petigura}, \& {von Braun}}]{Yee2017}
{Yee}, S.~W., {Petigura}, E.~A., \& {von Braun}, K. 2017, \apj, 836, 77,
  \dodoi{10.3847/1538-4357/836/1/77}

\bibitem[{{Zagaria} {et~al.}(2022){Zagaria}, {Clarke}, {Rosotti}, \&
  {Manara}}]{2022Zagaria}
{Zagaria}, F., {Clarke}, C.~J., {Rosotti}, G.~P., \& {Manara}, C.~F. 2022,
  \mnras, 512, 3538, \dodoi{10.1093/mnras/stac621}

\bibitem[{{Ziegler} {et~al.}(2021){Ziegler}, {Tokovinin}, {Latiolais},
  {Brice{\~n}o}, {Law}, \& {Mann}}]{2021Ziegler}
{Ziegler}, C., {Tokovinin}, A., {Latiolais}, M., {et~al.} 2021, \aj, 162, 192,
  \dodoi{10.3847/1538-3881/ac17f6}

\bibitem[{{Ziegler} {et~al.}(2018){Ziegler}, {Law}, {Baranec}, {Howard},
  {Morton}, {Riddle}, {Duev}, {Salama}, {Jensen-Clem}, \&
  {Kulkarni}}]{2018Ziegler}
{Ziegler}, C., {Law}, N.~M., {Baranec}, C., {et~al.} 2018, \aj, 156, 83,
  \dodoi{10.3847/1538-3881/aace59}

\end{thebibliography}
\bibliographystyle{aasjournal}

\end{document}